\documentclass[twocolumn,english,aps,prb,twocolum,superscriptaddress,bibnotes,amsmath,amssymb,floatfix]{revtex4-2}
\usepackage[colorlinks=true,citecolor=blue,linkcolor=magenta]{hyperref}
\usepackage{xcolor}
\usepackage[markup=nocolor, authormarkupposition=left]{changes} 
\usepackage{soul}
\usepackage[utf8]{inputenc}
\usepackage[english]{babel}
\usepackage{amsmath,amsfonts,amssymb}
\usepackage[T1]{fontenc}
\usepackage{url}

\usepackage{amsmath}
\usepackage{amsfonts}
\usepackage{amssymb}

\usepackage{epstopdf}
\usepackage{graphicx}
\graphicspath{{./Figures/}}

\usepackage{changes}

\begin{document}
\title{Kramers--Kronig causality in integrated photonics:\\
The spectral tension between ultraviolet transition and mid-infrared absorption}

\author{Yue Hu}
\thanks{These authors contributed equally to this work.}
\affiliation{Southern University of Science and Technology, Shenzhen 518055, China}
\affiliation{International Quantum Academy and Shenzhen Futian SUSTech Institute for Quantum Technology and Engineering, Shenzhen 518048, China}

\author{Zhenyuan Shang}
\thanks{These authors contributed equally to this work.}
\affiliation{Southern University of Science and Technology, Shenzhen 518055, China}
\affiliation{International Quantum Academy and Shenzhen Futian SUSTech Institute for Quantum Technology and Engineering, Shenzhen 518048, China}

\author{Chenxi Zhang}
\affiliation{International Quantum Academy and Shenzhen Futian SUSTech Institute for Quantum Technology and Engineering, Shenzhen 518048, China}
\affiliation{College of Physics and Optoelectronic Engineering, Shenzhen University, Shenzhen 518060, China}

\author{Yuanjie Ning}
\affiliation{Shanghai Key Laboratory of High Temperature Superconductors, Institute for Quantum Science and Technology, Department of Physics, Shanghai University, Shanghai 200444, China}

\author{Weiqin Zheng}
\affiliation{International Quantum Academy and Shenzhen Futian SUSTech Institute for Quantum Technology and Engineering, Shenzhen 518048, China}
\affiliation{Institute of Advanced Photonics Technology, School of Information Engineering, Guangdong University of Technology, Guangzhou 510006, China}

\author{Dengke Chen}
\affiliation{Southern University of Science and Technology, Shenzhen 518055, China}
\affiliation{International Quantum Academy and Shenzhen Futian SUSTech Institute for Quantum Technology and Engineering, Shenzhen 518048, China}

\author{Sanli Huang}
\affiliation{International Quantum Academy and Shenzhen Futian SUSTech Institute for Quantum Technology and Engineering, Shenzhen 518048, China}
\affiliation{Hefei National Laboratory, University of Science and Technology of China, Hefei 230088, China}

\author{Baoqi Shi}
\affiliation{International Quantum Academy and Shenzhen Futian SUSTech Institute for Quantum Technology and Engineering, Shenzhen 518048, China}
\affiliation{Hefei National Laboratory, University of Science and Technology of China, Hefei 230088, China}

\author{Zeying Zhong}
\affiliation{Southern University of Science and Technology, Shenzhen 518055, China}
\affiliation{International Quantum Academy and Shenzhen Futian SUSTech Institute for Quantum Technology and Engineering, Shenzhen 518048, China}

\author{Hao Tan}
\affiliation{International Quantum Academy and Shenzhen Futian SUSTech Institute for Quantum Technology and Engineering, Shenzhen 518048, China}

\author{Wei Sun}
\affiliation{International Quantum Academy and Shenzhen Futian SUSTech Institute for Quantum Technology and Engineering, Shenzhen 518048, China}

\author{Yi-Han Luo}
\affiliation{International Quantum Academy and Shenzhen Futian SUSTech Institute for Quantum Technology and Engineering, Shenzhen 518048, China}

\author{Xinmao Yin}
\affiliation{Shanghai Key Laboratory of High Temperature Superconductors, Institute for Quantum Science and Technology, Department of Physics, Shanghai University, Shanghai 200444, China}

\author{Zhi-Chuan Niu}
\affiliation{International Quantum Academy and Shenzhen Futian SUSTech Institute for Quantum Technology and Engineering, Shenzhen 518048, China}
\affiliation{State Key Laboratory of Optoelectronic Materials and Devices, Institute of Semiconductors, Chinese Academy of Sciences, Beijing 100083, China}
\affiliation{Center of Materials Science and Optoelectronics Engineering, University of Chinese Academy of Sciences, Beijing 100049, China}

\author{Junqiu Liu}
\email[]{liujq@iqasz.cn}
\affiliation{International Quantum Academy and Shenzhen Futian SUSTech Institute for Quantum Technology and Engineering, Shenzhen 518048, China}
\affiliation{Hefei National Laboratory, University of Science and Technology of China, Hefei 230088, China}

\maketitle

\noindent\textbf{Dispersion engineering via geometric confinement is essential to integrated photonics, enabling phenomena such as soliton microcombs, supercontinua, parametric oscillators, and entangled photons.
However, prevailing methodologies rely on semi-empirical Sellmeier models that assume idealized material purity, neglecting the pronounced dispersion shifts induced by residual impurities like hydrogen-related bonds.
Here, we demonstrate that these residual bonds fundamentally alter the dispersion landscape spanning from the ultraviolet (UV) to the mid-infrared (MIR) spectra. 
Specifically, they introduce MIR vibrational absorption while simultaneously modifying UV electronic transition, shifting the bandgap and UV pole.
We show that the spectral tension between these UV and MIR modifications dictates the group velocity dispersion from the visible to the near-infrared (NIR) via the Kramers--Kronig causality.  
We experimentally validate this phenomenon through systematic characterization of broadband loss and dispersion in ultralow-loss silicon nitride photonic integrated circuits.
By rigorously incorporating these effects, we bridge the gap between empirical fitting and predictive physical modelling.
Our study resolves long-standing discrepancies in dispersion engineering, providing precision control essential for next-generation integrated photonics.
}

\begin{figure*}[t!]
\centering
\includegraphics{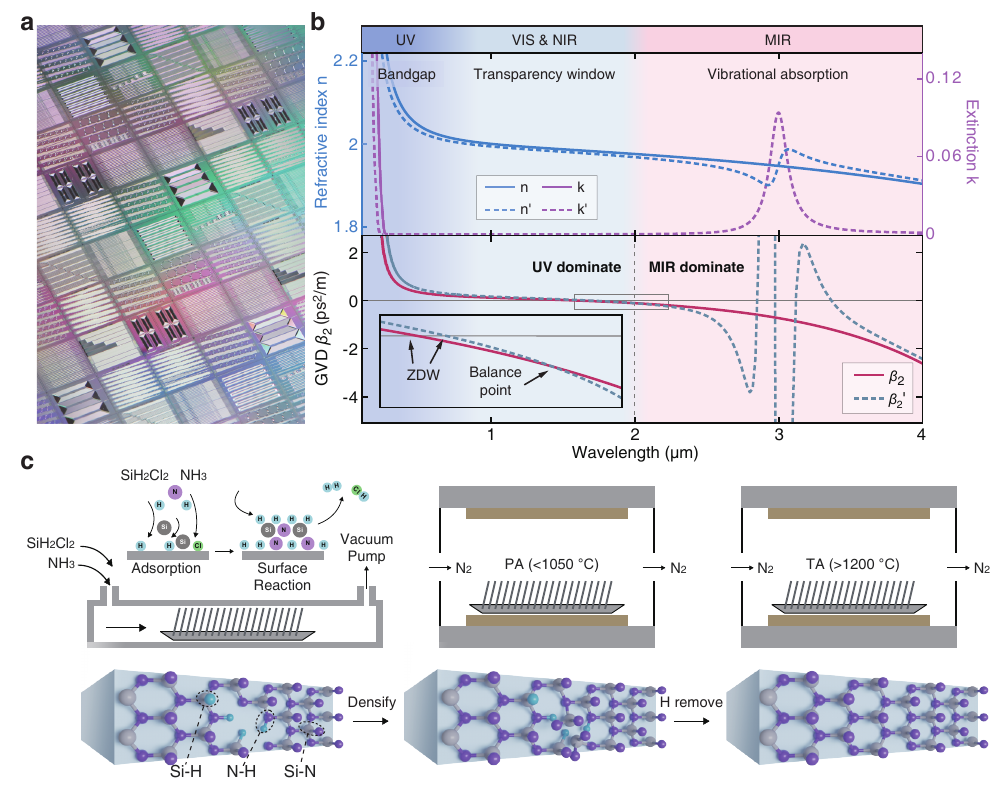}
\caption{
\textbf{Origin and non-local spectral impact of hydrogen impurities in silicon nitride}. 
\textbf{a}. 
Photograph of Si$_3$N$_4$ chips on a 6-inch wafer.
\textbf{b}. 
Combined spectral landscape illustrating the refractive index, extinction and GVD of Si$_3$N$_4$ with ($n'$, $k'$, $\beta_2'$, dashed curves) and without ($n$, $k$, $\beta_2$, solid curves) the N--H  bonds. 
Residual N--H  bonds induce blue-shifted bandgap edge in the UV and an absorption peak in the MIR. 
Linked by the KK relations, these variations in $k$ shift $n$, which in turn alters $\beta_2$ and shifts the zero-dispersion wavelength (ZDW).
The discrepancy between these curves illustrates the spectral non-locality of the KK relations. 
The opposing dispersion contributions from the UV and MIR domains culminate in a dispersion balance point near 2 $\mu$m, where the GVD becomes insensitive to hydrogen content. 
\textbf{c}.
Schematic of the LPCVD process and structural evolution of the Si$_3$N$_4$ film matrix. 
Precursor residues (NH$_3$ and SiH$_2$Cl$_2$) introduce extensive N--H  and Si--H bonds in the film.
Partial annealing (PA, $<1050^\circ$C) desifies the film but leaves significant H content, whereas thorough annealing (TA, $>1200^\circ$C) effectively eliminates these bonds to reach the idealized material state. 
}
\label{Fig:1}
\end{figure*}

Integrated Kerr-nonlinear photonics \cite{Moss:13, Gaeta:19} has emerged as a vital frontier for advanced signal processing, frequency synthesis, and quantum information.
High-refractive-contrast waveguides---predominantly stoichiometric silicon nitride (Si$_3$N$_4$) due to its ultralow optical loss, broad transparency window, and moderate Kerr nonlinearity---enable tight optical confinement sufficient to trigger nonlinear interactions at sub-milliwatt power level \cite{Xuan:16, Ji:17, Liu:18a}.
Transformative applications, including soliton microcombs \cite{Brasch:16, Li:17, Pfeiffer:17, Yu:19}, supercontinua \cite{Halir:12, Porcel:17a, Epping:15a, Guo:20}, optical parametric oscillators \cite{Levy:10, Li:16, Lu:19b, Lu:20}, and bright entangled photon pairs \cite{Lu:19a, Fan:23, ChenR:24, Ramelow:15}, rely on efficient frequency conversion across octave-spanning ranges (e.g. 1000--2000 nm). 

Central to these advances is dispersion engineering, which tailors group velocity dispersion (GVD) of waveguides for broadband phase matching (PM) \cite{Turner:06, Okawachi:14}. 
Unlike bulk optics, integrated waveguides allow the effective refractive index to be spectrally shaped by manipulating the waveguide geometry, specifically its height and width.  This induces geometry dispersion $D_\text{wg}$ that compensates for intrinsic material dispersion $D_\text{mat}$. 
While geometry can be controlled with nanometer precision for $D_\text{wg}$, successful engineering requires an accurate physical model of $D_\text{mat}$.  

Fundamentally, $D_\text{mat}$ is governed by spectral absorption via the Kramers--Kronig (KK) relations: \cite{Kramers:27, Kronig:26, Lucarini:05} 
\begin{equation}
\begin{aligned} 
n(\omega) &= 1 + \frac{2}{\pi} \mathcal{P} \int_{0}^{\infty} \frac{\omega' \kappa(\omega')}{\omega'^2 - \omega^2} \text{d}\omega' \\
\kappa(\omega) &= -\frac{2\omega}{\pi} \mathcal{P} \int_{0}^{\infty} \frac{n(\omega') - 1}{\omega'^2 - \omega^2} \text{d}\omega' \\ 
\end{aligned}
\label{Eq. kk}
\end{equation}
where $\omega$ is the angular frequency of interest, 
$\omega'$ is the integration variable over the entire spectrum, 
and $\mathcal{P}$ denotes the Cauchy principal value.
These relations establish a rigorous consequence of causality \cite{Toll:56, Altarelli:72}, linking spectral absorption (imaginary part) to the refractive index (real part)---each part of the susceptibility is Hilbert transforms of the other---of the complex refractive index $\tilde{n}(\omega) = n(\omega) + i\kappa(\omega)$ across the entire spectrum. 
They are practical design constraints and powerful analytical tools for many integrated devices, e.g. silicon electro-optic modulators \cite{Soref:87, Xu:05, LiSY:25}, linewidth broadening in semiconductor lasers \cite{Henry:82, Vahala:83, Osinski:87}, electromagnetically induced transparency in coupled microresonators \cite{Smith:04, Xu:06, Totsuka:07, Huet:16}, and Kramers--Kronig receivers \cite{Mecozzi:16}. 

Despite their importance, the implications of KK non-locality---where \textit{local} absorption affects the \textit{global} refractive index---are often overlooked in standard dispersion modelling.
Conventional semi-empirical models (e.g. Sellmeier equations) typically prioritize UV electronic resonances while neglecting lower-frequency contributions and impurity-induced effects. 
For Si$_3$N$_4$, it is generally presumed that the monotonic index increase in the NIR is governed exclusively by the UV bandgap (approximately 5.0 eV). 
Figure~\ref{Fig:1}b presents the widely cited model of Si$_3$N$_4$ refractive index $n$ (blue solid curve) \cite{Liu:20b, Luke:15}, extinction coefficient $k$ (purple solid curve) \cite{Corato-Zanarella:24}, and GVD parameter $\beta_2$ (red solid curve, calculated based on $n$), from UV to MIR. 
Since the UV bandgap exhibits substantial absorption (large $\kappa$), it acts as the primary contributor to Eq. \ref{Eq. kk}. 
Consequently, the monotonic increase in $n(\omega)$ as $\omega$ approaches the bandgap edge constitutes the material's dispersion in the NIR.

Here, we reveal that molecular bond resonances---specifically pervasive N--H impurities---exert a significant and highly distinct non-local influence on NIR dispersion. 
This influence manifests through a dual mechanism: MIR absorption and UV bandgap distortion. 
We demonstrate that this critical, yet long overlooked factor can no longer be ignored in dispersion engineering.
Our study identifies, characterizes, and aims to rectify this historical oversight.

\begin{figure*}[t!]
\centering
\includegraphics{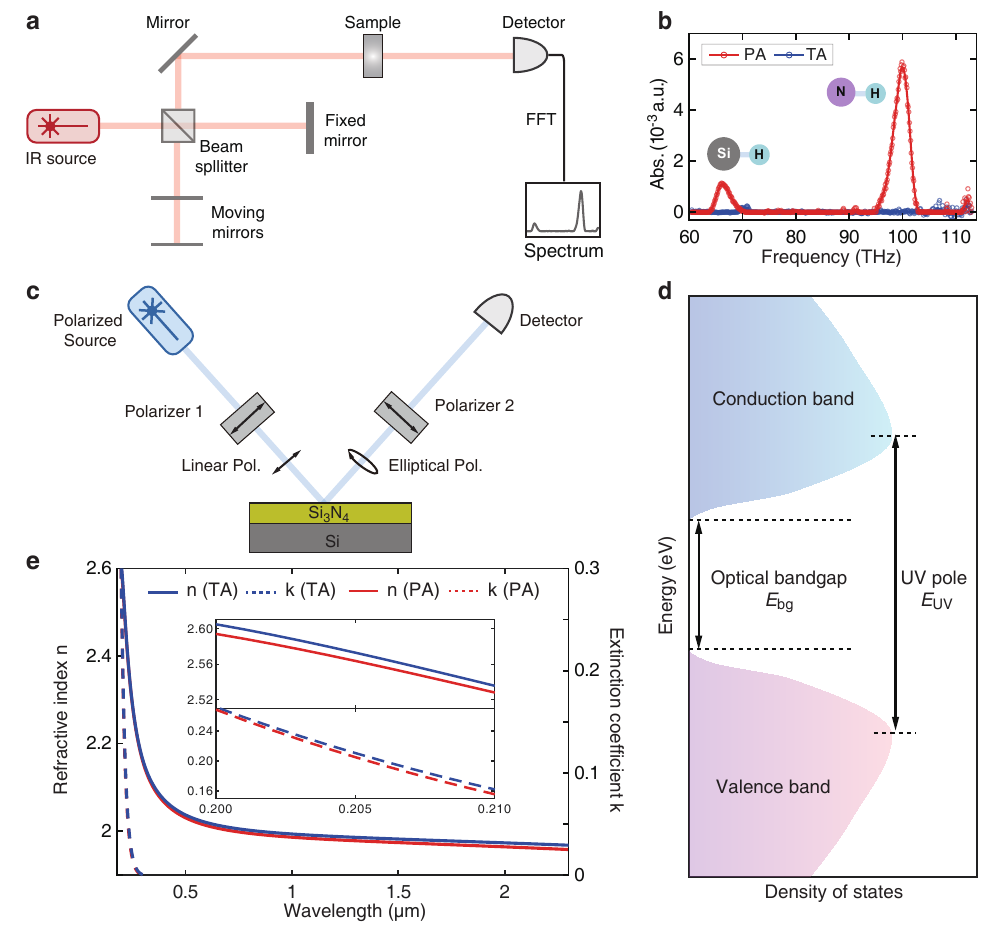}
\caption{
\textbf{Characterization of silicon nitride films}. 
\textbf{a, b}. 
Schematic of FTIR spectroscopy and the resulting measured spectra for blank Si$_3$N$_4$ films. 
While PA (red data) leaves significant H content, TA (blue data) effectively eliminates the N--H and Si--H fundamental stretching vibrational modes near 3.0 and 4.6 $\mu$m, respectively. 
\textbf{c}. 
Schematic of variable-angle spectroscopic ellipsometry to probe the material's dielectric response across the UV--NIR spectrum. 
\textbf{d}. 
Conceptual diagram of the Tauc-Lorentz model.
The optical bandgap energy $E_{\text{bg}}$ represents the minimum energy threshold for interband electronic transitions, while the UV pole $E_{\text{UV}}$ captures the dominant high-energy transition corresponding to the maximum density of states.
\textbf{e}. 
Extracted wavelength-dependent refractive index $n$ and extinction coefficient $k$ for Si$_3$N$_4$ in the PA (red) and TA (blue) states.
The inset highlights the higher $n$ and $k$ values for the TA state in 200--210 nm. 
}
\label{Fig:2}
\end{figure*}

\noindent \textbf{Film characterization}.
To quantify the impact of hydrogen (H) impurities, we first trace their origin to the standard low-pressure chemical vapor deposition (LPCVD) process.
Governed by the reaction $3\text{SiH}_2\text{Cl}_2+4\text{NH}_3 \rightarrow \text{Si}_3\text{N}_4+6\text{HCl}+6\text{H}_2$, the precursor residues intrinsically introduce extensive H content into the matrix, forming N--H  and Si--H bonds within the Si$_3$N$_4$ \cite{Mao:08, Bauters:11, Frigg:19, Wu:21}, as illustrated in Fig. \ref{Fig:1}c.
These bonds introduce pronounced MIR absorption, specifically near 3.0 and 4.6 $\mu$m for the fundamental N--H and Si--H stretching vibrations ($\Delta\nu=1$) \cite{Lanford:78, Luke:15, Bugaev:12}, respectively. 
To isolate the impact of these bonds, we track a film subjected to sequential ``partial annealing'' (PA, $<1050^\circ$C) and ``thorough annealing'' (TA, $>1200^\circ$C). 
Note that $1050^\circ$C represents the thermal limit for standard quartz-based LPCVD configurations. 

We employ Fourier-transform infrared spectroscopy (FTIR) to examine the presence and elimination of these H impurities. 
As illustrated in Fig. \ref{Fig:2}a, FTIR operates by transmitting a broadband infrared beam through the sample and applying a Fourier transform to the resulting interferogram to extract a continuous spectrum.
Because molecular bonds behave as quantum oscillators absorbing at distinct resonant frequencies, this technique directly maps optical absorption peaks to the concentration of N--H  and Si--H bonds within the matrix.
Figure \ref{Fig:2}b shows the measured FTIR spectra, confirming that while TA removes nearly all H bonds, PA leaves significant residuals, with H abundance remaining nearly identical to that of an unannealed film (see Supplementary Information Note 1). 
Given that N--H  absorption is substantially stronger and occurs at a higher frequency than Si--H absorption, we focus our subsequent analysis to the impact of N--H  bonds on the NIR dispersion.

Next, we characterize the complex refractive index of the Si$_3$N$_4$ films from 193 to 2300 nm using variable-angle spectroscopic ellipsometry (VASE) to investigate H-induced modification to UV electronic transition. 
Figure \ref{Fig:2}c illustrates the schematic of VASE, which probes the material's dielectric response by quantifying the amplitude ratio and relative phase shift between the $p$- and $s$-polarized light components upon reflection.
The measured spectra are parameterized using the Tauc-Lorentz oscillator model that describes interband transitions via a main oscillator (defined by its optical bandgap $E_{\text{bg}}$ and center energy $E_0$) combined with a high-energy UV pole (of energy $E_{\text{UV}}$).
Figure \ref{Fig:2}d conceptually illustrates that, $E_{\text{bg}}$ defines the minimum energy threshold for interband electronic transition, whereas $E_{\text{UV}}$ captures the dominant high-energy transition corresponding to the maximum density of states.
The extracted model parameters of our Si$_3$N$_4$ are detailed in Supplementary Information Note 2.

In the initial H-rich PA state, the film possesses a wider bandgap ($E_{\text{bg}} \approx 4.30$ eV, $E_0 \approx 6.75$ eV) compared to its H-free TA state ($E_{\text{bg}} \approx 4.24$ eV, $E_0 \approx 6.71$ eV).
This H-induced \textbf{bandgap blue-shift} occurs because N--H bonds possess higher bonding energies than Si--N bonds, effectively eliminating localized states near the band edges \cite{Robertson:1991, Lanford:1978, Karcher:1984}. 
Conversely, the PA state features a lower UV-pole energy ($E_{\text{UV}} \approx 10.04$ eV) than the TA state ($E_{\text{UV}} \approx 10.05$ eV). 
Because the UV pole reflects the average energy of deep interband transition across the continuous bulk network \cite{Robertson:1991}, this H-induced \textbf{pole red-shift} physically manifests structural relaxation and lower atomic density caused by H atoms acting as network terminators. 
Combined with a reduction in overall oscillator amplitudes---due to the substitution of highly polarizable Si--N bonds with less polarizable H bonds---these fundamental structural shifts profoundly modulate the broadband dielectric response. 
Consequently, as depicted in Fig. \ref{Fig:2}e, the H-free TA film exhibits a notably higher refractive index $n$ and extinction coefficient $k$ across the UV--NIR spectrum compared to its initial PA state.

While typical models---such as the solid curves in Fig. \ref{Fig:1}b---account for the fundamental Si--N stretching vibrations near 11 $\mu$m, they neglect H residuals and their associated non-local dispersion effects. 
Figure \ref{Fig:1}b exemplifies that the residual N--H  bonds introduce a MIR absorption peak near 3.0 $\mu$m ($k'$, purple dashed curve), which in turn modifies the NIR refractive index ($n'$, blue dashed curve).
This modification reduces the GVD ($\beta_2'$, cyan dashed curve), driving a blue-shift of the zero-dispersion wavelength (ZDW).

Crucially, this MIR-induced perturbation represents only a partial physical picture.
As established by our ellipsometry analysis, the presence of N--H  bonds concurrently modulates the high-energy UV electronic transition. 
These structural changes induce a blue-shift of the UV absorption edge, as further illustrated in Fig. \ref{Fig:1}b.
Through the non-local nature of the KK relations, this UV modification globally shifts $n'$ and exerts a positive contribution to the NIR GVD, competing with the negative GVD contribution originating from the MIR N--H  absorption.
The interplay of these opposing dispersion influences from the UV and MIR domains culminates in a dispersion balance point near 2 $\mu$m. 
Detailed calculation for $n'$ and $\beta_2'$ involving both MIR absorption and UV bandgap distortion are provided in Supplementary Information Note 3.

\begin{figure*}[t!]
\centering
\includegraphics{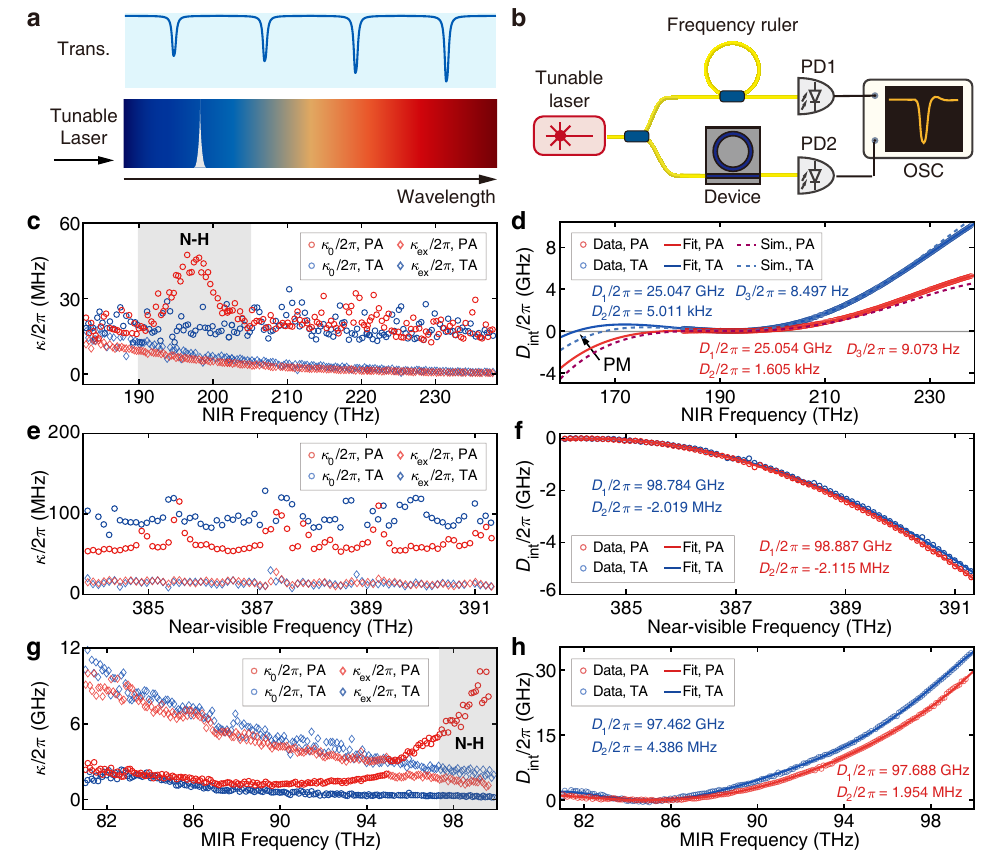}
\caption{
\textbf{Multi-band characterization of dispersion and loss in silicon nitride microresonators}. 
\textbf{a, b}. 
Schematic principle and experimental configuration of our VSAs.   
Tunable lasers sweep across the NIR (182.9--237.9 THz), near-visible(383.8--391.3 THz), and MIR (80.8--99.9 THz) bands to probe individual resonances. 
The instantaneous frequency is calibrated by a frequency ruler, and the device transmission spectrum is recorded synchronously.
PD, photodetector. 
OSC, oscilloscope. 
\textbf{c, e, g}. 
Measured $\kappa_0/2\pi$ and $\kappa_{ex}/2\pi$ values for microresonators in the NIR (c, 25-GHz-FSR), near-visible (e, 99-GHz-FSR), and MIR (g, 97-GHz-FSR) bands.
For visual clarity, only every third measured data point is displayed in c, d. 
Absorption features induced by N--H  bonds are gray-shaded. 
\textbf{d, f, h}. 
Measured (circles) and fitted (solid curves) $D_\text{int}$ profiles of the corresponding microresonators in c, e, g.
Panel d additionally includes simulated dispersion curves.
The fitted dispersion parameters $D_1$, $D_2$ and $D_3$ are denoted in each panel.
Throughout all data panels, the red and blue datasets represent the Si$_3$N$_4$ microresonators in the PA and TA states, respectively. 
}
\label{Fig:3}
\end{figure*}

\begin{figure*}[t!]
\centering
\includegraphics{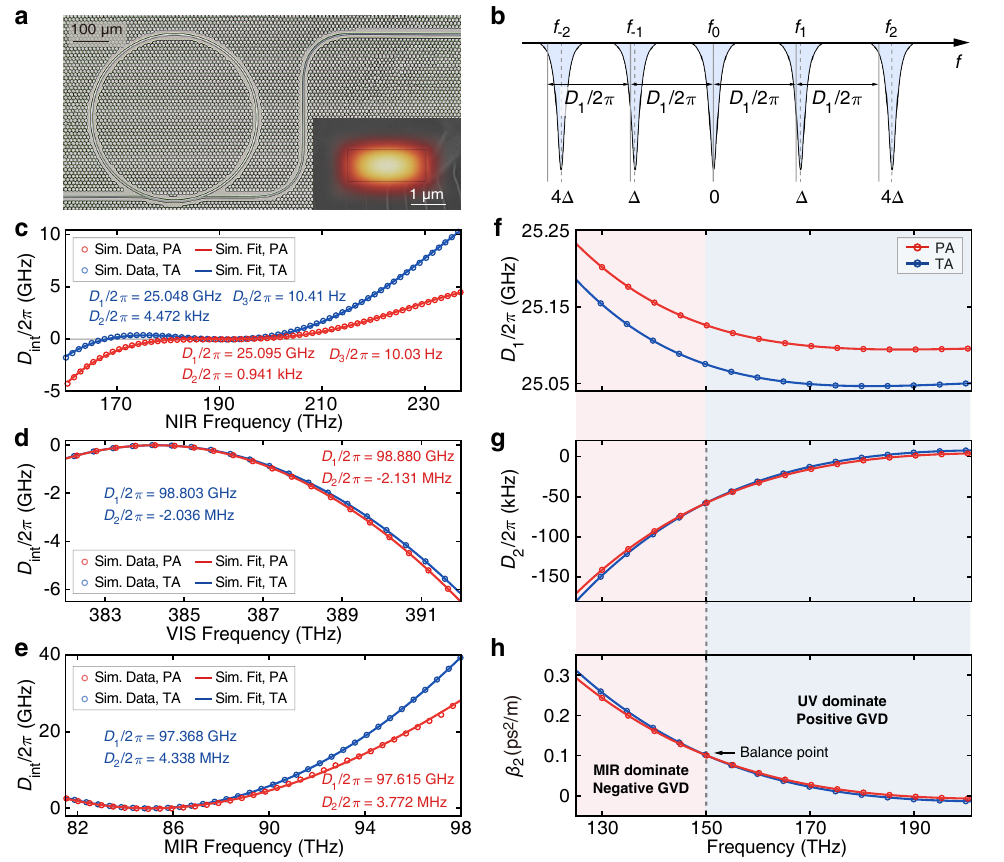}
\caption{
\textbf{Physical modelling and numerical simulation of the dispersion balance point and spectral tension}. 
\textbf{a}.
Optical micrograph of a Si$_3$N$_4$ microresonator.
The inset shows a cross-sectional SEM image of the waveguide core with the simulated fundamental TE$_{00}$ mode, demonstrating tight optical confinement. 
\textbf{b}.
Frequency-domain representation of microresonator dispersion.
Solid lines mark an ideal equidistant frequency grid with $D_1/2\pi$ spacing centered at $f_0$.
Actual resonances (dashed lines) at $f_{\pm1}$ and $f_{\pm2}$ deviate from this uniform grid. 
Considering only $D_2$ and neglecting higher-order dispersion terms, the detuning increases quadratically with the relative mode index $\mu$, giving an offset $\Delta\cdot\mu^2$.
\textbf{c, d, e}.
Simulated $D_\text{int}/2\pi$ in the NIR (c), near-visible (d) and MIR (e).
The simulations incorporate exact waveguide geometries and the comprehensive refractive index model, showing excellent agreement with experimental data for both PA (red) and TA (blue) states. 
\textbf{f, g, h}. 
Prediction of the dispersion balance point. 
Simulated $D_1/2\pi$ (f), $D_2/2\pi$ (g), and $\beta_2$ (h) reveal the interplay between UV and MIR perturbations. 
While the MIR N--H absorption imparts a negative dispersion pull, the blue-shifted UV bandgap and red-shifted UV pole exert a competing positive dispersion contribution.
These opposing influences culminate in a dispersion balance point near 150 THz (2 $\mu$m), where the GVD curves of both states intersect and become insensitive to H content. 
Throughout all data panels, the red and blue datasets represent the Si$_3$N$_4$ microresonators in the PA and TA states, respectively. 
}
\label{Fig:4}
\end{figure*}

\noindent \textbf{Device characterization}.   
We fabricate integrated Si$_3$N$_4$ microresonators using a deep-UV subtractive process based on 6-inch wafers \cite{Ye:23, Sun:25}. 
Figure \ref{Fig:1}a presents an image of Si$_3$N$_4$ chips on the wafer. 
Fabrication details are found in Supplementary Information Note 4. 
We first characterize the microresonators whose Si$_3$N$_4$ film had undergone PA. 
Using vector spectrum analyzers (VSA) \cite{Luo:24, Shi:25, Shi:26}, we characterize microresonator dispersion and optical loss (resonance linewidth) across the NIR (182.9--237.9 THz, 1260--1640 nm) \cite{Luo:24}, near-visible (383.8--391.3 THz, 766--781 nm) \cite{Shi:25} and MIR (80.8--99.9 THz, 3001--3711 nm) bands \cite{Shi:26}. 
Figure \ref{Fig:3}a,b depicts the principle and setup of our VSAs utilizing narrow-linewidth, widely tunable, mode-hop-free, continuous-wave (CW) lasers. 
By referencing these lasers to atomic transitions and calibration fiber cavities, our VSAs achieve sub-megahertz frequency resolution and megahertz accuracy \cite{Luo:24, Shi:25, Shi:26}. 

Light from the VSAs is coupled into and out of the Si$_3$N$_4$ chips using optimized lensed fibers \cite{Chen:26}, and the resulting microresonator transmission spectra are recorded. 
Experimental details are provided in Supplementary Information Notes 5--7.
The VSAs probe each resonance within the respective spectral ranges, enabling the measurement of integrated microresonator dispersion, defined as: 
\begin{equation}
D_\text{int}(\mu)=\omega_\mu-\omega_0-D_1\mu=\sum_{n=2}^{\cdots}\frac{D_n\mu^n}{n!}
\label{Eq.Dint}
\end{equation}
where $\omega_{\mu}/2\pi$ is the $\mu$-th resonance frequency relative to the reference resonance frequency $\omega_0/2\pi$, 
$D_1/2\pi$ represents the free spectral range (FSR), 
$D_2/2\pi=-v_gD_1^2\beta_2$ describes the GVD, with $v_g$ being the group velocity.   
Higher-order dispersion terms are captured by $D_3$, $D_4$ etc.  
Furthermore, each resonance is fitted with a mode-split Lorentzian profile \cite{Li:13} to extract the intrinsic loss $\kappa_0/2\pi$, external coupling strength $\kappa_\text{ex}/2\pi$, and the full (loaded) linewidth $\kappa/2\pi=(\kappa_0+\kappa_\text{ex})/2\pi$. 

Figure \ref{Fig:4}a presents an optical micrograph and a cross-sectional SEM image of a fabricated device, while Fig. \ref{Fig:4}b illustrates the physical picture of $D_\text{int}$. 
In an ideal microresonator with zero dispersion ($D_\text{int}=0$), the optical group index remains constant across the spectrum. 
This results in a uniform resonance grid spaced by a constant $D_1/2\pi$, as shown by Fig. \ref{Fig:4}b solid gray lines. 
In reality, material and geometry dispersion break this uniformity, causing resonances to shift away from this equidistant grid, as shown by Fig. \ref{Fig:4}b dashed gray lines.
This frequency offset for the $\mu$-th resonance is precisely described by $D_\text{int}(\mu)$.

Figure \ref{Fig:3}c red data present the measured $\kappa_0/2\pi$ and $\kappa_\text{ex}/2\pi$ in the NIR for a 25-GHz-FSR microresonator.
A distinct absorption peak of $\sim$30 MHz (corresponding to 5 dB/m loss) is observed near 197 THz (1520 nm).
This feature originates from the first overtone ($\Delta\nu=2$) and anharmonicity of the fundamental N--H  stretching vibration located near 3.0 $\mu$m \cite{Mao:08, Bauters:11}.
Resolving such a weak absorption signature is only possible in Si$_3$N$_4$ waveguides with ultralow background scattering loss.
The corresponding measured $D_\text{int}$ profile in the NIR is plotted by Fig. \ref{Fig:3}d red data, with the parameters $D_1$ and $D_2$ fitted and extracted using Eq. \ref{Eq.Dint}.
The device exhibits no PM point---i.e., $D_\text{int}=0$---in this band.
Crucially, the measured $D_\text{int}$ profile (Fig. \ref{Fig:3}d red solid curve) agrees with our simulation that incorporates the effects of N--H  bonds (Fig. \ref{Fig:3}d red dashed curve).

To remove the N--H  bonds after PA, we perform additional TA on the same chip. 
Figure \ref{Fig:3}c blue data show that the 197-THz absorption peak completely vanishes after TA, confirming the exhaustive removal of N--H bonds. 
Notably, $\kappa_0/2\pi$ values away from 1520 nm remain unchanged, indicating that TA does not further reduce optical scattering loss. 
This suggests that while PA leaves abundant H content, it effectively optimizes the film's structural quality to a point that is difficult to improve further. 
Strikingly, upon the removal of H content, the measured $D_\text{int}$ profile shifts dramatically and reveals the designed PM point, as shown by Fig. \ref{Fig:3}d blue data. 
The measured $D_\text{int}$ profile (Fig. \ref{Fig:3}d blue solid curve) agrees with our simulation that excludes N--H  bonds (Fig. \ref{Fig:3}d blue dashed curve). 

Having established the impact of N--H  bonds on NIR dispersion, we turn to the near-visible regime to investigate the concurrent non-local effects.
Figure \ref{Fig:3}e red data presents the measured $\kappa_0/2\pi$ and $\kappa_\text{ex}/2\pi$ for a 99-GHz-FSR microresonator in the near-visible, with its corresponding $D_\text{int}$ profile plotted in Fig. \ref{Fig:3}f red data and curve.
Following TA, $\kappa_0/2\pi$ exhibits a global increase of $\sim$28 MHz, as shown by Fig. \ref{Fig:3}e blue data.
This increased loss is likely attributed to partial crystallization, thermal stress, or micro-cracking within the Si$_3$N$_4$ waveguides induced by the high-temperature ($>1200~^\circ\text{C}$) annealing process \cite{Henry:87}.
Furthermore, as shown in Fig. \ref{Fig:3}f blue data and curve, the $D_\text{int}$ profile is altered, exhibiting an increase in $D_2/2\pi$.
This shift is primarily driven by the modulation of the UV absorption edge following the elimination of the N--H  bonds.

While FTIR effectively characterizes blank films, it fails to characterize MIR transmission in micro- and nano-structures such as waveguides. 
To provide further evidence that the NIR and near-visible dispersion are indeed distorted by N--H  bonds, we characterize a 97-GHz-FSR Si$_3$N$_4$ microresonator in the MIR. 
Figure \ref{Fig:3}g compares the measured $\kappa_0/2\pi$ and $\kappa_\text{ex}/2\pi$ of the same chip before (red data) and after (blue data) TA. 
After PA and before TA, the rising $\kappa_0/2\pi$ (red circle) towards higher frequency provides clear evidence of N--H  bonds near 3.0 $\mu$m. 
These bonds are successfully removed by TA, as evidenced by the significantly reduced $\kappa_0/2\pi$. 
Note that now the rising $\kappa_0/2\pi$ towards lower frequency is due to mode delocalization into the absorptive SiO$_2$ cladding and the associated onset of strong SiO$_2$ multi-phonon absorption \cite{Soref:06b, Kitamura:07, Lin:17, Miller:17}. 
Meanwhile, the discontinuity of red data around 95 THz arises because the resonances undergo a transition from over-coupling ($\kappa_\text{ex}>\kappa_0$) to under-coupling ($\kappa_\text{ex}<\kappa_0$), where the fit precision is compromised when $\kappa_\text{ex}\approx\kappa_0$ (critical coupling) \cite{Cai:00, Li:13}. 
Figure \ref{Fig:3}h compares the measured $D_\text{int}$ profiles before (red data) and after (blue data) TA, revealing a prominent difference.
More characterization results in the NIR, near-visible and MIR of other chips are provided in Supplementary Information Notes 5--7. 

We further exclude geometric deformation as a potential cause for these shifts, by performing extensive SEM characterization of the Si$_3$N$_4$ waveguide core and reflectometry measurement of the film thickness. 
These investigations, detailed in Supplementary Information Note 8, confirm that the 
Si$_3$N$_4$ film thickness and waveguide geometry remain unchanged following the additional TA.  
Furthermore, the excellent alignment of $\kappa_\text{ex}/2\pi$ values before and after TA in Fig. \ref{Fig:3}c, e, g supports the conclusion that the waveguides are not deformed, as any structural change would have modified the evanescent coupling. 

\noindent \textbf{Dispersion model and simulation}.
To quantitatively validate our physical analysis---specifically that the NIR dispersion shift is caused by the concurrent modulation of the UV electronic transition and MIR vibrational absorption---we establish a physical framework for our Si$_3$N$_4$ microresonators and perform numerical dispersion simulations.
This effort translates our material-level refractive index model into macroscopic device properties.
Building upon the standard Sellmeier model for the TA state \cite{Luke:15, Liu:20b}, we develop a comprehensive refractive index model for the PA state, as: 
\begin{widetext}
\begin{equation}
\begin{split}
n^2(\lambda) = 1 &+ 1.750893106708352 \cdot \textcolor{blue}{\boldsymbol{A_1}} \cdot \frac{\lambda^2}{\lambda^2-(0.159333426051559-\textcolor{blue}{\boldsymbol{\delta\lambda_{\mathrm{bg}}}})^2} \\
&+ 1.222742709909494 \cdot \textcolor{blue}{\boldsymbol{A_2}} \cdot \frac{\lambda^2}{\lambda^2-(0.054765813233156+\textcolor{blue}{\boldsymbol{\delta\lambda_{\mathrm{pole}}}})^2} \\
&+ 2.584602943737831 \cdot \frac{\lambda^2}{\lambda^2-11.600661120564506^2} \\
&+ \textcolor{blue}{\boldsymbol{A_3}} \cdot \frac{\lambda^2}{\lambda^2-\textcolor{blue}{\boldsymbol{\lambda_{\mathrm{N-H}}^2}}}
\label{Eq. n_PA}
\end{split}
\end{equation}
\end{widetext}
Here, $A_1 <1$ and $A_2 <1$ serve as the amplitude reduction coefficients for the bandgap and UV pole, 
while $\delta\lambda_{\mathrm{bg}}$ and $\delta\lambda_{\mathrm{pole}}$ define their respective wavelength shifts.
The final term phenomenologically captures the KK contribution of the fundamental N--H bond absorption at $\lambda_{\mathrm{N-H}} = 3$ $\mu$m, with an extracted amplitude coefficient $A_3$. 
Therefore, Eq. \ref{Eq. n_PA} explicitly incorporates the structural shifts of the UV bandgap and UV pole, alongside the MIR absorption from N--H stretching vibration near 3 $\mu$m. 
Utilizing this comprehensive index model and actual waveguide cross-sectional geometries measured via SEM (see Methods), our finite-element-method (FEM) simulations successfully reproduce the experimentally observed $D_\text{int}$ profiles across the NIR, near-visible, and MIR regimes.
Evidenced in Fig. \ref{Fig:4}c, d, e, these simulations replicate the broadband $D_\text{int}$ profiles, exhibiting excellent agreement with the experimental results in Fig. \ref{Fig:3}d, f, h.

Crucially, our model reveals a fundamental spectral tension governing the NIR GVD.
While the MIR N--H absorption inherently imparts a strong negative dispersion pull in the NIR, the blue-shifted UV bandgap and red-shifted UV pole exert a competing positive dispersion contribution.
To evaluate this interplay, we simulate $D_\text{int}$ with a fixed cross-section of 697 nm $\times$ 2.5 $\mu$m from 125 to 200 THz frequency.
The simulated $D_1/2\pi$, $D_2/2\pi$, and $\beta_2$ profiles are presented in Fig. \ref{Fig:4}f, g, h, respectively.
As illustrated in Fig. \ref{Fig:4}f, the $D_1$ of the PA state remains consistently larger than that of the TA state across the entire simulated spectrum.  
In contrast, the tension between the opposing UV and MIR influences leads to a distinctive dispersion balance point near 150 THz ($\sim$2 $\mu$m).
As depicted in Fig. \ref{Fig:4}g, h, the GVD curves of the H-rich (PA) and H-free (TA) states intersect precisely at this wavelength, marking a spectral region where the dispersion becomes insensitive to the H content.
More details are found in Supplementary Information Note 3.

\noindent \textbf{Broadband microcomb generation}.   
Finally, we validate the impact of this dispersion correction via microcomb generation in the chip described in Fig. \ref{Fig:3}c, d.  
The experimental setup is presented in Supplementary Information Note 9. 
Figure \ref{Fig:5} top presents the microcomb spectrum generated using the chip in its PA state, compared with the measured $D_\text{int}$ from Fig. \ref{Fig:3}d red curve. 
Because the N--H  bonds distort the NIR dispersion, the resulting flat GVD (small $D_2$) leads to a broadband, zero-dispersion microcomb \cite{Anderson:22}. 
Following TA, the same chip is re-evaluated, and the resulting microcomb spectrum is presented in Fig. \ref{Fig:5} middle, alongside the measured $D_\text{int}$ from Fig. \ref{Fig:3}d blue curve. 
Now PM-enhanced comb lines \cite{Okawachi:11} appear near 163 THz, aligning with our initial design and simulation shown in Fig. \ref{Fig:4}c blue curve. 
Figure \ref{Fig:5} bottom presents the simulated microcomb spectrum derived from the Lugiato-Lefever Equation \cite{Lugiato:87} using experimentally measured parameters.
The close agreement between the simulation and the experiment validates the efficacy of TA for dispersion correction. 

\begin{figure}[t!]
\centering
\includegraphics{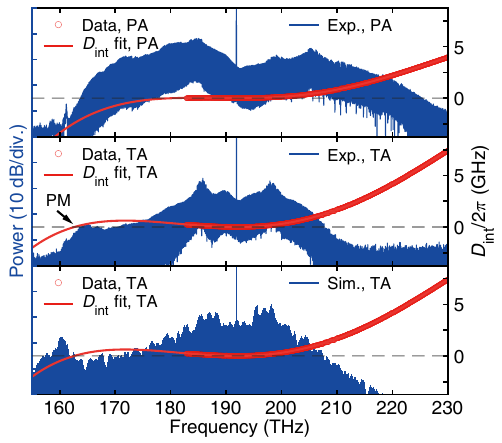}
\caption{
\textbf{Broadband microcomb generation.} 
Microcomb spectra in the Si$_3$N$_4$ microresonator before (top, PA) and after (middle, TA) the additional TA, compared with the simulation (bottom, TA).
The microresonator is the one characterized in Fig. \ref{Fig:3}c, d.
The measured $D_\text{int}$ profiles from Fig. \ref{Fig:3}d are overlaid. 
PM is revealed after TA. 
}
\label{Fig:5}
\end{figure}

\noindent \textbf{Discussion}.  
Our results underscore that standard refractive index models of Si$_3$N$_4$ are strictly valid only when high-temperature annealing ($>1200^\circ$C) has fully eliminated H impurities. 
Otherwise, neglecting H-mediated UV bandgap distortion and MIR vibrational absorption leads to catastrophic errors in broadband dispersion engineering. 
Consequently, rather than relying on standard models, material dispersion should be deliberately gauged via ellipsometry based on actual films and realistic annealing conditions. 

This conclusion extends beyond monolithic Si$_3$N$_4$. 
In heterogeneous integration, materials with rich vibrational spectra---such as 2D materials, polymers (e.g., SU-8, PMMA), and phase-change alloys---are frequently integrated onto waveguides. 
The KK relations dictate that their vibrational signatures can non-locally alter the global refractive index, shifting PM conditions and ultimately resulting in device failure.  

Furthermore, our study provides key guidelines for developing Si$_3$N$_4$ integrated photonics based on plasma-enhanced chemical vapor deposition (PECVD), which is entirely CMOS-compatible due to its low-temperature ($<400^\circ$C) operation \cite{Mao:08, Ji:23}. 
To avoid H-induced absorption in the NIR without high-temperature annealing, isotopic substitution via deuteration is widely employed---replacing $^1$H with deuterium ($^2$H or D) in precursor gases SiD$_4$ and ND$_3$. 
This red-shifts the fundamental vibrational frequencies of the Si--D and N--D bonds in deuterated silicon nitride (SiN:D) \cite{Chiles:18, Chia:23, Wu:21, Xie:22, Bose:24} and oxide (SiO$_2$:D) \cite{Jin:20}, creating an NIR window free from H-induced absorption. 
However, our study indicates that the refractive indices of as-deposited SiN:H and SiN:D can differ vastly.
While the former is governed by N--H  and Si--H bonds (fundamental stretching vibrations at 3.0 and 4.6 $\mu$m, respectively), the latter is impacted by N--D and Si--D bonds (fundamental stretching vibrations at 4.0 and 6.3 $\mu$m, respectively). 
Notably, the first overtone of N--D shifts to 2.0 $\mu$m---outside the standard telecommunication window---and its strength can serve to evaluate the fundamental tone.

\noindent \textbf{Summary}.  
In conclusion, we demonstrate that the Kramers--Kronig relations---a century-old consequence of causality---remain the governing principle for cutting-edge integrated photonics and precision engineered dispersion.
While conventional research focuses on minimizing localized optical loss, we reveal that spectral non-locality links NIR dispersion to profound, yet often overlooked, broadband structural perturbations. 
Specifically, residual H impurities introduce a dual macroscopic modulation: 
they generate fundamental vibrational resonances in the MIR while concurrently driving structural shifts in high-energy electronic transition, including a blue-shifted UV bandgap and a red-shifted UV pole. 
By rigorously accounting for these coupled phenomena, we establish a predictive framework where NIR dispersion is understood as a spectral tension between opposing UV electronic excitations and MIR molecular vibrations.
This holistic approach resolves critical discrepancies in broadband phase matching, paving the way for next-generation, octave-spanning nonlinear integrated photonics. 

\noindent \textbf{Methods}

\setcounter{figure}{0}
\begin{figure}[t!]
\renewcommand{\figurename}{Extended Data Figure}
\centering
\includegraphics{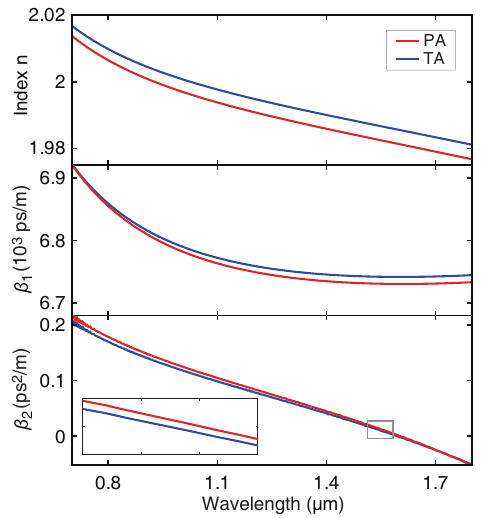}
\caption{
\textbf{Refractive index and dispersion of silicon nitride in the TA and PA states}.  
Calculated refractive index $n$ (top), first-order dispersion parameter $\beta_1$ (middle), and GVD parameter $\beta_2$ (bottom). 
The red and blue curves correspond to the PA and TA states, respectively.
}
\label{Fig:Extended 1}
\end{figure}

\noindent \textbf{Dispersion simulation and Si$_3$N$_4$ refractive index models}.
Microresonator dispersion simulations are performed using the FEM in COMSOL Multiphysics. 
The cross-sectional dimensions (thickness $\times$ width) of the simulated microresonators are precisely matched to the measured geometries via SEM:
696 nm $\times$ 2.5 $\mu$m for NIR characterization, 
297 nm $\times$ 1.6 $\mu$m for near-visible characterization, 
and 1150 nm $\times$ 3.0 $\mu$m for MIR characterization.
The refractive index of Si$_3$N$_4$ for the TA state is taken from an established model \cite{Liu:20b}, described by the following Sellmeier equation:
\begin{widetext}
\begin{equation}
\begin{split}
n^2(\lambda) = 1 &+ 1.750893106708352 \cdot \frac{\lambda^2}{\lambda^2-0.159333426051559^2}\\
&+ 1.222742709909494 \cdot \frac{\lambda^2}{\lambda^2-0.054765813233156^2}\\
&+ 2.584602943737831 \cdot \frac{\lambda^2}{\lambda^2-11.600661120564506^2}\\
\label{Eq. n_TA}
\end{split}
\end{equation}
\end{widetext}
where the first, second, and third terms represent the refractive index contributions from the bandgap, the UV pole, and the Si--N bond absorption at 11.6 $\mu$m, respectively.
For the H-rich PA state, this model is modified as shown in Eq. \ref{Eq. n_PA}. 
The amplitude reduction coefficients for the UV bandgap and UV pole are set to $A_1 = 0.997$ and $A_2 = 0.990$, respectively. 
The corresponding resonance wavelength shifts are $\delta\lambda_{\text{bg}} = 0.00073$ $\mu$m (blue-shift) and $\delta\lambda_{\text{pole}} = 0.02023$ $\mu$m (red-shift). 
Furthermore, we account for the refractive index perturbation from the N--H stretching vibration at $\lambda_{\text{N-H}}$ = 3.0 $\mu$m, yielding an extracted amplitude coefficient of $A_3 = 0.00105$.
Notably, by setting $A_1 = A_2 = 1$, $A_3 = 0$, and $\delta\lambda_{\text{bg}} = \delta\lambda_{\text{pole}} = 0$, the PA model reduces to the hydrogen-free TA base model, shown as Eq. \ref{Eq. n_TA}.

The refractive index $n$, first-order dispersion $\beta_1$, and the GVD $\beta_2$ for both the TA and PA states are derived from Eqs. \ref{Eq. n_TA} and \ref{Eq. n_PA}, as shown in Extended Data Fig. \ref{Fig:Extended 1}.
The $\beta_1$ and $\beta_2$ are calculated using: 
\begin{subequations}\label{Eq. beta1beta2}
\begin{align}
\beta_1 &= \frac{\text{d} k}{\text{d}\omega} = \frac{1}{c} \left( n + \omega \frac{\text{d}n}{\text{d}\omega} \right) \label{Eq. beta1}\\
\beta_2 &= \frac{\text{d}^2 k}{\text{d}\omega^2} = \frac{1}{c} \left( 2\frac{\text{d}n}{\text{d}\omega} + \omega \frac{\text{d}^2n}{\text{d}\omega^2} \right) \label{Eq. beta2}
\end{align}
\end{subequations}
Compared to the PA state, the TA state exhibits increased $\beta_1$ and decreased $\beta_2$ in the NIR and near-visible.
These changes correspond to decreased $D_1/2\pi$ and increased $D_2/2\pi$, consistent with the experimental results shown in Fig. \ref{Fig:3}d, f.

\medskip
\begin{footnotesize}

\vspace{0.1cm}
\noindent \textbf{Acknowledgments}: 
We thank Jiahao Sun and Chen Shen for assistance in fabricating the Si$_3$N$_4$ chips.
We thank Hairun Guo for fruitful discussion.
We acknowledge support from Quantum Science and Technology--National Science and Technology Major Project (Grant No. 2023ZD0301500 and 2021ZD0300800), 
National Natural Science Foundation of China (Grant No.U25D9005, 12404436, 12404417, 12494601),  
National Key R\&D Program of China (Grant No. 2024YFA1409300), 
Guangdong-Hong Kong Technology Cooperation Funding Scheme (Grant No. 2024A0505040008), 
Shenzhen Science and Technology Program (Grant No. RCJC20231211090042078), 
and Shenzhen-Hong Kong Cooperation Zone for Technology and Innovation (HZQB-KCZYB2020050). 

\vspace{0.1cm}
\noindent \textbf{Author contributions}: 
Z.S., S.H. and Z.Z. fabricated the Si$_3$N$_4$ chips.
Z.S., Y.H. and H.T. performed the FTIR characterization, supervised by Z.-C.N. and J.L..
Z.S. and Y.N. performed the ellipsometry characterization, supervised by X.Y. and J.L.. 
Y.H., C.Z., B.S. and W.S. characterized the chips with VSAs and performed microcomb experiments. 
Y.H., Z.S., D.C, W.Z. and Y.-H.L.  performed the numerical simulations. 
Y.H., Z.S., C.Z. and J.L. analyzed the data and wrote the manuscript, with input from others. 
J.L. supervised the project.

\vspace{0.1cm}
\noindent \textbf{Data Availability Statement}: 
The code and data used to produce the plots within this work will be released on the repository \texttt{Zenodo} upon publication of this preprint.

\end{footnotesize}
\bibliographystyle{apsrev4-1}
\bibliography{bibliography}

\begin{thebibliography}{73}%
\makeatletter
\providecommand \@ifxundefined [1]{%
 \@ifx{#1\undefined}
}%
\providecommand \@ifnum [1]{%
 \ifnum #1\expandafter \@firstoftwo
 \else \expandafter \@secondoftwo
 \fi
}%
\providecommand \@ifx [1]{%
 \ifx #1\expandafter \@firstoftwo
 \else \expandafter \@secondoftwo
 \fi
}%
\providecommand \natexlab [1]{#1}%
\providecommand \enquote  [1]{``#1''}%
\providecommand \bibnamefont  [1]{#1}%
\providecommand \bibfnamefont [1]{#1}%
\providecommand \citenamefont [1]{#1}%
\providecommand \href@noop [0]{\@secondoftwo}%
\providecommand \href [0]{\begingroup \@sanitize@url \@href}%
\providecommand \@href[1]{\@@startlink{#1}\@@href}%
\providecommand \@@href[1]{\endgroup#1\@@endlink}%
\providecommand \@sanitize@url [0]{\catcode `\\12\catcode `\$12\catcode
  `\&12\catcode `\#12\catcode `\^12\catcode `\_12\catcode `\%12\relax}%
\providecommand \@@startlink[1]{}%
\providecommand \@@endlink[0]{}%
\providecommand \url  [0]{\begingroup\@sanitize@url \@url }%
\providecommand \@url [1]{\endgroup\@href {#1}{\urlprefix }}%
\providecommand \urlprefix  [0]{URL }%
\providecommand \Eprint [0]{\href }%
\providecommand \doibase [0]{http://dx.doi.org/}%
\providecommand \selectlanguage [0]{\@gobble}%
\providecommand \bibinfo  [0]{\@secondoftwo}%
\providecommand \bibfield  [0]{\@secondoftwo}%
\providecommand \translation [1]{[#1]}%
\providecommand \BibitemOpen [0]{}%
\providecommand \bibitemStop [0]{}%
\providecommand \bibitemNoStop [0]{.\EOS\space}%
\providecommand \EOS [0]{\spacefactor3000\relax}%
\providecommand \BibitemShut  [1]{\csname bibitem#1\endcsname}%
\let\auto@bib@innerbib\@empty
\bibitem [{\citenamefont {Moss}\ \emph {et~al.}(2013)\citenamefont {Moss},
  \citenamefont {Morandotti}, \citenamefont {Gaeta},\ and\ \citenamefont
  {Lipson}}]{Moss:13}%
  \BibitemOpen
  \bibfield  {author} {\bibinfo {author} {\bibfnamefont {D.~J.}\ \bibnamefont
  {Moss}}, \bibinfo {author} {\bibfnamefont {R.}~\bibnamefont {Morandotti}},
  \bibinfo {author} {\bibfnamefont {A.~L.}\ \bibnamefont {Gaeta}}, \ and\
  \bibinfo {author} {\bibfnamefont {M.}~\bibnamefont {Lipson}},\ }\href
  {https://doi.org/10.1038/nphoton.2013.183} {\bibfield  {journal} {\bibinfo
  {journal} {Nature Photonics}\ }\textbf {\bibinfo {volume} {7}},\ \bibinfo
  {pages} {597} (\bibinfo {year} {2013})}\BibitemShut {NoStop}%
\bibitem [{\citenamefont {Gaeta}\ \emph {et~al.}(2019)\citenamefont {Gaeta},
  \citenamefont {Lipson},\ and\ \citenamefont {Kippenberg}}]{Gaeta:19}%
  \BibitemOpen
  \bibfield  {author} {\bibinfo {author} {\bibfnamefont {A.~L.}\ \bibnamefont
  {Gaeta}}, \bibinfo {author} {\bibfnamefont {M.}~\bibnamefont {Lipson}}, \
  and\ \bibinfo {author} {\bibfnamefont {T.~J.}\ \bibnamefont {Kippenberg}},\
  }\href {\doibase 10.1038/s41566-019-0358-x} {\bibfield  {journal} {\bibinfo
  {journal} {Nature Photonics}\ }\textbf {\bibinfo {volume} {13}},\ \bibinfo
  {pages} {158} (\bibinfo {year} {2019})}\BibitemShut {NoStop}%
\bibitem [{\citenamefont {Xuan}\ \emph {et~al.}(2016)\citenamefont {Xuan},
  \citenamefont {Liu}, \citenamefont {Varghese}, \citenamefont {Metcalf},
  \citenamefont {Xue}, \citenamefont {Wang}, \citenamefont {Han}, \citenamefont
  {Jaramillo-Villegas}, \citenamefont {Noman}, \citenamefont {Wang},
  \citenamefont {Kim}, \citenamefont {Teng}, \citenamefont {Lee}, \citenamefont
  {Niu}, \citenamefont {Fan}, \citenamefont {Wang}, \citenamefont {Leaird},
  \citenamefont {Weiner},\ and\ \citenamefont {Qi}}]{Xuan:16}%
  \BibitemOpen
  \bibfield  {author} {\bibinfo {author} {\bibfnamefont {Y.}~\bibnamefont
  {Xuan}}, \bibinfo {author} {\bibfnamefont {Y.}~\bibnamefont {Liu}}, \bibinfo
  {author} {\bibfnamefont {L.~T.}\ \bibnamefont {Varghese}}, \bibinfo {author}
  {\bibfnamefont {A.~J.}\ \bibnamefont {Metcalf}}, \bibinfo {author}
  {\bibfnamefont {X.}~\bibnamefont {Xue}}, \bibinfo {author} {\bibfnamefont
  {P.-H.}\ \bibnamefont {Wang}}, \bibinfo {author} {\bibfnamefont
  {K.}~\bibnamefont {Han}}, \bibinfo {author} {\bibfnamefont {J.~A.}\
  \bibnamefont {Jaramillo-Villegas}}, \bibinfo {author} {\bibfnamefont {A.~A.}\
  \bibnamefont {Noman}}, \bibinfo {author} {\bibfnamefont {C.}~\bibnamefont
  {Wang}}, \bibinfo {author} {\bibfnamefont {S.}~\bibnamefont {Kim}}, \bibinfo
  {author} {\bibfnamefont {M.}~\bibnamefont {Teng}}, \bibinfo {author}
  {\bibfnamefont {Y.~J.}\ \bibnamefont {Lee}}, \bibinfo {author} {\bibfnamefont
  {B.}~\bibnamefont {Niu}}, \bibinfo {author} {\bibfnamefont {L.}~\bibnamefont
  {Fan}}, \bibinfo {author} {\bibfnamefont {J.}~\bibnamefont {Wang}}, \bibinfo
  {author} {\bibfnamefont {D.~E.}\ \bibnamefont {Leaird}}, \bibinfo {author}
  {\bibfnamefont {A.~M.}\ \bibnamefont {Weiner}}, \ and\ \bibinfo {author}
  {\bibfnamefont {M.}~\bibnamefont {Qi}},\ }\href {\doibase
  10.1364/OPTICA.3.001171} {\bibfield  {journal} {\bibinfo  {journal} {Optica}\
  }\textbf {\bibinfo {volume} {3}},\ \bibinfo {pages} {1171} (\bibinfo {year}
  {2016})}\BibitemShut {NoStop}%
\bibitem [{\citenamefont {Ji}\ \emph {et~al.}(2017)\citenamefont {Ji},
  \citenamefont {Barbosa}, \citenamefont {Roberts}, \citenamefont {Dutt},
  \citenamefont {Cardenas}, \citenamefont {Okawachi}, \citenamefont {Bryant},
  \citenamefont {Gaeta},\ and\ \citenamefont {Lipson}}]{Ji:17}%
  \BibitemOpen
  \bibfield  {author} {\bibinfo {author} {\bibfnamefont {X.}~\bibnamefont
  {Ji}}, \bibinfo {author} {\bibfnamefont {F.~A.~S.}\ \bibnamefont {Barbosa}},
  \bibinfo {author} {\bibfnamefont {S.~P.}\ \bibnamefont {Roberts}}, \bibinfo
  {author} {\bibfnamefont {A.}~\bibnamefont {Dutt}}, \bibinfo {author}
  {\bibfnamefont {J.}~\bibnamefont {Cardenas}}, \bibinfo {author}
  {\bibfnamefont {Y.}~\bibnamefont {Okawachi}}, \bibinfo {author}
  {\bibfnamefont {A.}~\bibnamefont {Bryant}}, \bibinfo {author} {\bibfnamefont
  {A.~L.}\ \bibnamefont {Gaeta}}, \ and\ \bibinfo {author} {\bibfnamefont
  {M.}~\bibnamefont {Lipson}},\ }\href {\doibase 10.1364/OPTICA.4.000619}
  {\bibfield  {journal} {\bibinfo  {journal} {Optica}\ }\textbf {\bibinfo
  {volume} {4}},\ \bibinfo {pages} {619} (\bibinfo {year} {2017})}\BibitemShut
  {NoStop}%
\bibitem [{\citenamefont {Liu}\ \emph {et~al.}(2018)\citenamefont {Liu},
  \citenamefont {Raja}, \citenamefont {Karpov}, \citenamefont {Ghadiani},
  \citenamefont {Pfeiffer}, \citenamefont {Du}, \citenamefont {Engelsen},
  \citenamefont {Guo}, \citenamefont {Zervas},\ and\ \citenamefont
  {Kippenberg}}]{Liu:18a}%
  \BibitemOpen
  \bibfield  {author} {\bibinfo {author} {\bibfnamefont {J.}~\bibnamefont
  {Liu}}, \bibinfo {author} {\bibfnamefont {A.~S.}\ \bibnamefont {Raja}},
  \bibinfo {author} {\bibfnamefont {M.}~\bibnamefont {Karpov}}, \bibinfo
  {author} {\bibfnamefont {B.}~\bibnamefont {Ghadiani}}, \bibinfo {author}
  {\bibfnamefont {M.~H.~P.}\ \bibnamefont {Pfeiffer}}, \bibinfo {author}
  {\bibfnamefont {B.}~\bibnamefont {Du}}, \bibinfo {author} {\bibfnamefont
  {N.~J.}\ \bibnamefont {Engelsen}}, \bibinfo {author} {\bibfnamefont
  {H.}~\bibnamefont {Guo}}, \bibinfo {author} {\bibfnamefont {M.}~\bibnamefont
  {Zervas}}, \ and\ \bibinfo {author} {\bibfnamefont {T.~J.}\ \bibnamefont
  {Kippenberg}},\ }\href {\doibase 10.1364/OPTICA.5.001347} {\bibfield
  {journal} {\bibinfo  {journal} {Optica}\ }\textbf {\bibinfo {volume} {5}},\
  \bibinfo {pages} {1347} (\bibinfo {year} {2018})}\BibitemShut {NoStop}%
\bibitem [{\citenamefont {Brasch}\ \emph {et~al.}(2016)\citenamefont {Brasch},
  \citenamefont {Geiselmann}, \citenamefont {Pfeiffer},\ and\ \citenamefont
  {Kippenberg}}]{Brasch:16}%
  \BibitemOpen
  \bibfield  {author} {\bibinfo {author} {\bibfnamefont {V.}~\bibnamefont
  {Brasch}}, \bibinfo {author} {\bibfnamefont {M.}~\bibnamefont {Geiselmann}},
  \bibinfo {author} {\bibfnamefont {M.~H.~P.}\ \bibnamefont {Pfeiffer}}, \ and\
  \bibinfo {author} {\bibfnamefont {T.~J.}\ \bibnamefont {Kippenberg}},\ }\href
  {\doibase 10.1364/OE.24.029312} {\bibfield  {journal} {\bibinfo  {journal}
  {Opt. Express}\ }\textbf {\bibinfo {volume} {24}},\ \bibinfo {pages} {29312}
  (\bibinfo {year} {2016})}\BibitemShut {NoStop}%
\bibitem [{\citenamefont {Li}\ \emph {et~al.}(2017)\citenamefont {Li},
  \citenamefont {Briles}, \citenamefont {Westly}, \citenamefont {Drake},
  \citenamefont {Stone}, \citenamefont {Ilic}, \citenamefont {Diddams},
  \citenamefont {Papp},\ and\ \citenamefont {Srinivasan}}]{Li:17}%
  \BibitemOpen
  \bibfield  {author} {\bibinfo {author} {\bibfnamefont {Q.}~\bibnamefont
  {Li}}, \bibinfo {author} {\bibfnamefont {T.~C.}\ \bibnamefont {Briles}},
  \bibinfo {author} {\bibfnamefont {D.~A.}\ \bibnamefont {Westly}}, \bibinfo
  {author} {\bibfnamefont {T.~E.}\ \bibnamefont {Drake}}, \bibinfo {author}
  {\bibfnamefont {J.~R.}\ \bibnamefont {Stone}}, \bibinfo {author}
  {\bibfnamefont {B.~R.}\ \bibnamefont {Ilic}}, \bibinfo {author}
  {\bibfnamefont {S.~A.}\ \bibnamefont {Diddams}}, \bibinfo {author}
  {\bibfnamefont {S.~B.}\ \bibnamefont {Papp}}, \ and\ \bibinfo {author}
  {\bibfnamefont {K.}~\bibnamefont {Srinivasan}},\ }\href {\doibase
  10.1364/OPTICA.4.000193} {\bibfield  {journal} {\bibinfo  {journal} {Optica}\
  }\textbf {\bibinfo {volume} {4}},\ \bibinfo {pages} {193} (\bibinfo {year}
  {2017})}\BibitemShut {NoStop}%
\bibitem [{\citenamefont {Pfeiffer}\ \emph {et~al.}(2017)\citenamefont
  {Pfeiffer}, \citenamefont {Herkommer}, \citenamefont {Liu}, \citenamefont
  {Guo}, \citenamefont {Karpov}, \citenamefont {Lucas}, \citenamefont
  {Zervas},\ and\ \citenamefont {Kippenberg}}]{Pfeiffer:17}%
  \BibitemOpen
  \bibfield  {author} {\bibinfo {author} {\bibfnamefont {M.~H.~P.}\
  \bibnamefont {Pfeiffer}}, \bibinfo {author} {\bibfnamefont {C.}~\bibnamefont
  {Herkommer}}, \bibinfo {author} {\bibfnamefont {J.}~\bibnamefont {Liu}},
  \bibinfo {author} {\bibfnamefont {H.}~\bibnamefont {Guo}}, \bibinfo {author}
  {\bibfnamefont {M.}~\bibnamefont {Karpov}}, \bibinfo {author} {\bibfnamefont
  {E.}~\bibnamefont {Lucas}}, \bibinfo {author} {\bibfnamefont
  {M.}~\bibnamefont {Zervas}}, \ and\ \bibinfo {author} {\bibfnamefont {T.~J.}\
  \bibnamefont {Kippenberg}},\ }\href {\doibase 10.1364/OPTICA.4.000684}
  {\bibfield  {journal} {\bibinfo  {journal} {Optica}\ }\textbf {\bibinfo
  {volume} {4}},\ \bibinfo {pages} {684} (\bibinfo {year} {2017})}\BibitemShut
  {NoStop}%
\bibitem [{\citenamefont {Yu}\ \emph {et~al.}(2019)\citenamefont {Yu},
  \citenamefont {Briles}, \citenamefont {Moille}, \citenamefont {Lu},
  \citenamefont {Diddams}, \citenamefont {Srinivasan},\ and\ \citenamefont
  {Papp}}]{Yu:19}%
  \BibitemOpen
  \bibfield  {author} {\bibinfo {author} {\bibfnamefont {S.-P.}\ \bibnamefont
  {Yu}}, \bibinfo {author} {\bibfnamefont {T.~C.}\ \bibnamefont {Briles}},
  \bibinfo {author} {\bibfnamefont {G.~T.}\ \bibnamefont {Moille}}, \bibinfo
  {author} {\bibfnamefont {X.}~\bibnamefont {Lu}}, \bibinfo {author}
  {\bibfnamefont {S.~A.}\ \bibnamefont {Diddams}}, \bibinfo {author}
  {\bibfnamefont {K.}~\bibnamefont {Srinivasan}}, \ and\ \bibinfo {author}
  {\bibfnamefont {S.~B.}\ \bibnamefont {Papp}},\ }\href {\doibase
  10.1103/PhysRevApplied.11.044017} {\bibfield  {journal} {\bibinfo  {journal}
  {Phys. Rev. Applied}\ }\textbf {\bibinfo {volume} {11}},\ \bibinfo {pages}
  {044017} (\bibinfo {year} {2019})}\BibitemShut {NoStop}%
\bibitem [{\citenamefont {Halir}\ \emph {et~al.}(2012)\citenamefont {Halir},
  \citenamefont {Okawachi}, \citenamefont {Levy}, \citenamefont {Foster},
  \citenamefont {Lipson},\ and\ \citenamefont {Gaeta}}]{Halir:12}%
  \BibitemOpen
  \bibfield  {author} {\bibinfo {author} {\bibfnamefont {R.}~\bibnamefont
  {Halir}}, \bibinfo {author} {\bibfnamefont {Y.}~\bibnamefont {Okawachi}},
  \bibinfo {author} {\bibfnamefont {J.~S.}\ \bibnamefont {Levy}}, \bibinfo
  {author} {\bibfnamefont {M.~A.}\ \bibnamefont {Foster}}, \bibinfo {author}
  {\bibfnamefont {M.}~\bibnamefont {Lipson}}, \ and\ \bibinfo {author}
  {\bibfnamefont {A.~L.}\ \bibnamefont {Gaeta}},\ }\href {\doibase
  10.1364/OL.37.001685} {\bibfield  {journal} {\bibinfo  {journal} {Opt.
  Lett.}\ }\textbf {\bibinfo {volume} {37}},\ \bibinfo {pages} {1685} (\bibinfo
  {year} {2012})}\BibitemShut {NoStop}%
\bibitem [{\citenamefont {Porcel}\ \emph {et~al.}(2017)\citenamefont {Porcel},
  \citenamefont {Schepers}, \citenamefont {Epping}, \citenamefont {Hellwig},
  \citenamefont {Hoekman}, \citenamefont {Heideman}, \citenamefont {van~der
  Slot}, \citenamefont {Lee}, \citenamefont {Schmidt}, \citenamefont
  {Bratschitsch}, \citenamefont {Fallnich},\ and\ \citenamefont
  {Boller}}]{Porcel:17a}%
  \BibitemOpen
  \bibfield  {author} {\bibinfo {author} {\bibfnamefont {M.~A.~G.}\
  \bibnamefont {Porcel}}, \bibinfo {author} {\bibfnamefont {F.}~\bibnamefont
  {Schepers}}, \bibinfo {author} {\bibfnamefont {J.~P.}\ \bibnamefont
  {Epping}}, \bibinfo {author} {\bibfnamefont {T.}~\bibnamefont {Hellwig}},
  \bibinfo {author} {\bibfnamefont {M.}~\bibnamefont {Hoekman}}, \bibinfo
  {author} {\bibfnamefont {R.~G.}\ \bibnamefont {Heideman}}, \bibinfo {author}
  {\bibfnamefont {P.~J.~M.}\ \bibnamefont {van~der Slot}}, \bibinfo {author}
  {\bibfnamefont {C.~J.}\ \bibnamefont {Lee}}, \bibinfo {author} {\bibfnamefont
  {R.}~\bibnamefont {Schmidt}}, \bibinfo {author} {\bibfnamefont
  {R.}~\bibnamefont {Bratschitsch}}, \bibinfo {author} {\bibfnamefont
  {C.}~\bibnamefont {Fallnich}}, \ and\ \bibinfo {author} {\bibfnamefont
  {K.-J.}\ \bibnamefont {Boller}},\ }\href {\doibase 10.1364/OE.25.001542}
  {\bibfield  {journal} {\bibinfo  {journal} {Opt. Express}\ }\textbf {\bibinfo
  {volume} {25}},\ \bibinfo {pages} {1542} (\bibinfo {year}
  {2017})}\BibitemShut {NoStop}%
\bibitem [{\citenamefont {Epping}\ \emph {et~al.}(2015)\citenamefont {Epping},
  \citenamefont {Hellwig}, \citenamefont {Hoekman}, \citenamefont {Mateman},
  \citenamefont {Leinse}, \citenamefont {Heideman}, \citenamefont {van Rees},
  \citenamefont {van~der Slot}, \citenamefont {Lee}, \citenamefont {Fallnich},\
  and\ \citenamefont {Boller}}]{Epping:15a}%
  \BibitemOpen
  \bibfield  {author} {\bibinfo {author} {\bibfnamefont {J.~P.}\ \bibnamefont
  {Epping}}, \bibinfo {author} {\bibfnamefont {T.}~\bibnamefont {Hellwig}},
  \bibinfo {author} {\bibfnamefont {M.}~\bibnamefont {Hoekman}}, \bibinfo
  {author} {\bibfnamefont {R.}~\bibnamefont {Mateman}}, \bibinfo {author}
  {\bibfnamefont {A.}~\bibnamefont {Leinse}}, \bibinfo {author} {\bibfnamefont
  {R.~G.}\ \bibnamefont {Heideman}}, \bibinfo {author} {\bibfnamefont
  {A.}~\bibnamefont {van Rees}}, \bibinfo {author} {\bibfnamefont {P.~J.}\
  \bibnamefont {van~der Slot}}, \bibinfo {author} {\bibfnamefont {C.~J.}\
  \bibnamefont {Lee}}, \bibinfo {author} {\bibfnamefont {C.}~\bibnamefont
  {Fallnich}}, \ and\ \bibinfo {author} {\bibfnamefont {K.-J.}\ \bibnamefont
  {Boller}},\ }\href {\doibase 10.1364/OE.23.019596} {\bibfield  {journal}
  {\bibinfo  {journal} {Opt. Express}\ }\textbf {\bibinfo {volume} {23}},\
  \bibinfo {pages} {19596} (\bibinfo {year} {2015})}\BibitemShut {NoStop}%
\bibitem [{\citenamefont {Guo}\ \emph {et~al.}(2020)\citenamefont {Guo},
  \citenamefont {Weng}, \citenamefont {Liu}, \citenamefont {Yang},
  \citenamefont {H\"{a}nsel}, \citenamefont {Br\`{e}s}, \citenamefont
  {Th\'{e}venaz}, \citenamefont {Holzwarth},\ and\ \citenamefont
  {Kippenberg}}]{Guo:20}%
  \BibitemOpen
  \bibfield  {author} {\bibinfo {author} {\bibfnamefont {H.}~\bibnamefont
  {Guo}}, \bibinfo {author} {\bibfnamefont {W.}~\bibnamefont {Weng}}, \bibinfo
  {author} {\bibfnamefont {J.}~\bibnamefont {Liu}}, \bibinfo {author}
  {\bibfnamefont {F.}~\bibnamefont {Yang}}, \bibinfo {author} {\bibfnamefont
  {W.}~\bibnamefont {H\"{a}nsel}}, \bibinfo {author} {\bibfnamefont {C.~S.}\
  \bibnamefont {Br\`{e}s}}, \bibinfo {author} {\bibfnamefont {L.}~\bibnamefont
  {Th\'{e}venaz}}, \bibinfo {author} {\bibfnamefont {R.}~\bibnamefont
  {Holzwarth}}, \ and\ \bibinfo {author} {\bibfnamefont {T.~J.}\ \bibnamefont
  {Kippenberg}},\ }\href {\doibase 10.1364/OPTICA.396542} {\bibfield  {journal}
  {\bibinfo  {journal} {Optica}\ }\textbf {\bibinfo {volume} {7}},\ \bibinfo
  {pages} {1181} (\bibinfo {year} {2020})}\BibitemShut {NoStop}%
\bibitem [{\citenamefont {Levy}\ \emph {et~al.}(2010)\citenamefont {Levy},
  \citenamefont {Gondarenko}, \citenamefont {Foster}, \citenamefont
  {Turner-Foster}, \citenamefont {Gaeta},\ and\ \citenamefont
  {Lipson}}]{Levy:10}%
  \BibitemOpen
  \bibfield  {author} {\bibinfo {author} {\bibfnamefont {J.~S.}\ \bibnamefont
  {Levy}}, \bibinfo {author} {\bibfnamefont {A.}~\bibnamefont {Gondarenko}},
  \bibinfo {author} {\bibfnamefont {M.~A.}\ \bibnamefont {Foster}}, \bibinfo
  {author} {\bibfnamefont {A.~C.}\ \bibnamefont {Turner-Foster}}, \bibinfo
  {author} {\bibfnamefont {A.~L.}\ \bibnamefont {Gaeta}}, \ and\ \bibinfo
  {author} {\bibfnamefont {M.}~\bibnamefont {Lipson}},\ }\href
  {http://dx.doi.org/10.1038/nphoton.2009.259} {\bibfield  {journal} {\bibinfo
  {journal} {Nature Photonics}\ }\textbf {\bibinfo {volume} {4}},\ \bibinfo
  {pages} {37} (\bibinfo {year} {2010})}\BibitemShut {NoStop}%
\bibitem [{\citenamefont {Li}\ \emph {et~al.}(2016)\citenamefont {Li},
  \citenamefont {Davan{\c c}o},\ and\ \citenamefont {Srinivasan}}]{Li:16}%
  \BibitemOpen
  \bibfield  {author} {\bibinfo {author} {\bibfnamefont {Q.}~\bibnamefont
  {Li}}, \bibinfo {author} {\bibfnamefont {M.}~\bibnamefont {Davan{\c c}o}}, \
  and\ \bibinfo {author} {\bibfnamefont {K.}~\bibnamefont {Srinivasan}},\
  }\href {\doibase 10.1038/nphoton.2016.64} {\bibfield  {journal} {\bibinfo
  {journal} {Nature Photonics}\ }\textbf {\bibinfo {volume} {10}},\ \bibinfo
  {pages} {406} (\bibinfo {year} {2016})}\BibitemShut {NoStop}%
\bibitem [{\citenamefont {Lu}\ \emph {et~al.}(2019{\natexlab{a}})\citenamefont
  {Lu}, \citenamefont {Moille}, \citenamefont {Singh}, \citenamefont {Li},
  \citenamefont {Westly}, \citenamefont {Rao}, \citenamefont {Yu},
  \citenamefont {Briles}, \citenamefont {Papp},\ and\ \citenamefont
  {Srinivasan}}]{Lu:19b}%
  \BibitemOpen
  \bibfield  {author} {\bibinfo {author} {\bibfnamefont {X.}~\bibnamefont
  {Lu}}, \bibinfo {author} {\bibfnamefont {G.}~\bibnamefont {Moille}}, \bibinfo
  {author} {\bibfnamefont {A.}~\bibnamefont {Singh}}, \bibinfo {author}
  {\bibfnamefont {Q.}~\bibnamefont {Li}}, \bibinfo {author} {\bibfnamefont
  {D.~A.}\ \bibnamefont {Westly}}, \bibinfo {author} {\bibfnamefont
  {A.}~\bibnamefont {Rao}}, \bibinfo {author} {\bibfnamefont {S.-P.}\
  \bibnamefont {Yu}}, \bibinfo {author} {\bibfnamefont {T.~C.}\ \bibnamefont
  {Briles}}, \bibinfo {author} {\bibfnamefont {S.~B.}\ \bibnamefont {Papp}}, \
  and\ \bibinfo {author} {\bibfnamefont {K.}~\bibnamefont {Srinivasan}},\
  }\href {\doibase 10.1364/OPTICA.6.001535} {\bibfield  {journal} {\bibinfo
  {journal} {Optica}\ }\textbf {\bibinfo {volume} {6}},\ \bibinfo {pages}
  {1535} (\bibinfo {year} {2019}{\natexlab{a}})}\BibitemShut {NoStop}%
\bibitem [{\citenamefont {Lu}\ \emph {et~al.}(2020)\citenamefont {Lu},
  \citenamefont {Moille}, \citenamefont {Rao}, \citenamefont {Westly},\ and\
  \citenamefont {Srinivasan}}]{Lu:20}%
  \BibitemOpen
  \bibfield  {author} {\bibinfo {author} {\bibfnamefont {X.}~\bibnamefont
  {Lu}}, \bibinfo {author} {\bibfnamefont {G.}~\bibnamefont {Moille}}, \bibinfo
  {author} {\bibfnamefont {A.}~\bibnamefont {Rao}}, \bibinfo {author}
  {\bibfnamefont {D.~A.}\ \bibnamefont {Westly}}, \ and\ \bibinfo {author}
  {\bibfnamefont {K.}~\bibnamefont {Srinivasan}},\ }\href {\doibase
  10.1364/OPTICA.393810} {\bibfield  {journal} {\bibinfo  {journal} {Optica}\
  }\textbf {\bibinfo {volume} {7}},\ \bibinfo {pages} {1417} (\bibinfo {year}
  {2020})}\BibitemShut {NoStop}%
\bibitem [{\citenamefont {Lu}\ \emph {et~al.}(2019{\natexlab{b}})\citenamefont
  {Lu}, \citenamefont {Li}, \citenamefont {Westly}, \citenamefont {Moille},
  \citenamefont {Singh}, \citenamefont {Anant},\ and\ \citenamefont
  {Srinivasan}}]{Lu:19a}%
  \BibitemOpen
  \bibfield  {author} {\bibinfo {author} {\bibfnamefont {X.}~\bibnamefont
  {Lu}}, \bibinfo {author} {\bibfnamefont {Q.}~\bibnamefont {Li}}, \bibinfo
  {author} {\bibfnamefont {D.~A.}\ \bibnamefont {Westly}}, \bibinfo {author}
  {\bibfnamefont {G.}~\bibnamefont {Moille}}, \bibinfo {author} {\bibfnamefont
  {A.}~\bibnamefont {Singh}}, \bibinfo {author} {\bibfnamefont
  {V.}~\bibnamefont {Anant}}, \ and\ \bibinfo {author} {\bibfnamefont
  {K.}~\bibnamefont {Srinivasan}},\ }\href {\doibase 10.1038/s41567-018-0394-3}
  {\bibfield  {journal} {\bibinfo  {journal} {Nature Physics}\ }\textbf
  {\bibinfo {volume} {15}},\ \bibinfo {pages} {373} (\bibinfo {year}
  {2019}{\natexlab{b}})}\BibitemShut {NoStop}%
\bibitem [{\citenamefont {Fan}\ \emph {et~al.}(2023)\citenamefont {Fan},
  \citenamefont {Lyu}, \citenamefont {Yuan}, \citenamefont {Deng},
  \citenamefont {Zhou}, \citenamefont {Geng}, \citenamefont {Song},
  \citenamefont {Wang}, \citenamefont {Zhang}, \citenamefont {Jin},
  \citenamefont {Zhou}, \citenamefont {You}, \citenamefont {Wang},
  \citenamefont {Guo},\ and\ \citenamefont {Zhou}}]{Fan:23}%
  \BibitemOpen
  \bibfield  {author} {\bibinfo {author} {\bibfnamefont {Y.}~\bibnamefont
  {Fan}}, \bibinfo {author} {\bibfnamefont {C.}~\bibnamefont {Lyu}}, \bibinfo
  {author} {\bibfnamefont {C.}~\bibnamefont {Yuan}}, \bibinfo {author}
  {\bibfnamefont {G.}~\bibnamefont {Deng}}, \bibinfo {author} {\bibfnamefont
  {Z.}~\bibnamefont {Zhou}}, \bibinfo {author} {\bibfnamefont {Y.}~\bibnamefont
  {Geng}}, \bibinfo {author} {\bibfnamefont {H.}~\bibnamefont {Song}}, \bibinfo
  {author} {\bibfnamefont {Y.}~\bibnamefont {Wang}}, \bibinfo {author}
  {\bibfnamefont {Y.}~\bibnamefont {Zhang}}, \bibinfo {author} {\bibfnamefont
  {R.}~\bibnamefont {Jin}}, \bibinfo {author} {\bibfnamefont {H.}~\bibnamefont
  {Zhou}}, \bibinfo {author} {\bibfnamefont {L.}~\bibnamefont {You}}, \bibinfo
  {author} {\bibfnamefont {Z.}~\bibnamefont {Wang}}, \bibinfo {author}
  {\bibfnamefont {G.}~\bibnamefont {Guo}}, \ and\ \bibinfo {author}
  {\bibfnamefont {Q.}~\bibnamefont {Zhou}},\ }\href {\doibase
  10.1002/lpor.202300172} {\bibfield  {journal} {\bibinfo  {journal} {Laser
  Photonics Rev.}\ ,\ \bibinfo {pages} {2300172}} (\bibinfo {year}
  {2023})}\BibitemShut {NoStop}%
\bibitem [{\citenamefont {Chen}\ \emph {et~al.}(2024)\citenamefont {Chen},
  \citenamefont {Luo}, \citenamefont {Long}, \citenamefont {Shi}, \citenamefont
  {Shen},\ and\ \citenamefont {Liu}}]{ChenR:24}%
  \BibitemOpen
  \bibfield  {author} {\bibinfo {author} {\bibfnamefont {R.}~\bibnamefont
  {Chen}}, \bibinfo {author} {\bibfnamefont {Y.-H.}\ \bibnamefont {Luo}},
  \bibinfo {author} {\bibfnamefont {J.}~\bibnamefont {Long}}, \bibinfo {author}
  {\bibfnamefont {B.}~\bibnamefont {Shi}}, \bibinfo {author} {\bibfnamefont
  {C.}~\bibnamefont {Shen}}, \ and\ \bibinfo {author} {\bibfnamefont
  {J.}~\bibnamefont {Liu}},\ }\href {\doibase 10.1103/PhysRevLett.133.083803}
  {\bibfield  {journal} {\bibinfo  {journal} {Phys. Rev. Lett.}\ }\textbf
  {\bibinfo {volume} {133}},\ \bibinfo {pages} {083803} (\bibinfo {year}
  {2024})}\BibitemShut {NoStop}%
\bibitem [{\citenamefont {Ramelow}\ \emph {et~al.}(2015)\citenamefont
  {Ramelow}, \citenamefont {Farsi}, \citenamefont {Clemmen}, \citenamefont
  {Orquiza}, \citenamefont {Luke}, \citenamefont {Lipson},\ and\ \citenamefont
  {Gaeta}}]{Ramelow:15}%
  \BibitemOpen
  \bibfield  {author} {\bibinfo {author} {\bibfnamefont {S.}~\bibnamefont
  {Ramelow}}, \bibinfo {author} {\bibfnamefont {A.}~\bibnamefont {Farsi}},
  \bibinfo {author} {\bibfnamefont {S.}~\bibnamefont {Clemmen}}, \bibinfo
  {author} {\bibfnamefont {D.}~\bibnamefont {Orquiza}}, \bibinfo {author}
  {\bibfnamefont {K.}~\bibnamefont {Luke}}, \bibinfo {author} {\bibfnamefont
  {M.}~\bibnamefont {Lipson}}, \ and\ \bibinfo {author} {\bibfnamefont {A.~L.}\
  \bibnamefont {Gaeta}},\ }\href@noop {} {\enquote {\bibinfo {title}
  {Silicon-nitride platform for narrowband entangled photon generation},}\ }
  (\bibinfo {year} {2015}),\ \Eprint {http://arxiv.org/abs/1508.04358}
  {arXiv:1508.04358 [quant-ph]} \BibitemShut {NoStop}%
\bibitem [{\citenamefont {Turner}\ \emph {et~al.}(2006)\citenamefont {Turner},
  \citenamefont {Manolatou}, \citenamefont {Schmidt}, \citenamefont {Lipson},
  \citenamefont {Foster}, \citenamefont {Sharping},\ and\ \citenamefont
  {Gaeta}}]{Turner:06}%
  \BibitemOpen
  \bibfield  {author} {\bibinfo {author} {\bibfnamefont {A.~C.}\ \bibnamefont
  {Turner}}, \bibinfo {author} {\bibfnamefont {C.}~\bibnamefont {Manolatou}},
  \bibinfo {author} {\bibfnamefont {B.~S.}\ \bibnamefont {Schmidt}}, \bibinfo
  {author} {\bibfnamefont {M.}~\bibnamefont {Lipson}}, \bibinfo {author}
  {\bibfnamefont {M.~A.}\ \bibnamefont {Foster}}, \bibinfo {author}
  {\bibfnamefont {J.~E.}\ \bibnamefont {Sharping}}, \ and\ \bibinfo {author}
  {\bibfnamefont {A.~L.}\ \bibnamefont {Gaeta}},\ }\href {\doibase
  10.1364/OE.14.004357} {\bibfield  {journal} {\bibinfo  {journal} {Opt.
  Express}\ }\textbf {\bibinfo {volume} {14}},\ \bibinfo {pages} {4357}
  (\bibinfo {year} {2006})}\BibitemShut {NoStop}%
\bibitem [{\citenamefont {Okawachi}\ \emph {et~al.}(2014)\citenamefont
  {Okawachi}, \citenamefont {Lamont}, \citenamefont {Luke}, \citenamefont
  {Carvalho}, \citenamefont {Yu}, \citenamefont {Lipson},\ and\ \citenamefont
  {Gaeta}}]{Okawachi:14}%
  \BibitemOpen
  \bibfield  {author} {\bibinfo {author} {\bibfnamefont {Y.}~\bibnamefont
  {Okawachi}}, \bibinfo {author} {\bibfnamefont {M.~R.~E.}\ \bibnamefont
  {Lamont}}, \bibinfo {author} {\bibfnamefont {K.}~\bibnamefont {Luke}},
  \bibinfo {author} {\bibfnamefont {D.~O.}\ \bibnamefont {Carvalho}}, \bibinfo
  {author} {\bibfnamefont {M.}~\bibnamefont {Yu}}, \bibinfo {author}
  {\bibfnamefont {M.}~\bibnamefont {Lipson}}, \ and\ \bibinfo {author}
  {\bibfnamefont {A.~L.}\ \bibnamefont {Gaeta}},\ }\href {\doibase
  10.1364/OL.39.003535} {\bibfield  {journal} {\bibinfo  {journal} {Opt.
  Lett.}\ }\textbf {\bibinfo {volume} {39}},\ \bibinfo {pages} {3535} (\bibinfo
  {year} {2014})}\BibitemShut {NoStop}%
\bibitem [{\citenamefont {Kramers}(1927)}]{Kramers:27}%
  \BibitemOpen
  \bibfield  {author} {\bibinfo {author} {\bibfnamefont {H.~A.}\ \bibnamefont
  {Kramers}},\ }\href {https://cir.nii.ac.jp/crid/1571135649980336384}
  {\bibfield  {journal} {\bibinfo  {journal} {Atti Cong. Intern. Fisica
  (Transactions of Volta Centenary Congress) Como}\ }\textbf {\bibinfo {volume}
  {2}},\ \bibinfo {pages} {545} (\bibinfo {year} {1927})}\BibitemShut {NoStop}%
\bibitem [{\citenamefont {de~L.~Kronig}(1926)}]{Kronig:26}%
  \BibitemOpen
  \bibfield  {author} {\bibinfo {author} {\bibfnamefont {R.}~\bibnamefont
  {de~L.~Kronig}},\ }\href {\doibase 10.1364/JOSA.12.000547} {\bibfield
  {journal} {\bibinfo  {journal} {J. Opt. Soc. Am.}\ }\textbf {\bibinfo
  {volume} {12}},\ \bibinfo {pages} {547} (\bibinfo {year} {1926})}\BibitemShut
  {NoStop}%
\bibitem [{\citenamefont {Lucarini}\ \emph {et~al.}(2005)\citenamefont
  {Lucarini}, \citenamefont {Peiponen}, \citenamefont {Saarinen},\ and\
  \citenamefont {Vartiainen}}]{Lucarini:05}%
  \BibitemOpen
  \bibfield  {author} {\bibinfo {author} {\bibfnamefont {V.}~\bibnamefont
  {Lucarini}}, \bibinfo {author} {\bibfnamefont {K.-E.}\ \bibnamefont
  {Peiponen}}, \bibinfo {author} {\bibfnamefont {J.~J.}\ \bibnamefont
  {Saarinen}}, \ and\ \bibinfo {author} {\bibfnamefont {E.~M.}\ \bibnamefont
  {Vartiainen}},\ }\href@noop {} {\emph {\bibinfo {title} {Kramers-Kronig
  Relations and Sum Rules in Linear Optics}}}\ (\bibinfo  {publisher} {Springer
  Berlin Heidelberg},\ \bibinfo {year} {2005})\BibitemShut {NoStop}%
\bibitem [{\citenamefont {Toll}(1956)}]{Toll:56}%
  \BibitemOpen
  \bibfield  {author} {\bibinfo {author} {\bibfnamefont {J.~S.}\ \bibnamefont
  {Toll}},\ }\href {\doibase 10.1103/PhysRev.104.1760} {\bibfield  {journal}
  {\bibinfo  {journal} {Phys. Rev.}\ }\textbf {\bibinfo {volume} {104}},\
  \bibinfo {pages} {1760} (\bibinfo {year} {1956})}\BibitemShut {NoStop}%
\bibitem [{\citenamefont {Altarelli}\ \emph {et~al.}(1972)\citenamefont
  {Altarelli}, \citenamefont {Dexter}, \citenamefont {Nussenzveig},\ and\
  \citenamefont {Smith}}]{Altarelli:72}%
  \BibitemOpen
  \bibfield  {author} {\bibinfo {author} {\bibfnamefont {M.}~\bibnamefont
  {Altarelli}}, \bibinfo {author} {\bibfnamefont {D.~L.}\ \bibnamefont
  {Dexter}}, \bibinfo {author} {\bibfnamefont {H.~M.}\ \bibnamefont
  {Nussenzveig}}, \ and\ \bibinfo {author} {\bibfnamefont {D.~Y.}\ \bibnamefont
  {Smith}},\ }\href {\doibase 10.1103/PhysRevB.6.4502} {\bibfield  {journal}
  {\bibinfo  {journal} {Phys. Rev. B}\ }\textbf {\bibinfo {volume} {6}},\
  \bibinfo {pages} {4502} (\bibinfo {year} {1972})}\BibitemShut {NoStop}%
\bibitem [{\citenamefont {Soref}\ and\ \citenamefont
  {Bennett}(1987)}]{Soref:87}%
  \BibitemOpen
  \bibfield  {author} {\bibinfo {author} {\bibfnamefont {R.}~\bibnamefont
  {Soref}}\ and\ \bibinfo {author} {\bibfnamefont {B.}~\bibnamefont
  {Bennett}},\ }\href {\doibase 10.1109/JQE.1987.1073206} {\bibfield  {journal}
  {\bibinfo  {journal} {IEEE Journal of Quantum Electronics}\ }\textbf
  {\bibinfo {volume} {23}},\ \bibinfo {pages} {123} (\bibinfo {year}
  {1987})}\BibitemShut {NoStop}%
\bibitem [{\citenamefont {Xu}\ \emph {et~al.}(2005)\citenamefont {Xu},
  \citenamefont {Schmidt}, \citenamefont {Pradhan},\ and\ \citenamefont
  {Lipson}}]{Xu:05}%
  \BibitemOpen
  \bibfield  {author} {\bibinfo {author} {\bibfnamefont {Q.}~\bibnamefont
  {Xu}}, \bibinfo {author} {\bibfnamefont {B.}~\bibnamefont {Schmidt}},
  \bibinfo {author} {\bibfnamefont {S.}~\bibnamefont {Pradhan}}, \ and\
  \bibinfo {author} {\bibfnamefont {M.}~\bibnamefont {Lipson}},\ }\href
  {\doibase 10.1038/nature03569} {\bibfield  {journal} {\bibinfo  {journal}
  {Nature}\ }\textbf {\bibinfo {volume} {435}},\ \bibinfo {pages} {325}
  (\bibinfo {year} {2005})}\BibitemShut {NoStop}%
\bibitem [{\citenamefont {Li}\ \emph {et~al.}(2025)\citenamefont {Li},
  \citenamefont {Luo}, \citenamefont {Li},\ and\ \citenamefont
  {Liu}}]{LiSY:25}%
  \BibitemOpen
  \bibfield  {author} {\bibinfo {author} {\bibfnamefont {S.}~\bibnamefont
  {Li}}, \bibinfo {author} {\bibfnamefont {W.}~\bibnamefont {Luo}}, \bibinfo
  {author} {\bibfnamefont {Z.}~\bibnamefont {Li}}, \ and\ \bibinfo {author}
  {\bibfnamefont {J.}~\bibnamefont {Liu}},\ }\href {\doibase 10.1103/f5v4-n5dw}
  {\bibfield  {journal} {\bibinfo  {journal} {Phys. Rev. Appl.}\ }\textbf
  {\bibinfo {volume} {24}},\ \bibinfo {pages} {014021} (\bibinfo {year}
  {2025})}\BibitemShut {NoStop}%
\bibitem [{\citenamefont {Henry}(1982)}]{Henry:82}%
  \BibitemOpen
  \bibfield  {author} {\bibinfo {author} {\bibfnamefont {C.}~\bibnamefont
  {Henry}},\ }\href {\doibase 10.1109/JQE.1982.1071522} {\bibfield  {journal}
  {\bibinfo  {journal} {IEEE Journal of Quantum Electronics}\ }\textbf
  {\bibinfo {volume} {18}},\ \bibinfo {pages} {259} (\bibinfo {year}
  {1982})}\BibitemShut {NoStop}%
\bibitem [{\citenamefont {Vahala}\ and\ \citenamefont
  {Yariv}(1983)}]{Vahala:83}%
  \BibitemOpen
  \bibfield  {author} {\bibinfo {author} {\bibfnamefont {K.}~\bibnamefont
  {Vahala}}\ and\ \bibinfo {author} {\bibfnamefont {A.}~\bibnamefont {Yariv}},\
  }\href {\doibase 10.1109/JQE.1983.1071986} {\bibfield  {journal} {\bibinfo
  {journal} {IEEE Journal of Quantum Electronics}\ }\textbf {\bibinfo {volume}
  {19}},\ \bibinfo {pages} {1096} (\bibinfo {year} {1983})}\BibitemShut
  {NoStop}%
\bibitem [{\citenamefont {Osinski}\ and\ \citenamefont
  {Buus}(1987)}]{Osinski:87}%
  \BibitemOpen
  \bibfield  {author} {\bibinfo {author} {\bibfnamefont {M.}~\bibnamefont
  {Osinski}}\ and\ \bibinfo {author} {\bibfnamefont {J.}~\bibnamefont {Buus}},\
  }\href {\doibase 10.1109/JQE.1987.1073204} {\bibfield  {journal} {\bibinfo
  {journal} {IEEE Journal of Quantum Electronics}\ }\textbf {\bibinfo {volume}
  {23}},\ \bibinfo {pages} {9} (\bibinfo {year} {1987})}\BibitemShut {NoStop}%
\bibitem [{\citenamefont {Smith}\ \emph {et~al.}(2004)\citenamefont {Smith},
  \citenamefont {Chang}, \citenamefont {Fuller}, \citenamefont {Rosenberger},\
  and\ \citenamefont {Boyd}}]{Smith:04}%
  \BibitemOpen
  \bibfield  {author} {\bibinfo {author} {\bibfnamefont {D.~D.}\ \bibnamefont
  {Smith}}, \bibinfo {author} {\bibfnamefont {H.}~\bibnamefont {Chang}},
  \bibinfo {author} {\bibfnamefont {K.~A.}\ \bibnamefont {Fuller}}, \bibinfo
  {author} {\bibfnamefont {A.~T.}\ \bibnamefont {Rosenberger}}, \ and\ \bibinfo
  {author} {\bibfnamefont {R.~W.}\ \bibnamefont {Boyd}},\ }\href {\doibase
  10.1103/PhysRevA.69.063804} {\bibfield  {journal} {\bibinfo  {journal} {Phys.
  Rev. A}\ }\textbf {\bibinfo {volume} {69}},\ \bibinfo {pages} {063804}
  (\bibinfo {year} {2004})}\BibitemShut {NoStop}%
\bibitem [{\citenamefont {Xu}\ \emph {et~al.}(2006)\citenamefont {Xu},
  \citenamefont {Sandhu}, \citenamefont {Povinelli}, \citenamefont {Shakya},
  \citenamefont {Fan},\ and\ \citenamefont {Lipson}}]{Xu:06}%
  \BibitemOpen
  \bibfield  {author} {\bibinfo {author} {\bibfnamefont {Q.}~\bibnamefont
  {Xu}}, \bibinfo {author} {\bibfnamefont {S.}~\bibnamefont {Sandhu}}, \bibinfo
  {author} {\bibfnamefont {M.~L.}\ \bibnamefont {Povinelli}}, \bibinfo {author}
  {\bibfnamefont {J.}~\bibnamefont {Shakya}}, \bibinfo {author} {\bibfnamefont
  {S.}~\bibnamefont {Fan}}, \ and\ \bibinfo {author} {\bibfnamefont
  {M.}~\bibnamefont {Lipson}},\ }\href {\doibase 10.1103/PhysRevLett.96.123901}
  {\bibfield  {journal} {\bibinfo  {journal} {Phys. Rev. Lett.}\ }\textbf
  {\bibinfo {volume} {96}},\ \bibinfo {pages} {123901} (\bibinfo {year}
  {2006})}\BibitemShut {NoStop}%
\bibitem [{\citenamefont {Totsuka}\ \emph {et~al.}(2007)\citenamefont
  {Totsuka}, \citenamefont {Kobayashi},\ and\ \citenamefont
  {Tomita}}]{Totsuka:07}%
  \BibitemOpen
  \bibfield  {author} {\bibinfo {author} {\bibfnamefont {K.}~\bibnamefont
  {Totsuka}}, \bibinfo {author} {\bibfnamefont {N.}~\bibnamefont {Kobayashi}},
  \ and\ \bibinfo {author} {\bibfnamefont {M.}~\bibnamefont {Tomita}},\ }\href
  {\doibase 10.1103/PhysRevLett.98.213904} {\bibfield  {journal} {\bibinfo
  {journal} {Phys. Rev. Lett.}\ }\textbf {\bibinfo {volume} {98}},\ \bibinfo
  {pages} {213904} (\bibinfo {year} {2007})}\BibitemShut {NoStop}%
\bibitem [{\citenamefont {Huet}\ \emph {et~al.}(2016)\citenamefont {Huet},
  \citenamefont {Rasoloniaina}, \citenamefont {Guillem\'e}, \citenamefont
  {Rochard}, \citenamefont {F\'eron}, \citenamefont {Mortier}, \citenamefont
  {Levenson}, \citenamefont {Bencheikh}, \citenamefont {Yacomotti},\ and\
  \citenamefont {Dumeige}}]{Huet:16}%
  \BibitemOpen
  \bibfield  {author} {\bibinfo {author} {\bibfnamefont {V.}~\bibnamefont
  {Huet}}, \bibinfo {author} {\bibfnamefont {A.}~\bibnamefont {Rasoloniaina}},
  \bibinfo {author} {\bibfnamefont {P.}~\bibnamefont {Guillem\'e}}, \bibinfo
  {author} {\bibfnamefont {P.}~\bibnamefont {Rochard}}, \bibinfo {author}
  {\bibfnamefont {P.}~\bibnamefont {F\'eron}}, \bibinfo {author} {\bibfnamefont
  {M.}~\bibnamefont {Mortier}}, \bibinfo {author} {\bibfnamefont
  {A.}~\bibnamefont {Levenson}}, \bibinfo {author} {\bibfnamefont
  {K.}~\bibnamefont {Bencheikh}}, \bibinfo {author} {\bibfnamefont
  {A.}~\bibnamefont {Yacomotti}}, \ and\ \bibinfo {author} {\bibfnamefont
  {Y.}~\bibnamefont {Dumeige}},\ }\href {\doibase
  10.1103/PhysRevLett.116.133902} {\bibfield  {journal} {\bibinfo  {journal}
  {Phys. Rev. Lett.}\ }\textbf {\bibinfo {volume} {116}},\ \bibinfo {pages}
  {133902} (\bibinfo {year} {2016})}\BibitemShut {NoStop}%
\bibitem [{\citenamefont {Mecozzi}\ \emph {et~al.}(2016)\citenamefont
  {Mecozzi}, \citenamefont {Antonelli},\ and\ \citenamefont
  {Shtaif}}]{Mecozzi:16}%
  \BibitemOpen
  \bibfield  {author} {\bibinfo {author} {\bibfnamefont {A.}~\bibnamefont
  {Mecozzi}}, \bibinfo {author} {\bibfnamefont {C.}~\bibnamefont {Antonelli}},
  \ and\ \bibinfo {author} {\bibfnamefont {M.}~\bibnamefont {Shtaif}},\ }\href
  {\doibase 10.1364/OPTICA.3.001220} {\bibfield  {journal} {\bibinfo  {journal}
  {Optica}\ }\textbf {\bibinfo {volume} {3}},\ \bibinfo {pages} {1220}
  (\bibinfo {year} {2016})}\BibitemShut {NoStop}%
\bibitem [{\citenamefont {Liu}(2020)}]{Liu:20b}%
  \BibitemOpen
  \bibfield  {author} {\bibinfo {author} {\bibfnamefont {J.}~\bibnamefont
  {Liu}},\ }{\selectlanguage {english}\emph {\bibinfo {title} {Silicon Nitride
  Integrated Nonlinear Photonics}}},\ \href {\doibase
  10.5075/epfl-thesis-10343} {Ph.D. thesis},\ \bibinfo  {school} {EPFL},
  \bibinfo {address} {Lausanne} (\bibinfo {year} {2020})\BibitemShut {NoStop}%
\bibitem [{\citenamefont {Luke}\ \emph {et~al.}(2015)\citenamefont {Luke},
  \citenamefont {Okawachi}, \citenamefont {Lamont}, \citenamefont {Gaeta},\
  and\ \citenamefont {Lipson}}]{Luke:15}%
  \BibitemOpen
  \bibfield  {author} {\bibinfo {author} {\bibfnamefont {K.}~\bibnamefont
  {Luke}}, \bibinfo {author} {\bibfnamefont {Y.}~\bibnamefont {Okawachi}},
  \bibinfo {author} {\bibfnamefont {M.~R.~E.}\ \bibnamefont {Lamont}}, \bibinfo
  {author} {\bibfnamefont {A.~L.}\ \bibnamefont {Gaeta}}, \ and\ \bibinfo
  {author} {\bibfnamefont {M.}~\bibnamefont {Lipson}},\ }\href {\doibase
  10.1364/OL.40.004823} {\bibfield  {journal} {\bibinfo  {journal} {Opt.
  Lett.}\ }\textbf {\bibinfo {volume} {40}},\ \bibinfo {pages} {4823} (\bibinfo
  {year} {2015})}\BibitemShut {NoStop}%
\bibitem [{\citenamefont {Corato-Zanarella}\ \emph {et~al.}(2024)\citenamefont
  {Corato-Zanarella}, \citenamefont {Ji}, \citenamefont {Mohanty},\ and\
  \citenamefont {Lipson}}]{Corato-Zanarella:24}%
  \BibitemOpen
  \bibfield  {author} {\bibinfo {author} {\bibfnamefont {M.}~\bibnamefont
  {Corato-Zanarella}}, \bibinfo {author} {\bibfnamefont {X.}~\bibnamefont
  {Ji}}, \bibinfo {author} {\bibfnamefont {A.}~\bibnamefont {Mohanty}}, \ and\
  \bibinfo {author} {\bibfnamefont {M.}~\bibnamefont {Lipson}},\ }\href
  {\doibase 10.1364/OE.505892} {\bibfield  {journal} {\bibinfo  {journal} {Opt.
  Express}\ }\textbf {\bibinfo {volume} {32}},\ \bibinfo {pages} {5718}
  (\bibinfo {year} {2024})}\BibitemShut {NoStop}%
\bibitem [{\citenamefont {Mao}\ \emph {et~al.}(2008)\citenamefont {Mao},
  \citenamefont {Tao}, \citenamefont {Xu}, \citenamefont {Sun}, \citenamefont
  {Yu}, \citenamefont {Lo},\ and\ \citenamefont {Kwong}}]{Mao:08}%
  \BibitemOpen
  \bibfield  {author} {\bibinfo {author} {\bibfnamefont {S.~C.}\ \bibnamefont
  {Mao}}, \bibinfo {author} {\bibfnamefont {S.~H.}\ \bibnamefont {Tao}},
  \bibinfo {author} {\bibfnamefont {Y.~L.}\ \bibnamefont {Xu}}, \bibinfo
  {author} {\bibfnamefont {X.~W.}\ \bibnamefont {Sun}}, \bibinfo {author}
  {\bibfnamefont {M.~B.}\ \bibnamefont {Yu}}, \bibinfo {author} {\bibfnamefont
  {G.~Q.}\ \bibnamefont {Lo}}, \ and\ \bibinfo {author} {\bibfnamefont {D.~L.}\
  \bibnamefont {Kwong}},\ }\href {\doibase 10.1364/OE.16.020809} {\bibfield
  {journal} {\bibinfo  {journal} {Opt. Express}\ }\textbf {\bibinfo {volume}
  {16}},\ \bibinfo {pages} {20809} (\bibinfo {year} {2008})}\BibitemShut
  {NoStop}%
\bibitem [{\citenamefont {Bauters}\ \emph {et~al.}(2011)\citenamefont
  {Bauters}, \citenamefont {Heck}, \citenamefont {John}, \citenamefont
  {Barton}, \citenamefont {Bruinink}, \citenamefont {Leinse}, \citenamefont
  {Heideman}, \citenamefont {Blumenthal},\ and\ \citenamefont
  {Bowers}}]{Bauters:11}%
  \BibitemOpen
  \bibfield  {author} {\bibinfo {author} {\bibfnamefont {J.~F.}\ \bibnamefont
  {Bauters}}, \bibinfo {author} {\bibfnamefont {M.~J.~R.}\ \bibnamefont
  {Heck}}, \bibinfo {author} {\bibfnamefont {D.~D.}\ \bibnamefont {John}},
  \bibinfo {author} {\bibfnamefont {J.~S.}\ \bibnamefont {Barton}}, \bibinfo
  {author} {\bibfnamefont {C.~M.}\ \bibnamefont {Bruinink}}, \bibinfo {author}
  {\bibfnamefont {A.}~\bibnamefont {Leinse}}, \bibinfo {author} {\bibfnamefont
  {R.~G.}\ \bibnamefont {Heideman}}, \bibinfo {author} {\bibfnamefont {D.~J.}\
  \bibnamefont {Blumenthal}}, \ and\ \bibinfo {author} {\bibfnamefont {J.~E.}\
  \bibnamefont {Bowers}},\ }\href {\doibase 10.1364/OE.19.024090} {\bibfield
  {journal} {\bibinfo  {journal} {Opt. Express}\ }\textbf {\bibinfo {volume}
  {19}},\ \bibinfo {pages} {24090} (\bibinfo {year} {2011})}\BibitemShut
  {NoStop}%
\bibitem [{\citenamefont {Frigg}\ \emph {et~al.}(2019)\citenamefont {Frigg},
  \citenamefont {Boes}, \citenamefont {Ren}, \citenamefont {Abdo},
  \citenamefont {Choi}, \citenamefont {Gees},\ and\ \citenamefont
  {Mitchell}}]{Frigg:19}%
  \BibitemOpen
  \bibfield  {author} {\bibinfo {author} {\bibfnamefont {A.}~\bibnamefont
  {Frigg}}, \bibinfo {author} {\bibfnamefont {A.}~\bibnamefont {Boes}},
  \bibinfo {author} {\bibfnamefont {G.}~\bibnamefont {Ren}}, \bibinfo {author}
  {\bibfnamefont {I.}~\bibnamefont {Abdo}}, \bibinfo {author} {\bibfnamefont
  {D.-Y.}\ \bibnamefont {Choi}}, \bibinfo {author} {\bibfnamefont
  {S.}~\bibnamefont {Gees}}, \ and\ \bibinfo {author} {\bibfnamefont
  {A.}~\bibnamefont {Mitchell}},\ }\href {\doibase 10.1364/OE.380758}
  {\bibfield  {journal} {\bibinfo  {journal} {Opt. Express}\ }\textbf {\bibinfo
  {volume} {27}},\ \bibinfo {pages} {37795} (\bibinfo {year}
  {2019})}\BibitemShut {NoStop}%
\bibitem [{\citenamefont {Wu}\ \emph {et~al.}(2021)\citenamefont {Wu},
  \citenamefont {Zhang}, \citenamefont {Zeng}, \citenamefont {Li},
  \citenamefont {Xie}, \citenamefont {Chen},\ and\ \citenamefont {Yu}}]{Wu:21}%
  \BibitemOpen
  \bibfield  {author} {\bibinfo {author} {\bibfnamefont {Z.}~\bibnamefont
  {Wu}}, \bibinfo {author} {\bibfnamefont {Y.}~\bibnamefont {Zhang}}, \bibinfo
  {author} {\bibfnamefont {S.}~\bibnamefont {Zeng}}, \bibinfo {author}
  {\bibfnamefont {J.}~\bibnamefont {Li}}, \bibinfo {author} {\bibfnamefont
  {Y.}~\bibnamefont {Xie}}, \bibinfo {author} {\bibfnamefont {Y.}~\bibnamefont
  {Chen}}, \ and\ \bibinfo {author} {\bibfnamefont {S.}~\bibnamefont {Yu}},\
  }\href {\doibase 10.1364/OE.438436} {\bibfield  {journal} {\bibinfo
  {journal} {Opt. Express}\ }\textbf {\bibinfo {volume} {29}},\ \bibinfo
  {pages} {29557} (\bibinfo {year} {2021})}\BibitemShut {NoStop}%
\bibitem [{\citenamefont {Lanford}\ and\ \citenamefont
  {Rand}(1978{\natexlab{a}})}]{Lanford:78}%
  \BibitemOpen
  \bibfield  {author} {\bibinfo {author} {\bibfnamefont {W.~A.}\ \bibnamefont
  {Lanford}}\ and\ \bibinfo {author} {\bibfnamefont {M.~J.}\ \bibnamefont
  {Rand}},\ }\href {\doibase 10.1063/1.325095} {\bibfield  {journal} {\bibinfo
  {journal} {Journal of Applied Physics}\ }\textbf {\bibinfo {volume} {49}},\
  \bibinfo {pages} {2473} (\bibinfo {year} {1978}{\natexlab{a}})}\BibitemShut
  {NoStop}%
\bibitem [{\citenamefont {Bugaev}\ \emph {et~al.}(2012)\citenamefont {Bugaev},
  \citenamefont {Zelenina},\ and\ \citenamefont {Volodin}}]{Bugaev:12}%
  \BibitemOpen
  \bibfield  {author} {\bibinfo {author} {\bibfnamefont {K.~O.}\ \bibnamefont
  {Bugaev}}, \bibinfo {author} {\bibfnamefont {A.~A.}\ \bibnamefont
  {Zelenina}}, \ and\ \bibinfo {author} {\bibfnamefont {V.~A.}\ \bibnamefont
  {Volodin}},\ }\href {\doibase https://doi.org/10.1155/2012/281851} {\bibfield
   {journal} {\bibinfo  {journal} {International Journal of Spectroscopy}\
  }\textbf {\bibinfo {volume} {2012}},\ \bibinfo {pages} {281851} (\bibinfo
  {year} {2012})}\BibitemShut {NoStop}%
\bibitem [{\citenamefont {Robertson}(1991)}]{Robertson:1991}%
  \BibitemOpen
  \bibfield  {author} {\bibinfo {author} {\bibfnamefont {J.}~\bibnamefont
  {Robertson}},\ }\href@noop {} {\bibfield  {journal} {\bibinfo  {journal}
  {Philosophical Magazine B}\ }\textbf {\bibinfo {volume} {63}},\ \bibinfo
  {pages} {47} (\bibinfo {year} {1991})}\BibitemShut {NoStop}%
\bibitem [{\citenamefont {Lanford}\ and\ \citenamefont
  {Rand}(1978{\natexlab{b}})}]{Lanford:1978}%
  \BibitemOpen
  \bibfield  {author} {\bibinfo {author} {\bibfnamefont {W.~A.}\ \bibnamefont
  {Lanford}}\ and\ \bibinfo {author} {\bibfnamefont {M.}~\bibnamefont {Rand}},\
  }\href@noop {} {\bibfield  {journal} {\bibinfo  {journal} {Journal of applied
  physics}\ }\textbf {\bibinfo {volume} {49}},\ \bibinfo {pages} {2473}
  (\bibinfo {year} {1978}{\natexlab{b}})}\BibitemShut {NoStop}%
\bibitem [{\citenamefont {K{\"a}rcher}\ \emph {et~al.}(1984)\citenamefont
  {K{\"a}rcher}, \citenamefont {Ley},\ and\ \citenamefont
  {Johnson}}]{Karcher:1984}%
  \BibitemOpen
  \bibfield  {author} {\bibinfo {author} {\bibfnamefont {R.}~\bibnamefont
  {K{\"a}rcher}}, \bibinfo {author} {\bibfnamefont {L.}~\bibnamefont {Ley}}, \
  and\ \bibinfo {author} {\bibfnamefont {R.}~\bibnamefont {Johnson}},\
  }\href@noop {} {\bibfield  {journal} {\bibinfo  {journal} {Physical Review
  B}\ }\textbf {\bibinfo {volume} {30}},\ \bibinfo {pages} {1896} (\bibinfo
  {year} {1984})}\BibitemShut {NoStop}%
\bibitem [{\citenamefont {Ye}\ \emph {et~al.}(2023)\citenamefont {Ye},
  \citenamefont {Jia}, \citenamefont {Huang}, \citenamefont {Shen},
  \citenamefont {Long}, \citenamefont {Shi}, \citenamefont {Luo}, \citenamefont
  {Gao}, \citenamefont {Sun}, \citenamefont {Guo}, \citenamefont {He},\ and\
  \citenamefont {Liu}}]{Ye:23}%
  \BibitemOpen
  \bibfield  {author} {\bibinfo {author} {\bibfnamefont {Z.}~\bibnamefont
  {Ye}}, \bibinfo {author} {\bibfnamefont {H.}~\bibnamefont {Jia}}, \bibinfo
  {author} {\bibfnamefont {Z.}~\bibnamefont {Huang}}, \bibinfo {author}
  {\bibfnamefont {C.}~\bibnamefont {Shen}}, \bibinfo {author} {\bibfnamefont
  {J.}~\bibnamefont {Long}}, \bibinfo {author} {\bibfnamefont {B.}~\bibnamefont
  {Shi}}, \bibinfo {author} {\bibfnamefont {Y.-H.}\ \bibnamefont {Luo}},
  \bibinfo {author} {\bibfnamefont {L.}~\bibnamefont {Gao}}, \bibinfo {author}
  {\bibfnamefont {W.}~\bibnamefont {Sun}}, \bibinfo {author} {\bibfnamefont
  {H.}~\bibnamefont {Guo}}, \bibinfo {author} {\bibfnamefont {J.}~\bibnamefont
  {He}}, \ and\ \bibinfo {author} {\bibfnamefont {J.}~\bibnamefont {Liu}},\
  }\href {\doibase 10.1364/PRJ.486379} {\bibfield  {journal} {\bibinfo
  {journal} {Photon. Res.}\ }\textbf {\bibinfo {volume} {11}},\ \bibinfo
  {pages} {558} (\bibinfo {year} {2023})}\BibitemShut {NoStop}%
\bibitem [{\citenamefont {Sun}\ \emph {et~al.}(2025)\citenamefont {Sun},
  \citenamefont {Chen}, \citenamefont {Li}, \citenamefont {Shen}, \citenamefont
  {Yu}, \citenamefont {Li}, \citenamefont {Long}, \citenamefont {Zheng},
  \citenamefont {Wang}, \citenamefont {Long}, \citenamefont {Chen},
  \citenamefont {Zhang}, \citenamefont {Shi}, \citenamefont {Gao},
  \citenamefont {Luo}, \citenamefont {Chen},\ and\ \citenamefont
  {Liu}}]{Sun:25}%
  \BibitemOpen
  \bibfield  {author} {\bibinfo {author} {\bibfnamefont {W.}~\bibnamefont
  {Sun}}, \bibinfo {author} {\bibfnamefont {Z.}~\bibnamefont {Chen}}, \bibinfo
  {author} {\bibfnamefont {L.}~\bibnamefont {Li}}, \bibinfo {author}
  {\bibfnamefont {C.}~\bibnamefont {Shen}}, \bibinfo {author} {\bibfnamefont
  {K.}~\bibnamefont {Yu}}, \bibinfo {author} {\bibfnamefont {S.}~\bibnamefont
  {Li}}, \bibinfo {author} {\bibfnamefont {J.}~\bibnamefont {Long}}, \bibinfo
  {author} {\bibfnamefont {H.}~\bibnamefont {Zheng}}, \bibinfo {author}
  {\bibfnamefont {L.}~\bibnamefont {Wang}}, \bibinfo {author} {\bibfnamefont
  {T.}~\bibnamefont {Long}}, \bibinfo {author} {\bibfnamefont {Q.}~\bibnamefont
  {Chen}}, \bibinfo {author} {\bibfnamefont {Z.}~\bibnamefont {Zhang}},
  \bibinfo {author} {\bibfnamefont {B.}~\bibnamefont {Shi}}, \bibinfo {author}
  {\bibfnamefont {L.}~\bibnamefont {Gao}}, \bibinfo {author} {\bibfnamefont
  {Y.-H.}\ \bibnamefont {Luo}}, \bibinfo {author} {\bibfnamefont
  {B.}~\bibnamefont {Chen}}, \ and\ \bibinfo {author} {\bibfnamefont
  {J.}~\bibnamefont {Liu}},\ }\href {\doibase 10.1038/s41377-025-01795-0}
  {\bibfield  {journal} {\bibinfo  {journal} {Light: Science \& Applications}\
  }\textbf {\bibinfo {volume} {14}},\ \bibinfo {pages} {179} (\bibinfo {year}
  {2025})}\BibitemShut {NoStop}%
\bibitem [{\citenamefont {Luo}\ \emph {et~al.}(2024)\citenamefont {Luo},
  \citenamefont {Shi}, \citenamefont {Sun}, \citenamefont {Chen}, \citenamefont
  {Huang}, \citenamefont {Wang}, \citenamefont {Long}, \citenamefont {Shen},
  \citenamefont {Ye}, \citenamefont {Guo},\ and\ \citenamefont {Liu}}]{Luo:24}%
  \BibitemOpen
  \bibfield  {author} {\bibinfo {author} {\bibfnamefont {Y.-H.}\ \bibnamefont
  {Luo}}, \bibinfo {author} {\bibfnamefont {B.}~\bibnamefont {Shi}}, \bibinfo
  {author} {\bibfnamefont {W.}~\bibnamefont {Sun}}, \bibinfo {author}
  {\bibfnamefont {R.}~\bibnamefont {Chen}}, \bibinfo {author} {\bibfnamefont
  {S.}~\bibnamefont {Huang}}, \bibinfo {author} {\bibfnamefont
  {Z.}~\bibnamefont {Wang}}, \bibinfo {author} {\bibfnamefont {J.}~\bibnamefont
  {Long}}, \bibinfo {author} {\bibfnamefont {C.}~\bibnamefont {Shen}}, \bibinfo
  {author} {\bibfnamefont {Z.}~\bibnamefont {Ye}}, \bibinfo {author}
  {\bibfnamefont {H.}~\bibnamefont {Guo}}, \ and\ \bibinfo {author}
  {\bibfnamefont {J.}~\bibnamefont {Liu}},\ }\href
  {https://doi.org/10.1038/s41377-024-01435-z} {\bibfield  {journal} {\bibinfo
  {journal} {Light: Science \& Applications}\ }\textbf {\bibinfo {volume} {13}}
  (\bibinfo {year} {2024})}\BibitemShut {NoStop}%
\bibitem [{\citenamefont {Shi}\ \emph {et~al.}(2025)\citenamefont {Shi},
  \citenamefont {Zheng}, \citenamefont {Hu}, \citenamefont {Zhao},
  \citenamefont {Shang}, \citenamefont {Zhong}, \citenamefont {Chen},
  \citenamefont {Luo}, \citenamefont {Long}, \citenamefont {Sun}, \citenamefont
  {Ma}, \citenamefont {Xie}, \citenamefont {Gao}, \citenamefont {Shen},
  \citenamefont {Wang}, \citenamefont {Liang}, \citenamefont {Zhang},\ and\
  \citenamefont {Liu}}]{Shi:25}%
  \BibitemOpen
  \bibfield  {author} {\bibinfo {author} {\bibfnamefont {B.}~\bibnamefont
  {Shi}}, \bibinfo {author} {\bibfnamefont {M.-Y.}\ \bibnamefont {Zheng}},
  \bibinfo {author} {\bibfnamefont {Y.}~\bibnamefont {Hu}}, \bibinfo {author}
  {\bibfnamefont {Y.}~\bibnamefont {Zhao}}, \bibinfo {author} {\bibfnamefont
  {Z.}~\bibnamefont {Shang}}, \bibinfo {author} {\bibfnamefont
  {Z.}~\bibnamefont {Zhong}}, \bibinfo {author} {\bibfnamefont
  {Z.}~\bibnamefont {Chen}}, \bibinfo {author} {\bibfnamefont {Y.-H.}\
  \bibnamefont {Luo}}, \bibinfo {author} {\bibfnamefont {J.}~\bibnamefont
  {Long}}, \bibinfo {author} {\bibfnamefont {W.}~\bibnamefont {Sun}}, \bibinfo
  {author} {\bibfnamefont {W.}~\bibnamefont {Ma}}, \bibinfo {author}
  {\bibfnamefont {X.-P.}\ \bibnamefont {Xie}}, \bibinfo {author} {\bibfnamefont
  {L.}~\bibnamefont {Gao}}, \bibinfo {author} {\bibfnamefont {C.}~\bibnamefont
  {Shen}}, \bibinfo {author} {\bibfnamefont {A.}~\bibnamefont {Wang}}, \bibinfo
  {author} {\bibfnamefont {W.}~\bibnamefont {Liang}}, \bibinfo {author}
  {\bibfnamefont {Q.}~\bibnamefont {Zhang}}, \ and\ \bibinfo {author}
  {\bibfnamefont {J.}~\bibnamefont {Liu}},\ }\href {\doibase
  10.1038/s41467-025-61970-0} {\bibfield  {journal} {\bibinfo  {journal}
  {Nature Communications}\ }\textbf {\bibinfo {volume} {16}},\ \bibinfo {pages}
  {7025} (\bibinfo {year} {2025})}\BibitemShut {NoStop}%
\bibitem [{\citenamefont {Shi}\ \emph {et~al.}(2026)\citenamefont {Shi},
  \citenamefont {Zhang}, \citenamefont {Zheng}, \citenamefont {Hu},
  \citenamefont {Zhong}, \citenamefont {Shang}, \citenamefont {Ma},
  \citenamefont {Xie}, \citenamefont {Bai}, \citenamefont {Luo}, \citenamefont
  {Wang}, \citenamefont {Guo}, \citenamefont {Zhang},\ and\ \citenamefont
  {Liu}}]{Shi:26}%
  \BibitemOpen
  \bibfield  {author} {\bibinfo {author} {\bibfnamefont {B.}~\bibnamefont
  {Shi}}, \bibinfo {author} {\bibfnamefont {C.}~\bibnamefont {Zhang}}, \bibinfo
  {author} {\bibfnamefont {M.-Y.}\ \bibnamefont {Zheng}}, \bibinfo {author}
  {\bibfnamefont {Y.}~\bibnamefont {Hu}}, \bibinfo {author} {\bibfnamefont
  {Z.}~\bibnamefont {Zhong}}, \bibinfo {author} {\bibfnamefont
  {Z.}~\bibnamefont {Shang}}, \bibinfo {author} {\bibfnamefont
  {W.}~\bibnamefont {Ma}}, \bibinfo {author} {\bibfnamefont {X.-P.}\
  \bibnamefont {Xie}}, \bibinfo {author} {\bibfnamefont {X.}~\bibnamefont
  {Bai}}, \bibinfo {author} {\bibfnamefont {Y.-H.}\ \bibnamefont {Luo}},
  \bibinfo {author} {\bibfnamefont {A.}~\bibnamefont {Wang}}, \bibinfo {author}
  {\bibfnamefont {H.}~\bibnamefont {Guo}}, \bibinfo {author} {\bibfnamefont
  {Q.}~\bibnamefont {Zhang}}, \ and\ \bibinfo {author} {\bibfnamefont
  {J.}~\bibnamefont {Liu}},\ }\href@noop {} {\enquote {\bibinfo {title}
  {Metrology-grade mid-infrared spectroscopy for multi-dimensional
  perception},}\ } (\bibinfo {year} {2026}),\ \Eprint
  {http://arxiv.org/abs/2602.00958} {arXiv:2602.00958 [physics.optics]}
  \BibitemShut {NoStop}%
\bibitem [{\citenamefont {Chen}\ \emph {et~al.}(2026)\citenamefont {Chen},
  \citenamefont {Zhong}, \citenamefont {Huang}, \citenamefont {Sun},
  \citenamefont {Zeng}, \citenamefont {Shi}, \citenamefont {Luo},\ and\
  \citenamefont {Liu}}]{Chen:26}%
  \BibitemOpen
  \bibfield  {author} {\bibinfo {author} {\bibfnamefont {D.}~\bibnamefont
  {Chen}}, \bibinfo {author} {\bibfnamefont {Z.}~\bibnamefont {Zhong}},
  \bibinfo {author} {\bibfnamefont {S.}~\bibnamefont {Huang}}, \bibinfo
  {author} {\bibfnamefont {J.}~\bibnamefont {Sun}}, \bibinfo {author}
  {\bibfnamefont {S.}~\bibnamefont {Zeng}}, \bibinfo {author} {\bibfnamefont
  {B.}~\bibnamefont {Shi}}, \bibinfo {author} {\bibfnamefont {Y.-H.}\
  \bibnamefont {Luo}}, \ and\ \bibinfo {author} {\bibfnamefont
  {J.}~\bibnamefont {Liu}},\ }\href {\doibase 10.1103/th1c-nml5} {\bibfield
  {journal} {\bibinfo  {journal} {Phys. Rev. Appl.}\ }\textbf {\bibinfo
  {volume} {25}},\ \bibinfo {pages} {014078} (\bibinfo {year}
  {2026})}\BibitemShut {NoStop}%
\bibitem [{\citenamefont {Li}\ \emph {et~al.}(2013)\citenamefont {Li},
  \citenamefont {Eftekhar}, \citenamefont {Xia},\ and\ \citenamefont
  {Adibi}}]{Li:13}%
  \BibitemOpen
  \bibfield  {author} {\bibinfo {author} {\bibfnamefont {Q.}~\bibnamefont
  {Li}}, \bibinfo {author} {\bibfnamefont {A.~A.}\ \bibnamefont {Eftekhar}},
  \bibinfo {author} {\bibfnamefont {Z.}~\bibnamefont {Xia}}, \ and\ \bibinfo
  {author} {\bibfnamefont {A.}~\bibnamefont {Adibi}},\ }\href {\doibase
  10.1103/PhysRevA.88.033816} {\bibfield  {journal} {\bibinfo  {journal} {Phys.
  Rev. A}\ }\textbf {\bibinfo {volume} {88}},\ \bibinfo {pages} {033816}
  (\bibinfo {year} {2013})}\BibitemShut {NoStop}%
\bibitem [{\citenamefont {Henry}\ \emph {et~al.}(1987)\citenamefont {Henry},
  \citenamefont {Kazarinov}, \citenamefont {Lee}, \citenamefont {Orlowsky},\
  and\ \citenamefont {Katz}}]{Henry:87}%
  \BibitemOpen
  \bibfield  {author} {\bibinfo {author} {\bibfnamefont {C.~H.}\ \bibnamefont
  {Henry}}, \bibinfo {author} {\bibfnamefont {R.~F.}\ \bibnamefont
  {Kazarinov}}, \bibinfo {author} {\bibfnamefont {H.~J.}\ \bibnamefont {Lee}},
  \bibinfo {author} {\bibfnamefont {K.~J.}\ \bibnamefont {Orlowsky}}, \ and\
  \bibinfo {author} {\bibfnamefont {L.~E.}\ \bibnamefont {Katz}},\ }\href
  {\doibase 10.1364/AO.26.002621} {\bibfield  {journal} {\bibinfo  {journal}
  {Appl. Opt.}\ }\textbf {\bibinfo {volume} {26}},\ \bibinfo {pages} {2621}
  (\bibinfo {year} {1987})}\BibitemShut {NoStop}%
\bibitem [{\citenamefont {Soref}\ \emph {et~al.}(2006)\citenamefont {Soref},
  \citenamefont {Emelett},\ and\ \citenamefont {Buchwald}}]{Soref:06b}%
  \BibitemOpen
  \bibfield  {author} {\bibinfo {author} {\bibfnamefont {R.~A.}\ \bibnamefont
  {Soref}}, \bibinfo {author} {\bibfnamefont {S.~J.}\ \bibnamefont {Emelett}},
  \ and\ \bibinfo {author} {\bibfnamefont {W.~R.}\ \bibnamefont {Buchwald}},\
  }\href {\doibase 10.1088/1464-4258/8/10/004} {\bibfield  {journal} {\bibinfo
  {journal} {Journal of Optics A: Pure and Applied Optics}\ }\textbf {\bibinfo
  {volume} {8}},\ \bibinfo {pages} {840} (\bibinfo {year} {2006})}\BibitemShut
  {NoStop}%
\bibitem [{\citenamefont {Kitamura}\ \emph {et~al.}(2007)\citenamefont
  {Kitamura}, \citenamefont {Pilon},\ and\ \citenamefont
  {Jonasz}}]{Kitamura:07}%
  \BibitemOpen
  \bibfield  {author} {\bibinfo {author} {\bibfnamefont {R.}~\bibnamefont
  {Kitamura}}, \bibinfo {author} {\bibfnamefont {L.}~\bibnamefont {Pilon}}, \
  and\ \bibinfo {author} {\bibfnamefont {M.}~\bibnamefont {Jonasz}},\ }\href
  {\doibase 10.1364/ao.46.008118} {\bibfield  {journal} {\bibinfo  {journal}
  {Applied Optics}\ }\textbf {\bibinfo {volume} {46}},\ \bibinfo {pages} {8118}
  (\bibinfo {year} {2007})}\BibitemShut {NoStop}%
\bibitem [{\citenamefont {Lin}\ \emph {et~al.}(2017)\citenamefont {Lin},
  \citenamefont {Luo}, \citenamefont {Gu}, \citenamefont {Kimerling},
  \citenamefont {Wada}, \citenamefont {Agarwal},\ and\ \citenamefont
  {Hu}}]{Lin:17}%
  \BibitemOpen
  \bibfield  {author} {\bibinfo {author} {\bibfnamefont {H.}~\bibnamefont
  {Lin}}, \bibinfo {author} {\bibfnamefont {Z.}~\bibnamefont {Luo}}, \bibinfo
  {author} {\bibfnamefont {T.}~\bibnamefont {Gu}}, \bibinfo {author}
  {\bibfnamefont {L.~C.}\ \bibnamefont {Kimerling}}, \bibinfo {author}
  {\bibfnamefont {K.}~\bibnamefont {Wada}}, \bibinfo {author} {\bibfnamefont
  {A.}~\bibnamefont {Agarwal}}, \ and\ \bibinfo {author} {\bibfnamefont
  {J.}~\bibnamefont {Hu}},\ }\href {\doibase 10.1515/nanoph-2017-0085}
  {\bibfield  {journal} {\bibinfo  {journal} {Nanophotonics}\ }\textbf
  {\bibinfo {volume} {7}},\ \bibinfo {pages} {393} (\bibinfo {year}
  {2017})}\BibitemShut {NoStop}%
\bibitem [{\citenamefont {Miller}\ \emph {et~al.}(2017)\citenamefont {Miller},
  \citenamefont {Yu}, \citenamefont {Ji}, \citenamefont {Griffith},
  \citenamefont {Cardenas}, \citenamefont {Gaeta},\ and\ \citenamefont
  {Lipson}}]{Miller:17}%
  \BibitemOpen
  \bibfield  {author} {\bibinfo {author} {\bibfnamefont {S.~A.}\ \bibnamefont
  {Miller}}, \bibinfo {author} {\bibfnamefont {M.}~\bibnamefont {Yu}}, \bibinfo
  {author} {\bibfnamefont {X.}~\bibnamefont {Ji}}, \bibinfo {author}
  {\bibfnamefont {A.~G.}\ \bibnamefont {Griffith}}, \bibinfo {author}
  {\bibfnamefont {J.}~\bibnamefont {Cardenas}}, \bibinfo {author}
  {\bibfnamefont {A.~L.}\ \bibnamefont {Gaeta}}, \ and\ \bibinfo {author}
  {\bibfnamefont {M.}~\bibnamefont {Lipson}},\ }\href {\doibase
  10.1364/optica.4.000707} {\bibfield  {journal} {\bibinfo  {journal} {Optica}\
  }\textbf {\bibinfo {volume} {4}},\ \bibinfo {pages} {707} (\bibinfo {year}
  {2017})}\BibitemShut {NoStop}%
\bibitem [{\citenamefont {Cai}\ \emph {et~al.}(2000)\citenamefont {Cai},
  \citenamefont {Painter},\ and\ \citenamefont {Vahala}}]{Cai:00}%
  \BibitemOpen
  \bibfield  {author} {\bibinfo {author} {\bibfnamefont {M.}~\bibnamefont
  {Cai}}, \bibinfo {author} {\bibfnamefont {O.}~\bibnamefont {Painter}}, \ and\
  \bibinfo {author} {\bibfnamefont {K.~J.}\ \bibnamefont {Vahala}},\ }\href
  {\doibase 10.1103/PhysRevLett.85.74} {\bibfield  {journal} {\bibinfo
  {journal} {Phys. Rev. Lett.}\ }\textbf {\bibinfo {volume} {85}},\ \bibinfo
  {pages} {74} (\bibinfo {year} {2000})}\BibitemShut {NoStop}%
\bibitem [{\citenamefont {Anderson}\ \emph {et~al.}(2022)\citenamefont
  {Anderson}, \citenamefont {Weng}, \citenamefont {Lihachev}, \citenamefont
  {Tikan}, \citenamefont {Liu},\ and\ \citenamefont
  {Kippenberg}}]{Anderson:22}%
  \BibitemOpen
  \bibfield  {author} {\bibinfo {author} {\bibfnamefont {M.~H.}\ \bibnamefont
  {Anderson}}, \bibinfo {author} {\bibfnamefont {W.}~\bibnamefont {Weng}},
  \bibinfo {author} {\bibfnamefont {G.}~\bibnamefont {Lihachev}}, \bibinfo
  {author} {\bibfnamefont {A.}~\bibnamefont {Tikan}}, \bibinfo {author}
  {\bibfnamefont {J.}~\bibnamefont {Liu}}, \ and\ \bibinfo {author}
  {\bibfnamefont {T.~J.}\ \bibnamefont {Kippenberg}},\ }\href {\doibase
  10.1038/s41467-022-31916-x} {\bibfield  {journal} {\bibinfo  {journal}
  {Nature Communications}\ }\textbf {\bibinfo {volume} {13}},\ \bibinfo {pages}
  {4764} (\bibinfo {year} {2022})}\BibitemShut {NoStop}%
\bibitem [{\citenamefont {Okawachi}\ \emph {et~al.}(2011)\citenamefont
  {Okawachi}, \citenamefont {Saha}, \citenamefont {Levy}, \citenamefont {Wen},
  \citenamefont {Lipson},\ and\ \citenamefont {Gaeta}}]{Okawachi:11}%
  \BibitemOpen
  \bibfield  {author} {\bibinfo {author} {\bibfnamefont {Y.}~\bibnamefont
  {Okawachi}}, \bibinfo {author} {\bibfnamefont {K.}~\bibnamefont {Saha}},
  \bibinfo {author} {\bibfnamefont {J.~S.}\ \bibnamefont {Levy}}, \bibinfo
  {author} {\bibfnamefont {Y.~H.}\ \bibnamefont {Wen}}, \bibinfo {author}
  {\bibfnamefont {M.}~\bibnamefont {Lipson}}, \ and\ \bibinfo {author}
  {\bibfnamefont {A.~L.}\ \bibnamefont {Gaeta}},\ }\href@noop {} {\bibfield
  {journal} {\bibinfo  {journal} {Opt. Lett.}\ }\textbf {\bibinfo {volume}
  {36}},\ \bibinfo {pages} {3398} (\bibinfo {year} {2011})}\BibitemShut
  {NoStop}%
\bibitem [{\citenamefont {Lugiato}\ and\ \citenamefont
  {Lefever}(1987)}]{Lugiato:87}%
  \BibitemOpen
  \bibfield  {author} {\bibinfo {author} {\bibfnamefont {L.~A.}\ \bibnamefont
  {Lugiato}}\ and\ \bibinfo {author} {\bibfnamefont {R.}~\bibnamefont
  {Lefever}},\ }\href {\doibase 10.1103/PhysRevLett.58.2209} {\bibfield
  {journal} {\bibinfo  {journal} {Phys. Rev. Lett.}\ }\textbf {\bibinfo
  {volume} {58}},\ \bibinfo {pages} {2209} (\bibinfo {year}
  {1987})}\BibitemShut {NoStop}%
\bibitem [{\citenamefont {Ji}\ \emph {et~al.}(2023)\citenamefont {Ji},
  \citenamefont {Okawachi}, \citenamefont {Gil-Molina}, \citenamefont
  {Corato-Zanarella}, \citenamefont {Roberts}, \citenamefont {Gaeta},\ and\
  \citenamefont {Lipson}}]{Ji:23}%
  \BibitemOpen
  \bibfield  {author} {\bibinfo {author} {\bibfnamefont {X.}~\bibnamefont
  {Ji}}, \bibinfo {author} {\bibfnamefont {Y.}~\bibnamefont {Okawachi}},
  \bibinfo {author} {\bibfnamefont {A.}~\bibnamefont {Gil-Molina}}, \bibinfo
  {author} {\bibfnamefont {M.}~\bibnamefont {Corato-Zanarella}}, \bibinfo
  {author} {\bibfnamefont {S.}~\bibnamefont {Roberts}}, \bibinfo {author}
  {\bibfnamefont {A.~L.}\ \bibnamefont {Gaeta}}, \ and\ \bibinfo {author}
  {\bibfnamefont {M.}~\bibnamefont {Lipson}},\ }\href {\doibase
  https://doi.org/10.1002/lpor.202200544} {\bibfield  {journal} {\bibinfo
  {journal} {Laser \& Photonics Reviews}\ }\textbf {\bibinfo {volume} {17}},\
  \bibinfo {pages} {2200544} (\bibinfo {year} {2023})}\BibitemShut {NoStop}%
\bibitem [{\citenamefont {Chiles}\ \emph {et~al.}(2018)\citenamefont {Chiles},
  \citenamefont {Nader}, \citenamefont {Hickstein}, \citenamefont {Yu},
  \citenamefont {Briles}, \citenamefont {Carlson}, \citenamefont {Jung},
  \citenamefont {Shainline}, \citenamefont {Diddams}, \citenamefont {Papp},
  \citenamefont {Nam},\ and\ \citenamefont {Mirin}}]{Chiles:18}%
  \BibitemOpen
  \bibfield  {author} {\bibinfo {author} {\bibfnamefont {J.}~\bibnamefont
  {Chiles}}, \bibinfo {author} {\bibfnamefont {N.}~\bibnamefont {Nader}},
  \bibinfo {author} {\bibfnamefont {D.~D.}\ \bibnamefont {Hickstein}}, \bibinfo
  {author} {\bibfnamefont {S.~P.}\ \bibnamefont {Yu}}, \bibinfo {author}
  {\bibfnamefont {T.~C.}\ \bibnamefont {Briles}}, \bibinfo {author}
  {\bibfnamefont {D.}~\bibnamefont {Carlson}}, \bibinfo {author} {\bibfnamefont
  {H.}~\bibnamefont {Jung}}, \bibinfo {author} {\bibfnamefont {J.~M.}\
  \bibnamefont {Shainline}}, \bibinfo {author} {\bibfnamefont {S.}~\bibnamefont
  {Diddams}}, \bibinfo {author} {\bibfnamefont {S.~B.}\ \bibnamefont {Papp}},
  \bibinfo {author} {\bibfnamefont {S.~W.}\ \bibnamefont {Nam}}, \ and\
  \bibinfo {author} {\bibfnamefont {R.~P.}\ \bibnamefont {Mirin}},\ }\href
  {\doibase 10.1364/OL.43.001527} {\bibfield  {journal} {\bibinfo  {journal}
  {Opt. Lett.}\ }\textbf {\bibinfo {volume} {43}},\ \bibinfo {pages} {1527}
  (\bibinfo {year} {2018})}\BibitemShut {NoStop}%
\bibitem [{\citenamefont {Chia}\ \emph {et~al.}(2023)\citenamefont {Chia},
  \citenamefont {Choi}, \citenamefont {Peng}, \citenamefont {Gao},
  \citenamefont {Chen}, \citenamefont {Ng},\ and\ \citenamefont
  {Tan}}]{Chia:23}%
  \BibitemOpen
  \bibfield  {author} {\bibinfo {author} {\bibfnamefont {X.~X.}\ \bibnamefont
  {Chia}}, \bibinfo {author} {\bibfnamefont {J.~W.}\ \bibnamefont {Choi}},
  \bibinfo {author} {\bibfnamefont {X.}~\bibnamefont {Peng}}, \bibinfo {author}
  {\bibfnamefont {H.}~\bibnamefont {Gao}}, \bibinfo {author} {\bibfnamefont
  {G.}~\bibnamefont {Chen}}, \bibinfo {author} {\bibfnamefont {D.~K.~T.}\
  \bibnamefont {Ng}}, \ and\ \bibinfo {author} {\bibfnamefont {D.}~\bibnamefont
  {Tan}},\ }\href {https://opg.optica.org/jlt/abstract.cfm?URI=jlt-41-10-3115}
  {\bibfield  {journal} {\bibinfo  {journal} {J. Lightwave Technol.}\ }\textbf
  {\bibinfo {volume} {41}},\ \bibinfo {pages} {3115} (\bibinfo {year}
  {2023})}\BibitemShut {NoStop}%
\bibitem [{\citenamefont {Xie}\ \emph {et~al.}(2022)\citenamefont {Xie},
  \citenamefont {Li}, \citenamefont {Zhang}, \citenamefont {Wu}, \citenamefont
  {Zeng}, \citenamefont {Lin}, \citenamefont {Wu}, \citenamefont {Zhou},
  \citenamefont {Chen},\ and\ \citenamefont {Yu}}]{Xie:22}%
  \BibitemOpen
  \bibfield  {author} {\bibinfo {author} {\bibfnamefont {Y.}~\bibnamefont
  {Xie}}, \bibinfo {author} {\bibfnamefont {J.}~\bibnamefont {Li}}, \bibinfo
  {author} {\bibfnamefont {Y.}~\bibnamefont {Zhang}}, \bibinfo {author}
  {\bibfnamefont {Z.}~\bibnamefont {Wu}}, \bibinfo {author} {\bibfnamefont
  {S.}~\bibnamefont {Zeng}}, \bibinfo {author} {\bibfnamefont {S.}~\bibnamefont
  {Lin}}, \bibinfo {author} {\bibfnamefont {Z.}~\bibnamefont {Wu}}, \bibinfo
  {author} {\bibfnamefont {W.}~\bibnamefont {Zhou}}, \bibinfo {author}
  {\bibfnamefont {Y.}~\bibnamefont {Chen}}, \ and\ \bibinfo {author}
  {\bibfnamefont {S.}~\bibnamefont {Yu}},\ }\href {\doibase 10.1364/PRJ.454816}
  {\bibfield  {journal} {\bibinfo  {journal} {Photon. Res.}\ }\textbf {\bibinfo
  {volume} {10}},\ \bibinfo {pages} {1290} (\bibinfo {year}
  {2022})}\BibitemShut {NoStop}%
\bibitem [{\citenamefont {Bose}\ \emph {et~al.}(2024)\citenamefont {Bose},
  \citenamefont {Harrington}, \citenamefont {Isichenko}, \citenamefont {Liu},
  \citenamefont {Wang}, \citenamefont {Chauhan}, \citenamefont {Newman},\ and\
  \citenamefont {Blumenthal}}]{Bose:24}%
  \BibitemOpen
  \bibfield  {author} {\bibinfo {author} {\bibfnamefont {D.}~\bibnamefont
  {Bose}}, \bibinfo {author} {\bibfnamefont {M.~W.}\ \bibnamefont
  {Harrington}}, \bibinfo {author} {\bibfnamefont {A.}~\bibnamefont
  {Isichenko}}, \bibinfo {author} {\bibfnamefont {K.}~\bibnamefont {Liu}},
  \bibinfo {author} {\bibfnamefont {J.}~\bibnamefont {Wang}}, \bibinfo {author}
  {\bibfnamefont {N.}~\bibnamefont {Chauhan}}, \bibinfo {author} {\bibfnamefont
  {Z.~L.}\ \bibnamefont {Newman}}, \ and\ \bibinfo {author} {\bibfnamefont
  {D.~J.}\ \bibnamefont {Blumenthal}},\ }\href {\doibase
  10.1038/s41377-024-01503-4} {\bibfield  {journal} {\bibinfo  {journal}
  {Light: Science \& Applications}\ }\textbf {\bibinfo {volume} {13}},\
  \bibinfo {pages} {156} (\bibinfo {year} {2024})}\BibitemShut {NoStop}%
\bibitem [{\citenamefont {Jin}\ \emph {et~al.}(2020)\citenamefont {Jin},
  \citenamefont {John}, \citenamefont {Bauters}, \citenamefont {Bosch},
  \citenamefont {Thibeault},\ and\ \citenamefont {Bowers}}]{Jin:20}%
  \BibitemOpen
  \bibfield  {author} {\bibinfo {author} {\bibfnamefont {W.}~\bibnamefont
  {Jin}}, \bibinfo {author} {\bibfnamefont {D.~D.}\ \bibnamefont {John}},
  \bibinfo {author} {\bibfnamefont {J.~F.}\ \bibnamefont {Bauters}}, \bibinfo
  {author} {\bibfnamefont {T.}~\bibnamefont {Bosch}}, \bibinfo {author}
  {\bibfnamefont {B.~J.}\ \bibnamefont {Thibeault}}, \ and\ \bibinfo {author}
  {\bibfnamefont {J.~E.}\ \bibnamefont {Bowers}},\ }\href {\doibase
  10.1364/OL.394121} {\bibfield  {journal} {\bibinfo  {journal} {Opt. Lett.}\
  }\textbf {\bibinfo {volume} {45}},\ \bibinfo {pages} {3340} (\bibinfo {year}
  {2020})}\BibitemShut {NoStop}%
\end{thebibliography}%


\begin{thebibliography}{20}%
\makeatletter
\providecommand \@ifxundefined [1]{%
 \@ifx{#1\undefined}
}%
\providecommand \@ifnum [1]{%
 \ifnum #1\expandafter \@firstoftwo
 \else \expandafter \@secondoftwo
 \fi
}%
\providecommand \@ifx [1]{%
 \ifx #1\expandafter \@firstoftwo
 \else \expandafter \@secondoftwo
 \fi
}%
\providecommand \natexlab [1]{#1}%
\providecommand \enquote  [1]{``#1''}%
\providecommand \bibnamefont  [1]{#1}%
\providecommand \bibfnamefont [1]{#1}%
\providecommand \citenamefont [1]{#1}%
\providecommand \href@noop [0]{\@secondoftwo}%
\providecommand \href [0]{\begingroup \@sanitize@url \@href}%
\providecommand \@href[1]{\@@startlink{#1}\@@href}%
\providecommand \@@href[1]{\endgroup#1\@@endlink}%
\providecommand \@sanitize@url [0]{\catcode `\\12\catcode `\$12\catcode
  `\&12\catcode `\#12\catcode `\^12\catcode `\_12\catcode `\%12\relax}%
\providecommand \@@startlink[1]{}%
\providecommand \@@endlink[0]{}%
\providecommand \url  [0]{\begingroup\@sanitize@url \@url }%
\providecommand \@url [1]{\endgroup\@href {#1}{\urlprefix }}%
\providecommand \urlprefix  [0]{URL }%
\providecommand \Eprint [0]{\href }%
\providecommand \doibase [0]{http://dx.doi.org/}%
\providecommand \selectlanguage [0]{\@gobble}%
\providecommand \bibinfo  [0]{\@secondoftwo}%
\providecommand \bibfield  [0]{\@secondoftwo}%
\providecommand \translation [1]{[#1]}%
\providecommand \BibitemOpen [0]{}%
\providecommand \bibitemStop [0]{}%
\providecommand \bibitemNoStop [0]{.\EOS\space}%
\providecommand \EOS [0]{\spacefactor3000\relax}%
\providecommand \BibitemShut  [1]{\csname bibitem#1\endcsname}%
\let\auto@bib@innerbib\@empty
\bibitem [{\citenamefont {Lanford}\ and\ \citenamefont
  {Rand}(1978{\natexlab{a}})}]{Lanford:78}%
  \BibitemOpen
  \bibfield  {author} {\bibinfo {author} {\bibfnamefont {W.~A.}\ \bibnamefont
  {Lanford}}\ and\ \bibinfo {author} {\bibfnamefont {M.~J.}\ \bibnamefont
  {Rand}},\ }\href {\doibase 10.1063/1.325095} {\bibfield  {journal} {\bibinfo
  {journal} {Journal of Applied Physics}\ }\textbf {\bibinfo {volume} {49}},\
  \bibinfo {pages} {2473} (\bibinfo {year} {1978}{\natexlab{a}})}\BibitemShut
  {NoStop}%
\bibitem [{\citenamefont {Jellison}\ and\ \citenamefont
  {Modine}(1996)}]{Jellison:96}%
  \BibitemOpen
  \bibfield  {author} {\bibinfo {author} {\bibfnamefont {J.}~\bibnamefont
  {Jellison}, \bibfnamefont {G.~E.}}\ and\ \bibinfo {author} {\bibfnamefont
  {F.~A.}\ \bibnamefont {Modine}},\ }\href {\doibase 10.1063/1.118064}
  {\bibfield  {journal} {\bibinfo  {journal} {Applied Physics Letters}\
  }\textbf {\bibinfo {volume} {69}},\ \bibinfo {pages} {371} (\bibinfo {year}
  {1996})}\BibitemShut {NoStop}%
\bibitem [{\citenamefont {Robertson}(1991)}]{Robertson:1991}%
  \BibitemOpen
  \bibfield  {author} {\bibinfo {author} {\bibfnamefont {J.}~\bibnamefont
  {Robertson}},\ }\href@noop {} {\bibfield  {journal} {\bibinfo  {journal}
  {Philosophical Magazine B}\ }\textbf {\bibinfo {volume} {63}},\ \bibinfo
  {pages} {47} (\bibinfo {year} {1991})}\BibitemShut {NoStop}%
\bibitem [{\citenamefont {Lanford}\ and\ \citenamefont
  {Rand}(1978{\natexlab{b}})}]{Lanford:1978}%
  \BibitemOpen
  \bibfield  {author} {\bibinfo {author} {\bibfnamefont {W.~A.}\ \bibnamefont
  {Lanford}}\ and\ \bibinfo {author} {\bibfnamefont {M.}~\bibnamefont {Rand}},\
  }\href@noop {} {\bibfield  {journal} {\bibinfo  {journal} {Journal of applied
  physics}\ }\textbf {\bibinfo {volume} {49}},\ \bibinfo {pages} {2473}
  (\bibinfo {year} {1978}{\natexlab{b}})}\BibitemShut {NoStop}%
\bibitem [{\citenamefont {K{\"a}rcher}\ \emph {et~al.}(1984)\citenamefont
  {K{\"a}rcher}, \citenamefont {Ley},\ and\ \citenamefont
  {Johnson}}]{Karcher:1984}%
  \BibitemOpen
  \bibfield  {author} {\bibinfo {author} {\bibfnamefont {R.}~\bibnamefont
  {K{\"a}rcher}}, \bibinfo {author} {\bibfnamefont {L.}~\bibnamefont {Ley}}, \
  and\ \bibinfo {author} {\bibfnamefont {R.}~\bibnamefont {Johnson}},\
  }\href@noop {} {\bibfield  {journal} {\bibinfo  {journal} {Physical Review
  B}\ }\textbf {\bibinfo {volume} {30}},\ \bibinfo {pages} {1896} (\bibinfo
  {year} {1984})}\BibitemShut {NoStop}%
\bibitem [{\citenamefont {Kramers}(1927)}]{Kramers:27}%
  \BibitemOpen
  \bibfield  {author} {\bibinfo {author} {\bibfnamefont {H.~A.}\ \bibnamefont
  {Kramers}},\ }\href {https://cir.nii.ac.jp/crid/1571135649980336384}
  {\bibfield  {journal} {\bibinfo  {journal} {Atti Cong. Intern. Fisica
  (Transactions of Volta Centenary Congress) Como}\ }\textbf {\bibinfo {volume}
  {2}},\ \bibinfo {pages} {545} (\bibinfo {year} {1927})}\BibitemShut {NoStop}%
\bibitem [{\citenamefont {de~L.~Kronig}(1926)}]{Kronig:26}%
  \BibitemOpen
  \bibfield  {author} {\bibinfo {author} {\bibfnamefont {R.}~\bibnamefont
  {de~L.~Kronig}},\ }\href {\doibase 10.1364/JOSA.12.000547} {\bibfield
  {journal} {\bibinfo  {journal} {J. Opt. Soc. Am.}\ }\textbf {\bibinfo
  {volume} {12}},\ \bibinfo {pages} {547} (\bibinfo {year} {1926})}\BibitemShut
  {NoStop}%
\bibitem [{\citenamefont {Lucarini}\ \emph {et~al.}(2005)\citenamefont
  {Lucarini}, \citenamefont {Peiponen}, \citenamefont {Saarinen},\ and\
  \citenamefont {Vartiainen}}]{Lucarini:05}%
  \BibitemOpen
  \bibfield  {author} {\bibinfo {author} {\bibfnamefont {V.}~\bibnamefont
  {Lucarini}}, \bibinfo {author} {\bibfnamefont {K.-E.}\ \bibnamefont
  {Peiponen}}, \bibinfo {author} {\bibfnamefont {J.~J.}\ \bibnamefont
  {Saarinen}}, \ and\ \bibinfo {author} {\bibfnamefont {E.~M.}\ \bibnamefont
  {Vartiainen}},\ }\href@noop {} {\emph {\bibinfo {title} {Kramers-Kronig
  Relations and Sum Rules in Linear Optics}}}\ (\bibinfo  {publisher} {Springer
  Berlin Heidelberg},\ \bibinfo {year} {2005})\BibitemShut {NoStop}%
\bibitem [{\citenamefont {Liu}(2020)}]{Liu:20b}%
  \BibitemOpen
  \bibfield  {author} {\bibinfo {author} {\bibfnamefont {J.}~\bibnamefont
  {Liu}},\ }{\selectlanguage {english}\emph {\bibinfo {title} {Silicon Nitride
  Integrated Nonlinear Photonics}}},\ \href {\doibase
  10.5075/epfl-thesis-10343} {Ph.D. thesis},\ \bibinfo  {school} {EPFL},
  \bibinfo {address} {Lausanne} (\bibinfo {year} {2020})\BibitemShut {NoStop}%
\bibitem [{\citenamefont {Ye}\ \emph {et~al.}(2023)\citenamefont {Ye},
  \citenamefont {Jia}, \citenamefont {Huang}, \citenamefont {Shen},
  \citenamefont {Long}, \citenamefont {Shi}, \citenamefont {Luo}, \citenamefont
  {Gao}, \citenamefont {Sun}, \citenamefont {Guo}, \citenamefont {He},\ and\
  \citenamefont {Liu}}]{Ye:23}%
  \BibitemOpen
  \bibfield  {author} {\bibinfo {author} {\bibfnamefont {Z.}~\bibnamefont
  {Ye}}, \bibinfo {author} {\bibfnamefont {H.}~\bibnamefont {Jia}}, \bibinfo
  {author} {\bibfnamefont {Z.}~\bibnamefont {Huang}}, \bibinfo {author}
  {\bibfnamefont {C.}~\bibnamefont {Shen}}, \bibinfo {author} {\bibfnamefont
  {J.}~\bibnamefont {Long}}, \bibinfo {author} {\bibfnamefont {B.}~\bibnamefont
  {Shi}}, \bibinfo {author} {\bibfnamefont {Y.-H.}\ \bibnamefont {Luo}},
  \bibinfo {author} {\bibfnamefont {L.}~\bibnamefont {Gao}}, \bibinfo {author}
  {\bibfnamefont {W.}~\bibnamefont {Sun}}, \bibinfo {author} {\bibfnamefont
  {H.}~\bibnamefont {Guo}}, \bibinfo {author} {\bibfnamefont {J.}~\bibnamefont
  {He}}, \ and\ \bibinfo {author} {\bibfnamefont {J.}~\bibnamefont {Liu}},\
  }\href {\doibase 10.1364/PRJ.486379} {\bibfield  {journal} {\bibinfo
  {journal} {Photon. Res.}\ }\textbf {\bibinfo {volume} {11}},\ \bibinfo
  {pages} {558} (\bibinfo {year} {2023})}\BibitemShut {NoStop}%
\bibitem [{\citenamefont {Luo}\ \emph {et~al.}(2024)\citenamefont {Luo},
  \citenamefont {Shi}, \citenamefont {Sun}, \citenamefont {Chen}, \citenamefont
  {Huang}, \citenamefont {Wang}, \citenamefont {Long}, \citenamefont {Shen},
  \citenamefont {Ye}, \citenamefont {Guo},\ and\ \citenamefont {Liu}}]{Luo:24}%
  \BibitemOpen
  \bibfield  {author} {\bibinfo {author} {\bibfnamefont {Y.-H.}\ \bibnamefont
  {Luo}}, \bibinfo {author} {\bibfnamefont {B.}~\bibnamefont {Shi}}, \bibinfo
  {author} {\bibfnamefont {W.}~\bibnamefont {Sun}}, \bibinfo {author}
  {\bibfnamefont {R.}~\bibnamefont {Chen}}, \bibinfo {author} {\bibfnamefont
  {S.}~\bibnamefont {Huang}}, \bibinfo {author} {\bibfnamefont
  {Z.}~\bibnamefont {Wang}}, \bibinfo {author} {\bibfnamefont {J.}~\bibnamefont
  {Long}}, \bibinfo {author} {\bibfnamefont {C.}~\bibnamefont {Shen}}, \bibinfo
  {author} {\bibfnamefont {Z.}~\bibnamefont {Ye}}, \bibinfo {author}
  {\bibfnamefont {H.}~\bibnamefont {Guo}}, \ and\ \bibinfo {author}
  {\bibfnamefont {J.}~\bibnamefont {Liu}},\ }\href
  {https://doi.org/10.1038/s41377-024-01435-z} {\bibfield  {journal} {\bibinfo
  {journal} {Light: Science \& Applications}\ }\textbf {\bibinfo {volume} {13}}
  (\bibinfo {year} {2024})}\BibitemShut {NoStop}%
\bibitem [{\citenamefont {Liu}\ \emph {et~al.}(2018)\citenamefont {Liu},
  \citenamefont {Raja}, \citenamefont {Pfeiffer}, \citenamefont {Herkommer},
  \citenamefont {Guo}, \citenamefont {Zervas}, \citenamefont {Geiselmann},\
  and\ \citenamefont {Kippenberg}}]{Liu:18}%
  \BibitemOpen
  \bibfield  {author} {\bibinfo {author} {\bibfnamefont {J.}~\bibnamefont
  {Liu}}, \bibinfo {author} {\bibfnamefont {A.~S.}\ \bibnamefont {Raja}},
  \bibinfo {author} {\bibfnamefont {M.~H.~P.}\ \bibnamefont {Pfeiffer}},
  \bibinfo {author} {\bibfnamefont {C.}~\bibnamefont {Herkommer}}, \bibinfo
  {author} {\bibfnamefont {H.}~\bibnamefont {Guo}}, \bibinfo {author}
  {\bibfnamefont {M.}~\bibnamefont {Zervas}}, \bibinfo {author} {\bibfnamefont
  {M.}~\bibnamefont {Geiselmann}}, \ and\ \bibinfo {author} {\bibfnamefont
  {T.~J.}\ \bibnamefont {Kippenberg}},\ }\href {\doibase 10.1364/OL.43.003200}
  {\bibfield  {journal} {\bibinfo  {journal} {Opt. Lett.}\ }\textbf {\bibinfo
  {volume} {43}},\ \bibinfo {pages} {3200} (\bibinfo {year}
  {2018})}\BibitemShut {NoStop}%
\bibitem [{\citenamefont {Pfeiffer}\ \emph {et~al.}(2017)\citenamefont
  {Pfeiffer}, \citenamefont {Herkommer}, \citenamefont {Liu}, \citenamefont
  {Guo}, \citenamefont {Karpov}, \citenamefont {Lucas}, \citenamefont
  {Zervas},\ and\ \citenamefont {Kippenberg}}]{Pfeiffer:17}%
  \BibitemOpen
  \bibfield  {author} {\bibinfo {author} {\bibfnamefont {M.~H.~P.}\
  \bibnamefont {Pfeiffer}}, \bibinfo {author} {\bibfnamefont {C.}~\bibnamefont
  {Herkommer}}, \bibinfo {author} {\bibfnamefont {J.}~\bibnamefont {Liu}},
  \bibinfo {author} {\bibfnamefont {H.}~\bibnamefont {Guo}}, \bibinfo {author}
  {\bibfnamefont {M.}~\bibnamefont {Karpov}}, \bibinfo {author} {\bibfnamefont
  {E.}~\bibnamefont {Lucas}}, \bibinfo {author} {\bibfnamefont
  {M.}~\bibnamefont {Zervas}}, \ and\ \bibinfo {author} {\bibfnamefont {T.~J.}\
  \bibnamefont {Kippenberg}},\ }\href {\doibase 10.1364/OPTICA.4.000684}
  {\bibfield  {journal} {\bibinfo  {journal} {Optica}\ }\textbf {\bibinfo
  {volume} {4}},\ \bibinfo {pages} {684} (\bibinfo {year} {2017})}\BibitemShut
  {NoStop}%
\bibitem [{\citenamefont {Chen}\ \emph {et~al.}(2026)\citenamefont {Chen},
  \citenamefont {Zhong}, \citenamefont {Huang}, \citenamefont {Sun},
  \citenamefont {Zeng}, \citenamefont {Shi}, \citenamefont {Luo},\ and\
  \citenamefont {Liu}}]{Chen:26}%
  \BibitemOpen
  \bibfield  {author} {\bibinfo {author} {\bibfnamefont {D.}~\bibnamefont
  {Chen}}, \bibinfo {author} {\bibfnamefont {Z.}~\bibnamefont {Zhong}},
  \bibinfo {author} {\bibfnamefont {S.}~\bibnamefont {Huang}}, \bibinfo
  {author} {\bibfnamefont {J.}~\bibnamefont {Sun}}, \bibinfo {author}
  {\bibfnamefont {S.}~\bibnamefont {Zeng}}, \bibinfo {author} {\bibfnamefont
  {B.}~\bibnamefont {Shi}}, \bibinfo {author} {\bibfnamefont {Y.-H.}\
  \bibnamefont {Luo}}, \ and\ \bibinfo {author} {\bibfnamefont
  {J.}~\bibnamefont {Liu}},\ }\href {\doibase 10.1103/th1c-nml5} {\bibfield
  {journal} {\bibinfo  {journal} {Phys. Rev. Appl.}\ }\textbf {\bibinfo
  {volume} {25}},\ \bibinfo {pages} {014078} (\bibinfo {year}
  {2026})}\BibitemShut {NoStop}%
\bibitem [{\citenamefont {Li}\ \emph {et~al.}(2013)\citenamefont {Li},
  \citenamefont {Eftekhar}, \citenamefont {Xia},\ and\ \citenamefont
  {Adibi}}]{Li:13}%
  \BibitemOpen
  \bibfield  {author} {\bibinfo {author} {\bibfnamefont {Q.}~\bibnamefont
  {Li}}, \bibinfo {author} {\bibfnamefont {A.~A.}\ \bibnamefont {Eftekhar}},
  \bibinfo {author} {\bibfnamefont {Z.}~\bibnamefont {Xia}}, \ and\ \bibinfo
  {author} {\bibfnamefont {A.}~\bibnamefont {Adibi}},\ }\href {\doibase
  10.1103/PhysRevA.88.033816} {\bibfield  {journal} {\bibinfo  {journal} {Phys.
  Rev. A}\ }\textbf {\bibinfo {volume} {88}},\ \bibinfo {pages} {033816}
  (\bibinfo {year} {2013})}\BibitemShut {NoStop}%
\bibitem [{\citenamefont {Shi}\ \emph {et~al.}(2025)\citenamefont {Shi},
  \citenamefont {Zheng}, \citenamefont {Hu}, \citenamefont {Zhao},
  \citenamefont {Shang}, \citenamefont {Zhong}, \citenamefont {Chen},
  \citenamefont {Luo}, \citenamefont {Long}, \citenamefont {Sun}, \citenamefont
  {Ma}, \citenamefont {Xie}, \citenamefont {Gao}, \citenamefont {Shen},
  \citenamefont {Wang}, \citenamefont {Liang}, \citenamefont {Zhang},\ and\
  \citenamefont {Liu}}]{Shi:25}%
  \BibitemOpen
  \bibfield  {author} {\bibinfo {author} {\bibfnamefont {B.}~\bibnamefont
  {Shi}}, \bibinfo {author} {\bibfnamefont {M.-Y.}\ \bibnamefont {Zheng}},
  \bibinfo {author} {\bibfnamefont {Y.}~\bibnamefont {Hu}}, \bibinfo {author}
  {\bibfnamefont {Y.}~\bibnamefont {Zhao}}, \bibinfo {author} {\bibfnamefont
  {Z.}~\bibnamefont {Shang}}, \bibinfo {author} {\bibfnamefont
  {Z.}~\bibnamefont {Zhong}}, \bibinfo {author} {\bibfnamefont
  {Z.}~\bibnamefont {Chen}}, \bibinfo {author} {\bibfnamefont {Y.-H.}\
  \bibnamefont {Luo}}, \bibinfo {author} {\bibfnamefont {J.}~\bibnamefont
  {Long}}, \bibinfo {author} {\bibfnamefont {W.}~\bibnamefont {Sun}}, \bibinfo
  {author} {\bibfnamefont {W.}~\bibnamefont {Ma}}, \bibinfo {author}
  {\bibfnamefont {X.-P.}\ \bibnamefont {Xie}}, \bibinfo {author} {\bibfnamefont
  {L.}~\bibnamefont {Gao}}, \bibinfo {author} {\bibfnamefont {C.}~\bibnamefont
  {Shen}}, \bibinfo {author} {\bibfnamefont {A.}~\bibnamefont {Wang}}, \bibinfo
  {author} {\bibfnamefont {W.}~\bibnamefont {Liang}}, \bibinfo {author}
  {\bibfnamefont {Q.}~\bibnamefont {Zhang}}, \ and\ \bibinfo {author}
  {\bibfnamefont {J.}~\bibnamefont {Liu}},\ }\href {\doibase
  10.1038/s41467-025-61970-0} {\bibfield  {journal} {\bibinfo  {journal}
  {Nature Communications}\ }\textbf {\bibinfo {volume} {16}},\ \bibinfo {pages}
  {7025} (\bibinfo {year} {2025})}\BibitemShut {NoStop}%
\bibitem [{\citenamefont {Henry}\ \emph {et~al.}(1987)\citenamefont {Henry},
  \citenamefont {Kazarinov}, \citenamefont {Lee}, \citenamefont {Orlowsky},\
  and\ \citenamefont {Katz}}]{Henry:87}%
  \BibitemOpen
  \bibfield  {author} {\bibinfo {author} {\bibfnamefont {C.~H.}\ \bibnamefont
  {Henry}}, \bibinfo {author} {\bibfnamefont {R.~F.}\ \bibnamefont
  {Kazarinov}}, \bibinfo {author} {\bibfnamefont {H.~J.}\ \bibnamefont {Lee}},
  \bibinfo {author} {\bibfnamefont {K.~J.}\ \bibnamefont {Orlowsky}}, \ and\
  \bibinfo {author} {\bibfnamefont {L.~E.}\ \bibnamefont {Katz}},\ }\href
  {\doibase 10.1364/AO.26.002621} {\bibfield  {journal} {\bibinfo  {journal}
  {Appl. Opt.}\ }\textbf {\bibinfo {volume} {26}},\ \bibinfo {pages} {2621}
  (\bibinfo {year} {1987})}\BibitemShut {NoStop}%
\bibitem [{\citenamefont {Shi}\ \emph {et~al.}(2026)\citenamefont {Shi},
  \citenamefont {Zhang}, \citenamefont {Zheng}, \citenamefont {Hu},
  \citenamefont {Zhong}, \citenamefont {Shang}, \citenamefont {Ma},
  \citenamefont {Xie}, \citenamefont {Bai}, \citenamefont {Luo}, \citenamefont
  {Wang}, \citenamefont {Guo}, \citenamefont {Zhang},\ and\ \citenamefont
  {Liu}}]{Shi:26}%
  \BibitemOpen
  \bibfield  {author} {\bibinfo {author} {\bibfnamefont {B.}~\bibnamefont
  {Shi}}, \bibinfo {author} {\bibfnamefont {C.}~\bibnamefont {Zhang}}, \bibinfo
  {author} {\bibfnamefont {M.-Y.}\ \bibnamefont {Zheng}}, \bibinfo {author}
  {\bibfnamefont {Y.}~\bibnamefont {Hu}}, \bibinfo {author} {\bibfnamefont
  {Z.}~\bibnamefont {Zhong}}, \bibinfo {author} {\bibfnamefont
  {Z.}~\bibnamefont {Shang}}, \bibinfo {author} {\bibfnamefont
  {W.}~\bibnamefont {Ma}}, \bibinfo {author} {\bibfnamefont {X.-P.}\
  \bibnamefont {Xie}}, \bibinfo {author} {\bibfnamefont {X.}~\bibnamefont
  {Bai}}, \bibinfo {author} {\bibfnamefont {Y.-H.}\ \bibnamefont {Luo}},
  \bibinfo {author} {\bibfnamefont {A.}~\bibnamefont {Wang}}, \bibinfo {author}
  {\bibfnamefont {H.}~\bibnamefont {Guo}}, \bibinfo {author} {\bibfnamefont
  {Q.}~\bibnamefont {Zhang}}, \ and\ \bibinfo {author} {\bibfnamefont
  {J.}~\bibnamefont {Liu}},\ }\href@noop {} {\enquote {\bibinfo {title}
  {Metrology-grade mid-infrared spectroscopy for multi-dimensional
  perception},}\ } (\bibinfo {year} {2026}),\ \Eprint
  {http://arxiv.org/abs/2602.00958} {arXiv:2602.00958 [physics.optics]}
  \BibitemShut {NoStop}%
\bibitem [{\citenamefont {Lugiato}\ and\ \citenamefont
  {Lefever}(1987)}]{Lugiato:87}%
  \BibitemOpen
  \bibfield  {author} {\bibinfo {author} {\bibfnamefont {L.~A.}\ \bibnamefont
  {Lugiato}}\ and\ \bibinfo {author} {\bibfnamefont {R.}~\bibnamefont
  {Lefever}},\ }\href {\doibase 10.1103/PhysRevLett.58.2209} {\bibfield
  {journal} {\bibinfo  {journal} {Phys. Rev. Lett.}\ }\textbf {\bibinfo
  {volume} {58}},\ \bibinfo {pages} {2209} (\bibinfo {year}
  {1987})}\BibitemShut {NoStop}%
\bibitem [{\citenamefont {Lugiato}\ \emph {et~al.}(2018)\citenamefont
  {Lugiato}, \citenamefont {Prati}, \citenamefont {Gorodetsky},\ and\
  \citenamefont {Kippenberg}}]{Lugiato:18}%
  \BibitemOpen
  \bibfield  {author} {\bibinfo {author} {\bibfnamefont {L.~A.}\ \bibnamefont
  {Lugiato}}, \bibinfo {author} {\bibfnamefont {F.}~\bibnamefont {Prati}},
  \bibinfo {author} {\bibfnamefont {M.~L.}\ \bibnamefont {Gorodetsky}}, \ and\
  \bibinfo {author} {\bibfnamefont {T.~J.}\ \bibnamefont {Kippenberg}},\ }\href
  {https://www.jstor.org/stable/26601869} {\bibfield  {journal} {\bibinfo
  {journal} {Philosophical Transactions: Mathematical, Physical and Engineering
  Sciences}\ }\textbf {\bibinfo {volume} {376}},\ \bibinfo {pages} {1}
  (\bibinfo {year} {2018})}\BibitemShut {NoStop}%
\end{thebibliography}%
\end{document}


\title{Supplementary Information for: Kramers--Kronig causality in integrated photonics:\\
The spectral tension between ultraviolet transition and mid-infrared absorption}

\author{Yue Hu}
\thanks{These authors contributed equally to this work.}
\affiliation{Southern University of Science and Technology, Shenzhen 518055, China}
\affiliation{International Quantum Academy and Shenzhen Futian SUSTech Institute for Quantum Technology and Engineering, Shenzhen 518048, China}

\author{Zhenyuan Shang}
\thanks{These authors contributed equally to this work.}
\affiliation{Southern University of Science and Technology, Shenzhen 518055, China}
\affiliation{International Quantum Academy and Shenzhen Futian SUSTech Institute for Quantum Technology and Engineering, Shenzhen 518048, China}

\author{Chenxi Zhang}
\affiliation{International Quantum Academy and Shenzhen Futian SUSTech Institute for Quantum Technology and Engineering, Shenzhen 518048, China}
\affiliation{College of Physics and Optoelectronic Engineering, Shenzhen University, Shenzhen 518060, China}

\author{Yuanjie Ning}
\affiliation{Shanghai Key Laboratory of High Temperature Superconductors, Institute for Quantum Science and Technology, Department of Physics, Shanghai University, Shanghai 200444, China}

\author{Weiqin Zheng}
\affiliation{International Quantum Academy and Shenzhen Futian SUSTech Institute for Quantum Technology and Engineering, Shenzhen 518048, China}
\affiliation{Institute of Advanced Photonics Technology, School of Information Engineering, Guangdong University of Technology, Guangzhou 510006, China}

\author{Dengke Chen}
\affiliation{Southern University of Science and Technology, Shenzhen 518055, China}
\affiliation{International Quantum Academy and Shenzhen Futian SUSTech Institute for Quantum Technology and Engineering, Shenzhen 518048, China}

\author{Sanli Huang}
\affiliation{International Quantum Academy and Shenzhen Futian SUSTech Institute for Quantum Technology and Engineering, Shenzhen 518048, China}
\affiliation{Hefei National Laboratory, University of Science and Technology of China, Hefei 230088, China}

\author{Baoqi Shi}
\affiliation{International Quantum Academy and Shenzhen Futian SUSTech Institute for Quantum Technology and Engineering, Shenzhen 518048, China}
\affiliation{Hefei National Laboratory, University of Science and Technology of China, Hefei 230088, China}

\author{Zeying Zhong}
\affiliation{Southern University of Science and Technology, Shenzhen 518055, China}
\affiliation{International Quantum Academy and Shenzhen Futian SUSTech Institute for Quantum Technology and Engineering, Shenzhen 518048, China}

\author{Hao Tan}
\affiliation{International Quantum Academy and Shenzhen Futian SUSTech Institute for Quantum Technology and Engineering, Shenzhen 518048, China}

\author{Wei Sun}
\affiliation{International Quantum Academy and Shenzhen Futian SUSTech Institute for Quantum Technology and Engineering, Shenzhen 518048, China}

\author{Yi-Han Luo}
\affiliation{International Quantum Academy and Shenzhen Futian SUSTech Institute for Quantum Technology and Engineering, Shenzhen 518048, China}

\author{Xinmao Yin}
\affiliation{Shanghai Key Laboratory of High Temperature Superconductors, Institute for Quantum Science and Technology, Department of Physics, Shanghai University, Shanghai 200444, China}

\author{Zhi-Chuan Niu}
\affiliation{International Quantum Academy and Shenzhen Futian SUSTech Institute for Quantum Technology and Engineering, Shenzhen 518048, China}
\affiliation{State Key Laboratory of Optoelectronic Materials and Devices, Institute of Semiconductors, Chinese Academy of Sciences, Beijing 100083, China}
\affiliation{Center of Materials Science and Optoelectronics Engineering, University of Chinese Academy of Sciences, Beijing 100049, China}

\author{Junqiu Liu}
\email[]{liujq@iqasz.cn}
\affiliation{International Quantum Academy and Shenzhen Futian SUSTech Institute for Quantum Technology and Engineering, Shenzhen 518048, China}
\affiliation{Hefei National Laboratory, University of Science and Technology of China, Hefei 230088, China}

\maketitle

\section{Characterization of silicon nitride films in the MIR}

\begin{figure*}[b!]
\renewcommand{\figurename}{Supplementary Figure}
\centering
\includegraphics{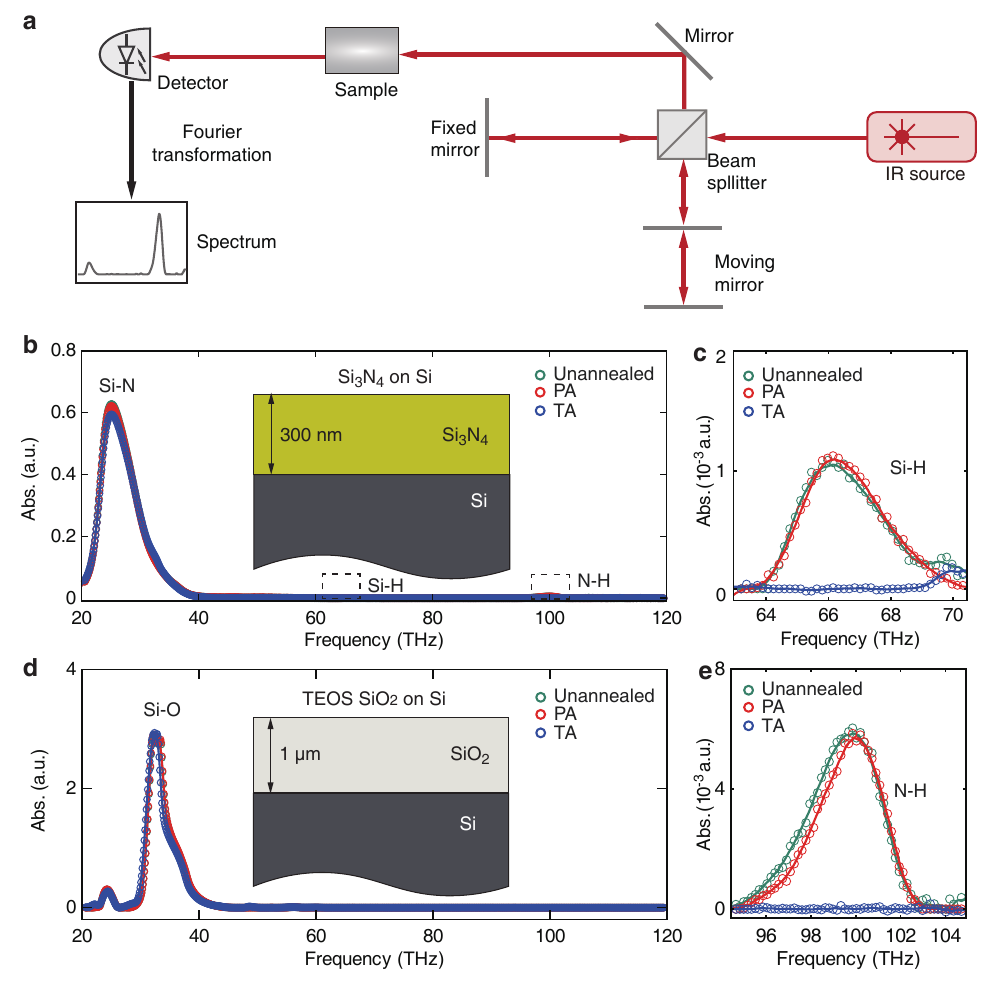}
\caption{
\textbf{Principle of FTIR spectroscopy and spectral characterization of Si$_3$N$_4$ and TEOS SiO$_2$ films.}
\textbf{a}. 
Schematic illustration of the FTIR operating principle. 
\textbf{b}. 
Absorption spectra of the Si$_3$N$_4$ film. 
Green, red, and blue curves correspond to the unannealed, partially annealed (PA), and thoroughly annealed (TA) states, respectively. 
Inset: Schematic of the Si$_3$N$_4$ film structure. 
\textbf{c, e}. 
Magnified views of the spectra around 67 THz and 100 THz, corresponding to Si--H and N--H bond vibrations, respectively. 
\textbf{d}.  
Absorption spectra of the TEOS SiO$_2$ film following PA and TA. 
Inset: Schematic diagram of the TEOS SiO$_2$ film structure.
}
\label{Fig:S1}
\end{figure*}

Fourier-transform infrared (FTIR) spectroscopy relies on interferometry coupled with mathematical transformation to acquire infrared absorption spectra. 
Figure~\ref{Fig:S1}a presents a schematic of the FTIR setup.  
In a typical configuration, broadband infrared radiation traverses a Michelson interferometer, generating a time-domain interference signal known as an interferogram\cite{Lanford:78}. 
This signal is recorded as a function of optical path difference and processed via a fast Fourier transform (FFT) to produce a spectrum of absorbance intensity versus wavenumber. Absorption occurs when the incident infrared frequency resonates with the vibrational modes of specific chemical bonds or functional groups, yielding characteristic peaks that serve as ``fingerprints'' for chemical identification and structural analysis. Due to its high signal-to-noise ratio, rapid data acquisition, and broad spectral coverage, FTIR is a critical tool in materials science, chemistry, and surface characterization.

To investigate whether the absorption characteristics of Si$_3$N$_4$ waveguides and SiO$_2$ cladding are modified by partial annealing (PA, $< 1050^\circ\text{C}$) and thorough annealing (TA, $> 1200^\circ\text{C}$), we performed FTIR measurements on blank films.
We deposited a 300-nm-thick Si$_3$N$_4$ film and a 1-$\mu\text{m}$-thick tetraethyl orthosilicate (TEOS) SiO$_2$ film on silicon substrates using LPCVD.
The Si$_3$N$_4$ film structure, depicted in Fig.~\ref{Fig:S1}b inset, was measured immediately after deposition. 
The FTIR measurement data is presented in Fig.~\ref{Fig:S1}b green curve. 
Figures~\ref{Fig:S1}c and e show the magnified views of the Si--H and N--H absorption peaks, respectively. 
Absorption peaks are found at approximately 27, 66 and 100 THz, corresponding to Si--N, Si--H, and N--H bond vibrations, respectively. 

Subsequently, the Si$_3$N$_4$ film underwent PA, while the absorption features remains unchanged as evidenced by the red curves in Fig. \ref{Fig:S1}b, c and e. 
However, following TA of the same Si$_3$N$_4$ film, the absorption peaks associated with Si--H and N--H bonds decrease significantly, as evidenced by the blue curves in Fig. \ref{Fig:S1}c and e. 
The Si--N peaks remains unchanged, as expected.

A similar procedure was applied to the TEOS SiO$_2$ film, whose structure is depicted in Fig.~\ref{Fig:S1}d inset. 
The FTIR measurement data presented in Fig.~\ref{Fig:S1}d shows that the Si--O absorption peak remains unchanged after TA. 

\section{Characterization of silicon nitride films in the UV and NIR}

\begin{figure*}[b!]
\renewcommand{\figurename}{Supplementary Figure}
\centering
\includegraphics{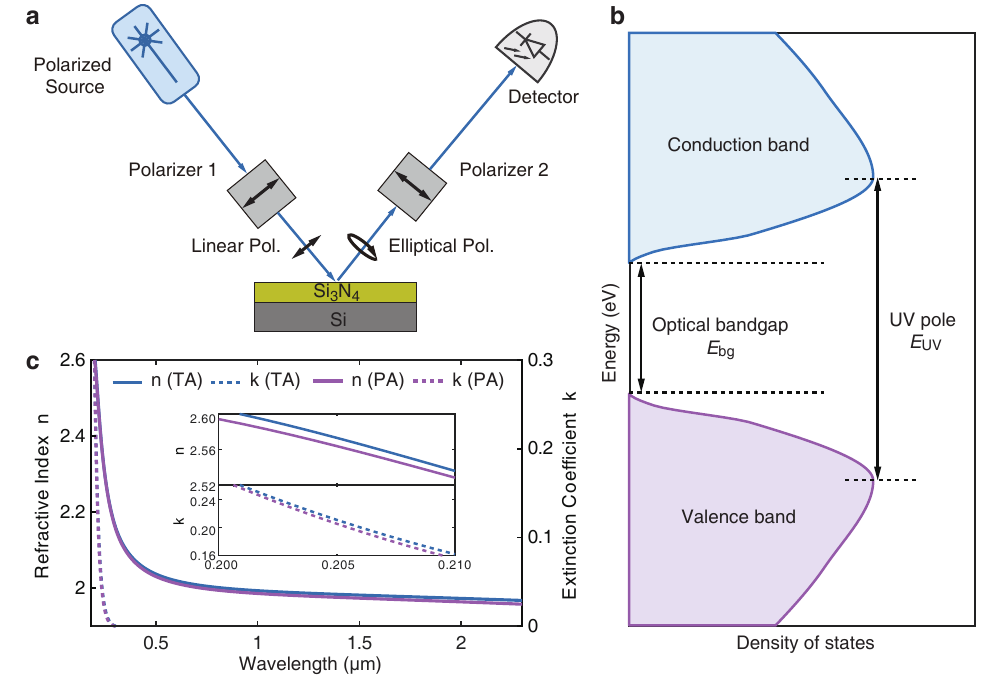}
\caption{
\textbf{Measurement principle of ellipsometry and extraction of complex refractive indices.} 
\textbf{a}. 
Schematic of the variable-angle spectroscopic ellipsometry (VASE) technique. 
The interaction of incident linearly polarized light with the sample induces a shift to elliptical polarization, quantified by the ellipsometric parameters ($\Psi, \Delta$).
\textbf{b}. 
Conceptual diagram of the Tauc-Lorentz model.
The optical bandgap acts as the minimum energy threshold for interband electronic transitions, whereas the UV pole represents the dominant high-energy transitions corresponding to the maximum density of states (DOS). 
\textbf{c}. 
Extracted wavelength-dependent refractive index $n$ and extinction coefficient $k$ for the Si$_3$N$_4$ film in the TA and PA states.
The inset highlights the notably higher $n$ and $k$ values of the TA film in 200--210 nm.
}
\label{Fig:S2_1}
\end{figure*}

The complex refractive index of Si$_3$N$_4$ were measured via variable-angle spectroscopic ellipsometry (VASE, Woollam RC2) over the 193--2300 nm (UV--NIR) spectral range.
As illustrated in Fig. \ref{Fig:S2_1}a, this technique relies on the change in the polarization state of light upon reflection. 
Specifically, when incident linearly polarized light interacts with the thin-film sample, the unequal reflection coefficients for the $p$- and $s$-polarized components transform the light into an elliptically polarized state. 
The instrument quantifies this change through the ellipsometric parameters $\Psi$ and $\Delta$, which represent the amplitude ratio and the relative phase shift between the $p$- and $s$-polarized reflected waves, respectively. 
These parameters fundamentally determine the complex reflectance ratio $\rho$ via the relation $\rho = \tan(\Psi)e^{i\Delta}$.

To extract the optical constants from $\rho$, a model-based regression analysis was performed. 
Specifically, the optical dispersion of Si$_3$N$_4$ was parameterized using the Tauc-Lorentz oscillator model \cite{Jellison:96}. 
This model combines the Tauc joint density of states with the classical Lorentz oscillator to describe interband electronic transitions.
The oscillator is described by four parameters: 
the optical bandgap energy $E_{\text{bg}}$, the Lorentz resonant center energy $E_0$, the broadening term $C$, and the oscillator amplitude $A$. 
The oscillator contributed imaginary part $\varepsilon_{2,\text{bg}}$ of the complex dielectric constant $\varepsilon = \varepsilon_1 + i(\varepsilon_{2,\text{bg}} + \varepsilon_{2,\text{pole}})$ is defined as:
\begin{equation}
    \varepsilon_{2,\text{bg}}(E) = \left \{
    \begin{aligned} 
     & \frac{A \cdot E_0 \cdot C \cdot (E - E_{\text{bg}})^2}{(E^2 - E_0^2)^2 + C^2 E^2} \cdot \frac{1}{E} & (E > E_{\text{bg}}) \\ 
    & 0 & (E \le E_{\text{bg}})
    \end{aligned}
    \right.
\label{eq:tauc_lorentz}
\end{equation}
where $E$ is the photon energy.
The $E_{\text{bg}}$ parameter defines the absorption threshold, enforcing that the extinction coefficient $k$ remains strictly zero for incident photon energies below the bandgap ($E < E_{\text{bg}}$), as depicted in Fig. \ref{Fig:S2_1}b.
This reflects the transparent nature of Si$_3$N$_4$ in the visible and NIR regimes.

Concurrently, to capture the dominant high-energy interband electronic transitions, an additional UV pole is introduced.
Mathematically, the UV pole is expressed as a delta function:
\begin{equation}
\varepsilon_{2,\text{pole}}=\frac{\pi}{2}A_{\text{UV}}\cdot \delta(E-E_{\text{UV}})
\end{equation}
where $A_{\text{UV}}$ is the amplitude and $E_{\text{UV}}$ is the energy of the UV pole.
This UV pole contributes to the baseline of the real part $\varepsilon_1$ of the complex dielectric constant, yielding the complete expression:
\begin{equation}
\begin{split}
\varepsilon_1(E) & = 1+\frac{2}{\pi}\mathcal{P} \int_{0}^{\infty} \frac{\xi \cdot [\varepsilon_{2,\text{bg}}(\xi)+\varepsilon_{2,\text{pole}}(\xi)]}{\xi^2 - E^2} \text{d}\xi\\
& =1+\frac{A_{\text{UV}}E_{\text{UV}}}{E_{\text{UV}}^2-E^2}+ \frac{2}{\pi}\mathcal{P} \int_{E_{\text{bg}}}^{\infty} \frac{\xi \cdot \varepsilon_{2,\text{bg}}(\xi)}{\xi^2 - E^2} \text{d}\xi
\end{split}
\end{equation}
By minimizing the mean squared error (MSE) between the experimental and generated spectra, the model coefficients were optimized.
This approach is anchored in fundamental physical boundaries, inherently satisfies KK causality, and ultimately yields the wavelength-dependent refractive index $n$ and extinction coefficient $k$.

\begin{table}[b!]
\centering
\caption{
\textbf{Fitted Tauc-Lorentz model parameters for the Si$_3$N$_4$ films in the TA and PA states.} 
The table summarizes the extracted amplitude ($A_{\text{UV}}$) and energy ($E_{\text{UV}}$) of the high-energy UV pole, alongside the amplitude ($A$), optical bandgap ($E_{\text{bg}}$), and resonant center energy ($E_0$) of the primary bandgap oscillator. 
To confirm film uniformity, measurements are conducted at three points on the film.
}

\setlength{\tabcolsep}{17pt} 
\begin{tabular}{l|c|c|c|c|c} 
\hline
\multirow{2}{*}{\textbf{Condition}} & \multicolumn{2}{c|}{\textbf{UV pole}} & \multicolumn{3}{c}{\textbf{Bandgap oscillator}} \\
\cline{2-3} \cline{4-6}
& $A_{\text{UV}}$ ($\text{eV}^2$) & $E_{\text{UV}}$ ($\text{eV}$) & $A$ ($\text{eV}^2$) & $E_{\text{bg}}$ ($\text{eV}$) & $E_0$ ($\text{eV}$) \\
\hline
TA (point 1) & 224.0036 & 10.049 & 36.1017 & 4.241 & 6.715 \\
TA (point 2) & 223.8792 & 10.051 & 36.5431 & 4.241 & 6.715 \\
TA (point 3) & 223.7952 & 10.051 & 36.9657 & 4.245 & 6.719 \\
\hline 
PA (point 1) & 217.5406 & 10.041 & 33.9602 & 4.303 & 6.750 \\
PA (point 2) & 218.4008 & 10.045 & 34.1365 & 4.309 & 6.759 \\
PA (point 3) & 218.4304 & 10.041 & 34.1047 & 4.306 & 6.755 \\
\hline
\end{tabular}
\label{tab:TL_params}
\end{table}

To quantify the optical properties of the Si$_3$N$_4$ film in the TA and PA states, the fitted parameters extracted from multiple points on our samples are summarized in Table \ref{tab:TL_params}.
The TA films exhibit a Lorentz center energy $E_0$ of 6.715--6.719 eV alongside a UV pole center energy $E_{\text{UV}}$ of 10.049--10.051 eV.
In contrast, \textbf{the hydrogen-rich PA films show a clear blue-shift of the bandgap center energy $E_0$ to 6.750--6.759 eV}.
This widening is primarily attributed to the higher bonding energies of the Si–H and N–H bonds compared to Si--N bonds, which effectively removes localized states near the band edges \cite{Robertson:1991, Lanford:1978, Karcher:1984}.
Conversely, \textbf{the PA films exhibit a red-shift of the UV pole $E_{\text{UV}}$ to 10.041--10.045 eV}.
Because the UV pole represents the average energy of deep interband transitions across the bulk continuous network, this red-shift reflects the structural relaxation and lower atomic density inherent to hydrogen-rich films \cite{Robertson:1991}. 
The hydrogen atoms act as network terminators, reducing the overall coordination number and disrupting the dense cross-linking of the Si–N matrix, which lowers the average bonding-antibonding energy splitting.
Furthermore, the amplitudes of both the bandgap oscillator ($A$) and the UV pole ($A_{\text{UV}}$) decrease in the PA state due to the replacement of Si–N bonds with less polarizable Si–H and N–H bonds.
Consequently, as depicted in Fig. \ref{Fig:S2_1}c, the extracted wavelength-dependent optical constants $n$ and $k$ reveal that the TA films exhibit a notably higher $n$ and $k$ than the PA films across the entire measured spectral range.

\section{Modification of refractive index and GVD by hydrogen}
Hydrogen exists within the Si$_3$N$_4$ waveguides as N--H and Si--H bonds, as detailed in Supplementary Note 1.
These chemical bonds not only affect the center frequency and amplitude of the bandgap and the UV pole (see Supplementary Note 2), but their absorption peaks also modify the refractive index via the KK relations \cite{Kramers:27,Kronig:26,Lucarini:05}, thereby altering the group velocity dispersion (GVD) primarily in the infrared regime.
The KK relations dictate that the real part of the refractive index $n(\omega)$ is linked to the extinction coefficient $\kappa(\omega)$, via the following integral transform:
\begin{equation}
\begin{split}
n(\omega) &= 1 + \frac{2}{\pi} \mathcal{P} \int_{0}^{\infty} \frac{\omega' \kappa(\omega')}{\omega'^2 - \omega^2} \rm{d} \omega'\\
\kappa(\omega) &= - \frac{2\omega}{\pi} \mathcal{P} \int_{0}^{\infty} \frac{n(\omega')-1}{\omega'^2 - \omega^2} \rm{d} \omega'
\label{Eq. kk}
\end{split}
\end{equation}
where $\omega$ is the angular frequency of interest, $\omega'$ is the integration variable over the entire spectrum, and $\mathcal{P}$ denotes the Cauchy principal value.
Based on Eq. \ref{Eq. kk}, we can isolate the perturbation to the refractive index $\Delta n(\omega)$, induced by a specific absorption feature using the following formulation:
\begin{equation}
\Delta n(\omega) = \frac{2}{\pi} \mathcal{P} \int_{\omega_{\rm{min}}}^{\omega_{\rm{max}}} \frac{\omega' \Delta \kappa(\omega')}{\omega'^2 - \omega^2} \rm{d} \omega'
\label{Eq. kk_dn}
\end{equation}
where $[\omega_{\rm{min}}, \omega_{\rm{max}}]$ represents the frequency range of the absorption peak and $\Delta \kappa(\omega)$ denotes the contribution of the peak to the extinction coefficient. 
In this approximation, the extinction coefficient outside this range is assumed to be unaffected. 

\begin{figure*}[b!]
\renewcommand{\figurename}{Supplementary Figure}
\centering
\includegraphics{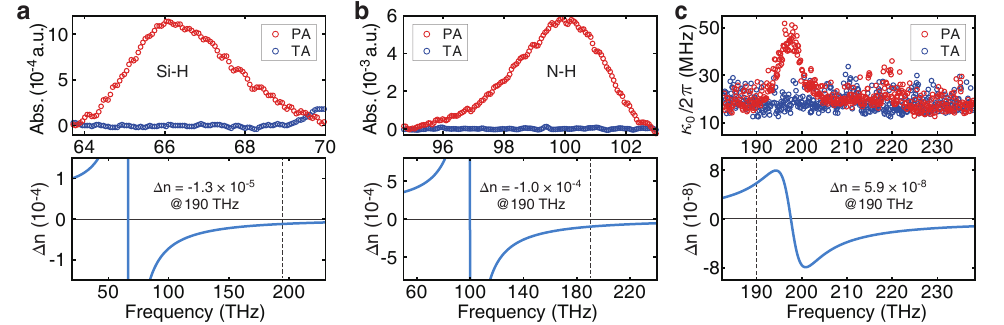}
\caption{
\textbf{Refractive index change induced by absorption peaks.}
Calculated $\Delta n$ contributions originating from the stretching vibrations of \textbf{a}. Si--H bonds, \textbf{b}. N--H bonds, and \textbf{c}. the first overtone of the fundamental N--H stretch.
The fundamental Si--H and N--H absorption peaks were characterized via FTIR spectroscopy, whereas the first overtone of the N--H stretch was measured using microresonators.
}
\label{Fig:S3}
\end{figure*}

Figure \ref{Fig:S3}a--c presents the calculated refractive index changes using Eq. \ref{Eq. kk_dn} for the absorption peaks at 66 THz (Si--H bond stretching), 100 THz (N--H bond stretching), and 197 THz (N--H first overtone).
The absorption profiles centered at 66 and 100 THz were derived from FTIR measurements (see Supplementary Note 1), while the peak at 197 THz was characterized via microresonator measurements (Fig. 2a in the main text).
In the spectral vicinity of each peak, the refractive index $n$ increases on the low-frequency side and decreases on the high-frequency side. 
The magnitude of this modulation scales with the absorption strength.
Specifically, the $\Delta n$ induced by the strong N--H fundamental vibration is on the order of $10^{-4}$, whereas the changes caused by the Si--H bond ($10^{-5}$) and the N--H overtone ($10^{-8}$) are significantly weaker in the NIR range. 
Consequently, the N--H bond stretching vibration is treated as the dominant perturbation in our analysis, while contributions from the Si--H bond and the N--H first overtone are neglected due to their minimal impact.

To quantitatively evaluate the hydrogen-induced modifications to the refractive index and GVD, we numerically simulated the dispersion of the Si$_3$N$_4$ microresonators in both the PA and TA states using FEM in COMSOL Multiphysics.
The refractive index model for the TA state is adopted from Ref. \cite{Liu:20b}:
\begin{equation}
\begin{split}
n^2(\lambda) = 1 &+ 1.750893106708352 \cdot \frac{\lambda^2}{\lambda^2-0.159333426051559^2}\\
&+ 1.222742709909494 \cdot \frac{\lambda^2}{\lambda^2-0.054765813233156^2}\\
&+ 2.584602943737831 \cdot \frac{\lambda^2}{\lambda^2-11.600661120564506^2}\\
\label{Eq. n_TA}
\end{split}
\end{equation}
where the first, second, and third terms represent the refractive index contributions from the bandgap, the UV pole, and the Si--N bond absorption at 11.6 $\mu$m, respectively.
In the hydrogen-rich PA state, the bandgap blue-shifts and the UV pole red-shifts, accompanied by a reduction in their oscillator amplitudes, as confirmed by the ellipsometry measurements (see Supplementary Note 2).
Accordingly, the refractive index model is modified for the PA state as follows:
\begin{equation}
\begin{split}
n^2(\lambda) = 1 &+ 1.750893106708352 \cdot A_1 \cdot \frac{\lambda^2}{\lambda^2-(0.159333426051559-\delta\lambda_{\rm{bg}})^2}\\
&+ 1.222742709909494 \cdot A_2 \cdot \frac{\lambda^2}{\lambda^2-(0.054765813233156+\delta\lambda_{\rm{pole}})^2}\\
&+ 2.584602943737831 \cdot \frac{\lambda^2}{\lambda^2-11.600661120564506^2}\\
&+ A_3 \cdot \frac{\lambda^2}{\lambda^2-\lambda_{\rm{N-H}}^2}\\
\label{Eq. n_PA}
\end{split}
\end{equation}
where $A_1=0.997$ and $A_2=0.990$ are the amplitude reduction coefficients for the bandgap and UV pole, respectively.
The parameters $\delta\lambda_{\rm{bg}}=0.00073$ $\mu$m and $\delta\lambda_{\rm{pole}}=0.02023$ $\mu$m denote the corresponding wavelength shifts.
Finally, the fourth term accounts for the refractive index contribution from the N--H bond absorption at $\lambda_{\mathrm{N-H}} = 3$ $\mu$m, with the amplitude coefficient $A_3 = 0.00105$ extracted by fitting the $\Delta n$ curve in Fig. \ref{Fig:S3}b.

\begin{figure*}[b!]
\renewcommand{\figurename}{Supplementary Figure}
\centering
\includegraphics{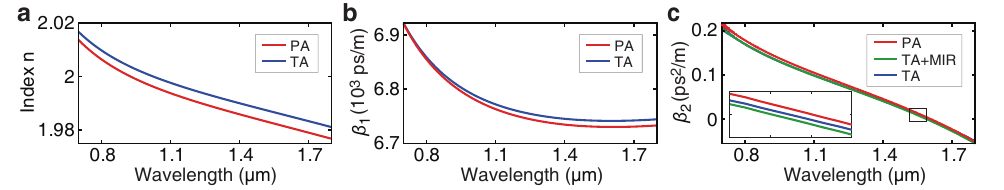}
\caption{
\textbf{Refractive index and dispersion for the Si$_3$N$_4$ film in the TA and PA states.} 
Calculated \textbf{a}. refractive index $n$, \textbf{b}. first-order dispersion parameter $\beta_1$, and \textbf{c}. GVD parameter $\beta_2$. 
The red and blue curves correspond to the PA and TA states, respectively.
The green curve in \textbf{c}. illustrates the $\beta_2$ calculated by considering solely the N--H bond absorption in MIR.
}
\label{Fig:n beta}
\end{figure*}

The refractive index $n$, the first-order dispersion parameter $\beta_1$, and the GVD parameter $\beta_2$, for the TA and PA states are presented in Fig. \ref{Fig:n beta}a--c, respectively.
The parameters $\beta_1$ and $\beta_2$ are defined as the first and second derivatives of the propagation constant $k(\omega)$ with respect to the angular frequency $\omega$:
\begin{equation}
\beta_1 = \frac{\text{d} k}{\rm{d}\omega} = \frac{1}{c} \left( n + \omega \frac{\text{d}n}{\rm{d}\omega} \right)
\end{equation}

\begin{equation}
\beta_2 = \frac{\text{d}^2 k}{\rm{d}\omega^2} = \frac{1}{c} \left( 2\frac{\text{d}n}{\rm{d}\omega} + \omega \frac{\text{d}^2n}{\rm{d}\omega^2} \right)
\end{equation}
Compared to the PA state, the TA state exhibits an increased $\beta_1$ and a decreased $\beta_2$ in the NIR and near-visible ranges.
These changes correspond to a decreased FSR ($D_1/2\pi$) and an increased $D_2/2\pi$ in the microresonators, consistent with the experimental results shown in Fig. 3 of the main text and Supplementary Notes 5--6.
The microresonator dispersion parameters $D_1$ and $D_2$ are related to $\beta_1$ and $\beta_2$ as follows:
\begin{subequations}\label{Eq. D1D2}
\begin{align}
 D_1 &= \frac{1}{R\beta_1} \label{Eq:D1}\\
 D_2 &= -D_1^3R\beta_2 \label{Eq:D2}
\end{align}
\end{subequations}
where $R$ is the radius of the microresonator. 
Notably, $\beta_2$ in Eq. \ref{Eq:D2} represents the total dispersion, which encompasses both the material dispersion (shown in Fig. \ref{Fig:n beta}c) and the geometric dispersion dictated by the thickness and width of the waveguide.

\begin{figure*}[b!]
\renewcommand{\figurename}{Supplementary Figure}
\centering
\includegraphics[width=\textwidth]{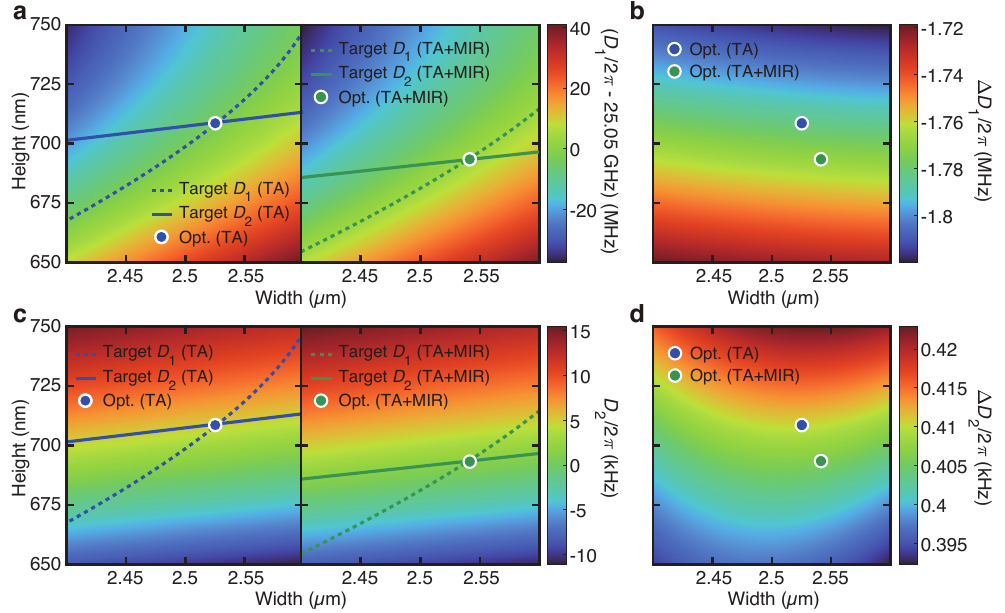}
\caption{
\textbf{Simulated dispersion landscapes of Si$_3$N$_4$ microresonators using the TA and MIR-only refractive index models.} 
\textbf{a}, Simulated $D_1/2\pi$ as a function of waveguide width ($w$) and height ($h$) for Si$_3$N$_4$ microresonators using the TA (left) and TA+MIR (right) refractive index models. 
The dashed contours represent the target $D_1/2\pi$ values (25.047 GHz for the TA model and 25.054 GHz for the MIR-only model). 
To highlight the variations, a baseline of 25.05 GHz is subtracted from the color scale. 
\textbf{b}, Calculated difference in $D_1/2\pi$ between the TA+MIR and TA models ($\Delta D_1/2\pi$). 
\textbf{c}, Simulated $D_2/2\pi$ landscapes for microresonators using the TA (left) and TA+MIR (right) models. 
The dashed contours represent the target $D_2/2\pi$ values (5.011 kHz for the TA model and 1.605 kHz for the TA+MIR model). 
\textbf{d}, Calculated difference in $D_2/2\pi$ between the MIR-only and TA models ($\Delta D_2/2\pi$). 
In all panels, the circles labeled ``Opt. (TA)'' and ``Opt. (TA+MIR)'' indicate the identified optimum geometric parameters for the TA and TA+MIR models, respectively, where the microresonator simultaneously satisfies the target $D_1$ and $D_2$ conditions. 
The extracted optimum dimensions are $(w, h) = (2.52~\mu\text{m}, 709~\text{nm})$ for the TA model and $(w, h) = (2.54~\mu\text{m}, 693~\text{nm})$ for the TA+MIR model.
}
\label{Fig:dispersion_sweep}
\end{figure*}

As the refractive index model of PA is that of PA with extra contribution of the UV bandgap, the UV pole, and the MIR absorption peak, in the following numerical simulation we define PA=TA+UV+MIR.
To systematically decouple the contributions of the UV bandgap, the UV pole, and the MIR absorption peak to the overall $\beta_2$, we first calculated the dispersion considering solely the MIR absorption (i.e. TA+MIR), which is presented as the green curve in Fig. \ref{Fig:n beta}c.
The green curve (TA+MIR) is lower than the blue curve (TA), indicating that the MIR absorption imparts a negative dispersion contribution. 
Conversely, when incorporating the effects of the UV bandgap and the UV pole alongside the MIR absorption, the resulting $\beta_2$ (red curve, PA) exceeds the blue curve (TA). 
This demonstrates that these UV features contribute positive dispersion.
Consequently, the dispersion of the PA state in the visible and NIR regimes is governed by the tension between the UV bandgap, the UV pole, and the MIR absorption.

Notably, the effects of the UV bandgap and the UV pole cannot be compensated for by geometric dispersion in dispersion engineering. 
By sweeping the waveguide width from 2.4 $\mu$m to 2.6 $\mu$m and the height from 650 nm to 750 nm, we mapped the simulated $D_1/2\pi$ and $D_2/2\pi$ landscapes for the Si$_3$N$_4$ microresonators using both the TA and TA+MIR refractive index models, as shown in Fig. \ref{Fig:dispersion_sweep}a,c. 
The corresponding differences ($\Delta D_1/2\pi$ and $\Delta D_2/2\pi$) between the TA+MIR and TA models are presented in Fig. \ref{Fig:dispersion_sweep}b,d.
Across the entire parameter space, the $D_1/2\pi$ of the TA+MIR index model is consistently smaller than that of the TA model, whereas its $D_2/2\pi$ is consistently larger.
Furthermore, we identified the optimal geometric parameters that simultaneously satisfy the target $D_1/2\pi$ and $D_2/2\pi$ values (corresponding to the measured PA and TA states shown in Fig. 3d of the main text) for both index models.
These optima are marked by the blue and green circles in Fig. \ref{Fig:dispersion_sweep}.
However, achieving these dispersion targets requires distinctly different geometric parameters for each model. 
This discrepancy clearly demonstrates that geometric tuning alone is insufficient to offset the intrinsic material dispersion modifications induced by the UV features.

\begin{figure*}[t!]
\renewcommand{\figurename}{Supplementary Figure}
\centering
\includegraphics{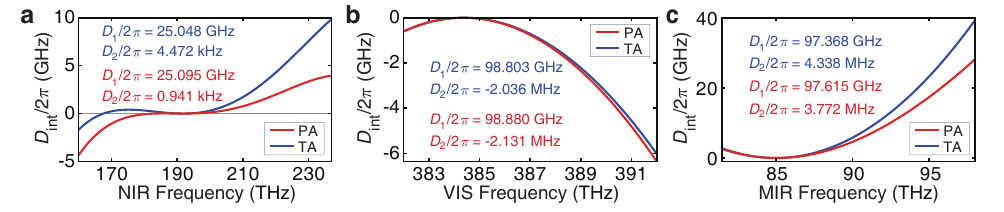}
\caption{
\textbf{Simulated microresonator dispersion in the PA and TA states.} 
Simulated integrated dispersion profile $D_\text{int}/2\pi$ in the \textbf{a}. NIR, \textbf{b}. visible (VIS), and \textbf{c}. MIR regimes. 
The red and blue curves correspond to the PA and TA states, respectively.
The cross-sectional dimensions (thickness $\times$ width) of the simulated microresonators are \textbf{a}. 696 nm $\times$ 2.5 $\mu$m, \textbf{b}. 297 nm $\times$ 1.6 $\mu$m, and \textbf{c}. 1150 nm $\times$ 3.0 $\mu$m.
}
\label{Fig:dispersion_sim}
\end{figure*}

Building upon the preceding analysis, both the UV and MIR dispersion effects are incorporated into the refractive index model of the PA state.
The simulated $D_{\text{int}}/2\pi$ of the Si$_3$N$_4$ microresonators are displayed in Fig. \ref{Fig:dispersion_sim}a--c.
The simulations were performed across the NIR (155--240 THz), visible (382--392 THz), and MIR (81--99 THz) regimes.
The cross-sectional dimensions (thickness $\times$ width) of the modeled microresonators are 696 nm $\times$ 2.5 $\mu$m for the NIR, 297 nm $\times$ 1.6 $\mu$m for the visible, and 1150 nm $\times$ 3.0 $\mu$m for the MIR regime.
Figures \ref{Fig:dispersion_sim}a,b show that, in the NIR and visible regimes, the PA state exhibits an increased $D_1/2\pi$ and a decreased $D_2/2\pi$ relative to the TA state.
These simulated $D_{\text{int}}/2\pi$ profiles shows the same trends as the experimental results presented in Fig. 2b,d in the main text.
However, in the MIR regime, the conventional Sellmeier equation breaks down near the strong absorption peaks.
To address this, the refractive index $n$ of the PA state around 3 $\mu$m was derived using the KK relations.
The corresponding simulated $D_{\text{int}}/2\pi$ profiles in the MIR regime are presented in Fig. \ref{Fig:dispersion_sim}c.

\begin{figure*}[b!]
\renewcommand{\figurename}{Supplementary Figure}
\centering
\includegraphics{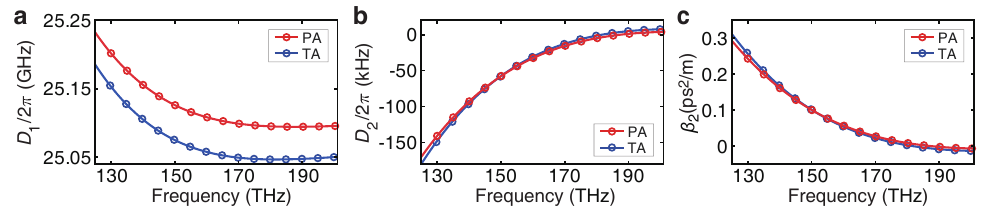}
\caption{
\textbf{Predicted dispersion balance point of the UV bandgap, the UV pole, and the MIR absorption.} 
Simulated \textbf{a}. $D_1/2\pi$, \textbf{b}. $D_2/2\pi$, and \textbf{c}. $\beta_2$ for a microresonator with cross-sectional dimensions of 697 nm $\times$ 2.5 $\mu$m. 
The red and blue curves correspond to the PA and TA states, respectively.
The $D_2/2\pi$ curves of the two states intersect at 149.1 THz (2.01 $\mu$m), while the $\beta_2$ curves intersect at 150.6 THz (1.99 $\mu$m).
}
\label{Fig:balance}
\end{figure*}

Furthermore, to predict the dispersion balance point of the UV bandgap, the UV pole, and the MIR absorption, we simulated the dispersion of a microresonator with cross-sectional dimensions of 697 nm $\times$ 2.5 $\mu$m.  
The simulated $D_1/2\pi$, $D_2/2\pi$, and $\beta_2$ profiles are presented in Fig. \ref{Fig:balance}a--c, respectively.
As shown in Fig. \ref{Fig:balance}a, the $D_1/2\pi$ of the TA state is consistently smaller than that of the PA state across the 125--200 THz frequency range.
However, the $D_2/2\pi$ curves intersect at 149.1 THz (2.01 $\mu$m), while the $\beta_2$ curves intersect at 150.6 THz (1.99 $\mu$m), revealing a distinct crossover.
These simulations indicate that the balance point between the UV and MIR dispersion contributions is located at approximately 2 $\mu$m.

\clearpage

\section{Fabrication processes of silicon nitride integrated waveguides}
\vspace{0.2cm}
\begin{figure*}[h!]
\renewcommand{\figurename}{Supplementary Figure}
\centering
\includegraphics{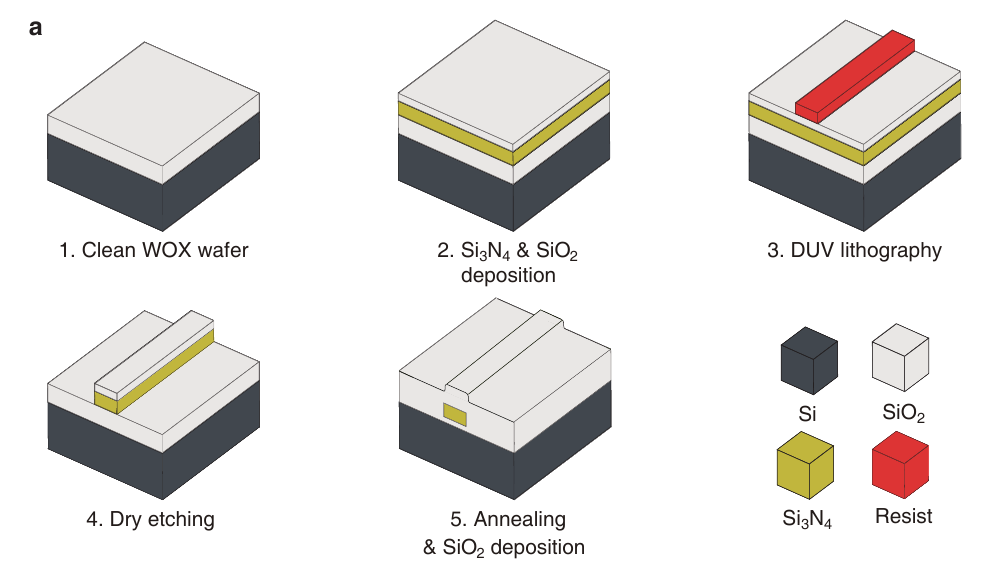}
\caption{
\textbf{Process flow of Si$_3$N$_4$ integrated waveguides.}
WOX, wet oxide.
DUV, deep-ultraviolet.
}
\label{Fig:S4}
\end{figure*}

The fabrication processes of Si$_3$N$_4$ integrated waveguides is described in the following:

\vspace{0.2cm}

The Si$_3$N$_4$ photonic integrated circuits (PIC) is fabricated using a deep-ultraviolet (DUV) subtractive process \cite{Ye:23}. 
The process flow is presented in Supplementary Fig. \ref{Fig:S4}. 
First, a 6-inch (150 mm) wafer with wet thermal SiO$_2$ bottom cladding undergoes a standard radio corporation of America (RCA) cleaning protocol to ensure optimal surface quality.

Subsequently, a 300-nm-thick Si$_3$N$_4$ stoichiometric silicon nitride (Si$_3$N$_4$) film is deposited via LPCVD at approximately 770°C, utilizing dichlorosilane (SiH$_2$Cl$_2$) and ammonia (NH$_3$) as precursor gases. 
A silicon dioxide (SiO$_2$) layer is then deposited via LPCVD to serve as a hard mask for pattern transfer.
The waveguide patterns are defined using DUV stepper photolithography (KrF source, 248 nm) with a resolution of 110 nm.
The pattern is subsequently transferred from the photoresist mask to the SiO$_2$ hardmask, and then into the Si$_3$N$_4$ layer to form waveguides, via inductively coupled plasma (ICP) dry etching. 
The ICP process parameters, including radio frequency (RF) power, chamber pressure, and gas flow ratios, are meticulously optimized to achieve vertical sidewall profiles and minimal surface roughness, both essential for reducing optical scattering losses.

Following resist removal, the device undergoes a critical high-temperature annealing step at approximately 1200$^\circ C$ in an inert nitrogen ambient. 
This annealing process serves to effectively eliminate hydrogen incorporation within the Si$_3$N$_4$ matrix, a common byproduct of LPCVD deposition that contributes significantly to optical absorption losses in the near-infrared spectrum, particularly through N--H and Si--H bond vibrations.

Finally, a 3800-nm-thick SiO$_2$ top cladding  is deposited via LPCVD at high temperature, completely encapsulating the waveguide structures. This cladding layer undergoes additional high-temperature annealing to enhance film density and reduce intrinsic stress, thereby improving long-term structural stability and optical performance. Then, the wafer is diced into chips for following characterization and experiment. 

\vspace{0.2cm}

\clearpage

\section{Characterization of microresonators in the NIR}

\begin{figure*}[b!]
\renewcommand{\figurename}{Supplementary Figure}
\centering
\includegraphics{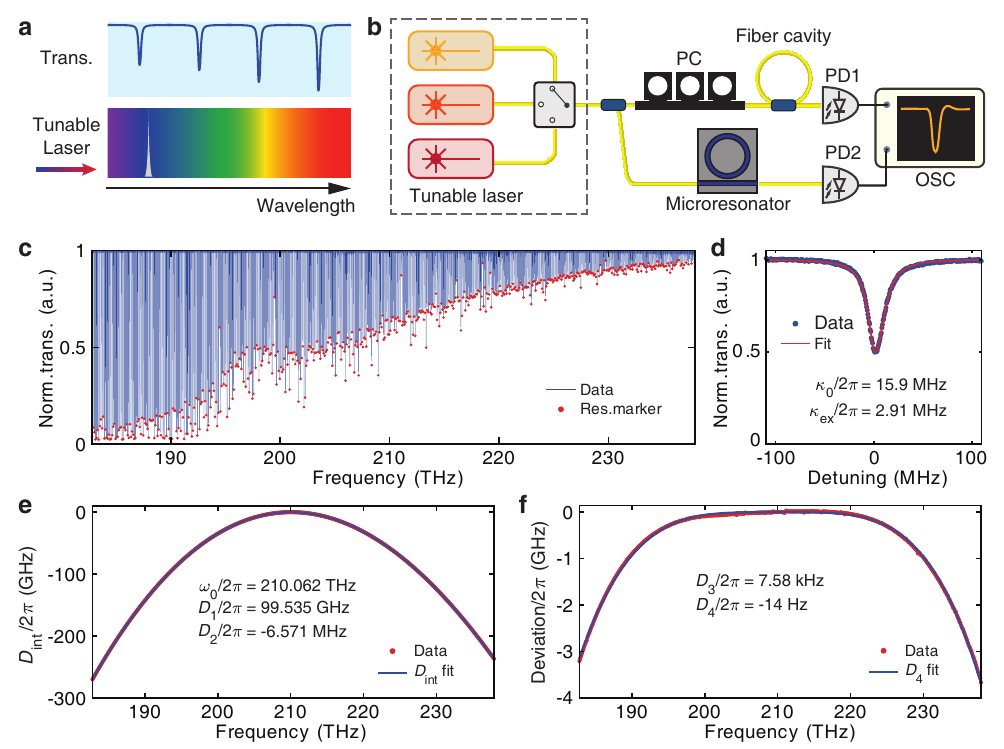}
\caption{
\textbf{Experimental setup to characterize microresonators' loss and dispersion in the near-infrared (NIR).} 
\textbf{a,b}. Schematic principle and experimental setup. 
A tunable laser chirps from 1260 to 1640 nm (182.9 to 237.9 THz), whose instantaneous frequency is calibrated on-the-fly by a pre-calibrated fiber cavity. A portion of the laser output is coupled into optical microresonators under study. 
During laser chirping, the frequency-calibrated microresonator transmission spectrum is recorded. 
PC, polarization controller.
PD, photodetector. 
OSC, oscilloscope 
\textbf{c}. A typical microresonator transmission spectrum. 
Each resonance is identified and fitted as shown in \textbf{d}, whose intrinsic loss $\kappa_0$/2$\pi$ is extracted and analyzed. 
\textbf{e}. Measured integrated microresonator's dispersion profile and fit up to the fourth order.
\textbf{f}. The corresponding high-order dispersion component derived from the fit in (e).
}
\label{Fig:S5}
\end{figure*}

The NIR characterization of Si$_3$N$_4$ microresonators is performed using a VSA operating across telecommunication bands\cite{Luo:24}.
The operating principle and experimental setup of the VSA are illustrated in Fig. \ref{Fig:S5}a and b, respectively. 
The system employs cascaded, mode-hop-free, external-cavity diode lasers (ECDL, Santec TSL 570) that chirp from 1260 to 1640 nm (182.9--237.9 THz).
During the sweep, the instantaneous laser frequency is calibrated in real-time using a reference fiber cavity\cite{Luo:24}. 
A portion of the laser output is coupled into the microresonators via lensed fibers that interface with inverse tapers and bus waveguides\cite{Liu:18,Pfeiffer:17,Chen:26}, allowing for the recording of frequency-calibrated transmission spectra.
Figure \ref{Fig:S5}c presents a typical transmission spectrum. 
We identified and fitted each resonance within the measurement range\cite{Li:13} to extract the intrinsic loss rate $\kappa_0/2\pi$, the external coupling rate $\kappa_{\rm{ex}}/2\pi$, and the total (loaded) linewidth $\kappa/2\pi=(\kappa_0 + \kappa_{\rm{ex}})/2\pi$ using Eq. \ref{fitres}:
\begin{equation}
 T(\Delta\omega) = \left|1-\frac{\kappa_{\rm{ex}}[i \Delta \omega + (\kappa_0 + \kappa_{\rm{ex}})/2]}{[i \Delta \omega + (\kappa_0 + \kappa_{\rm{ex}})/2]^2+\kappa_{\rm{c}}^2/4}\right|^2
 \label{fitres}
\end{equation}
where $T$ represents the transmission, $\Delta \omega/2\pi$ is the laser detuning relative to the resonance, and $\kappa_{\rm{c}}/2\pi$ is the complex coupling coefficient between the clockwise and counter-clockwise modes (accounting for mode splitting).
Figure \ref{Fig:S5}d shows a representative resonance fit alongside the extracted values for $\kappa_0/2\pi$ and $\kappa_{\rm{ex}}/2\pi$.

Furthermore, VSA characterizes the integrated dispersion profile $D_\text{int}/2\pi$ of each microresonator, defined as:
\begin{equation}
D_\text{int}(\mu)=\omega_{\mu}-\omega_{0}-D_{1}\mu=\sum\limits_{n=2}^{...} \frac{D_n \mu^n}{n!}
\label{dispersion}
\end{equation}
where $\omega_{\mu}/2\pi$ is the measured frequency of the $\mu$-th resonance relative to a reference resonance at $\omega_{0}/2\pi$, $D_1/2\pi$ is the microresonator free spectral range (FSR), $D_2/2\pi$ represents the second-order dispersion, and the $D_{\rm{n}}/2\pi$ terms (for $n\geqslant3$) correspond to high-order dispersion parameters.
Typical $D_\text{int}$ profiles for a 300-nm-thick Si$_3$N$_4$ microresonator, along with the corresponding high-order dispersion curves are fitted based on Eq. \ref{dispersion}, as shown in Fig. \ref{Fig:S5}e and f.

\begin{figure*}[t!]
\renewcommand{\figurename}{Supplementary Figure}
\centering
\includegraphics{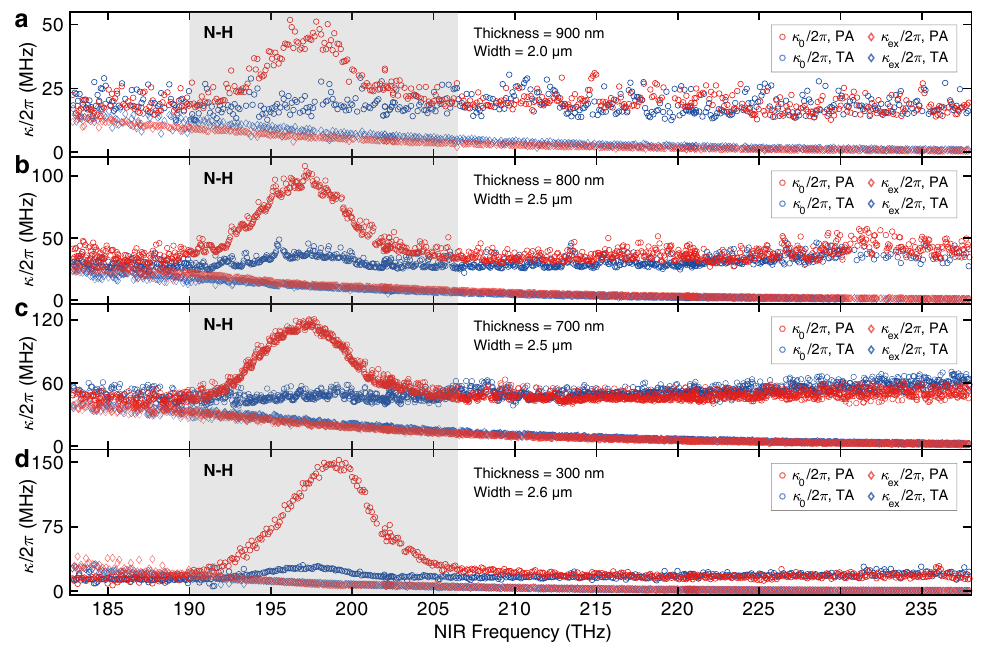}
\caption{
\textbf{Near-infrared characterization of Si$_3$N$_4$ microresonators' losses.}
Red and blue data points correspond to PA and TA states, respectively. 
Gray-shaded regions indicate absorption features induced by the first overtone of N--H bond stretching vibrations.
Circles represent the intrinsic loss rate ($\kappa_0/2\pi$), while diamonds represent the external coupling rate ($\kappa_{\text{ex}}/2\pi$).
The measurements are performed on microresonators with the following dimensions (thickness $\times$ width):
\textbf{a}. 1050 nm $\times$ 2.0 $\mu$m;
\textbf{b}. 800 nm $\times$ 2.5 $\mu$m;
\textbf{c}. 700 nm $\times$ 2.5 $\mu$m; and
\textbf{d}. 300 nm $\times$ 2.6 $\mu$m.
}
\label{Fig:S6}
\end{figure*}

\begin{figure*}[t!]
\renewcommand{\figurename}{Supplementary Figure}
\centering
\includegraphics{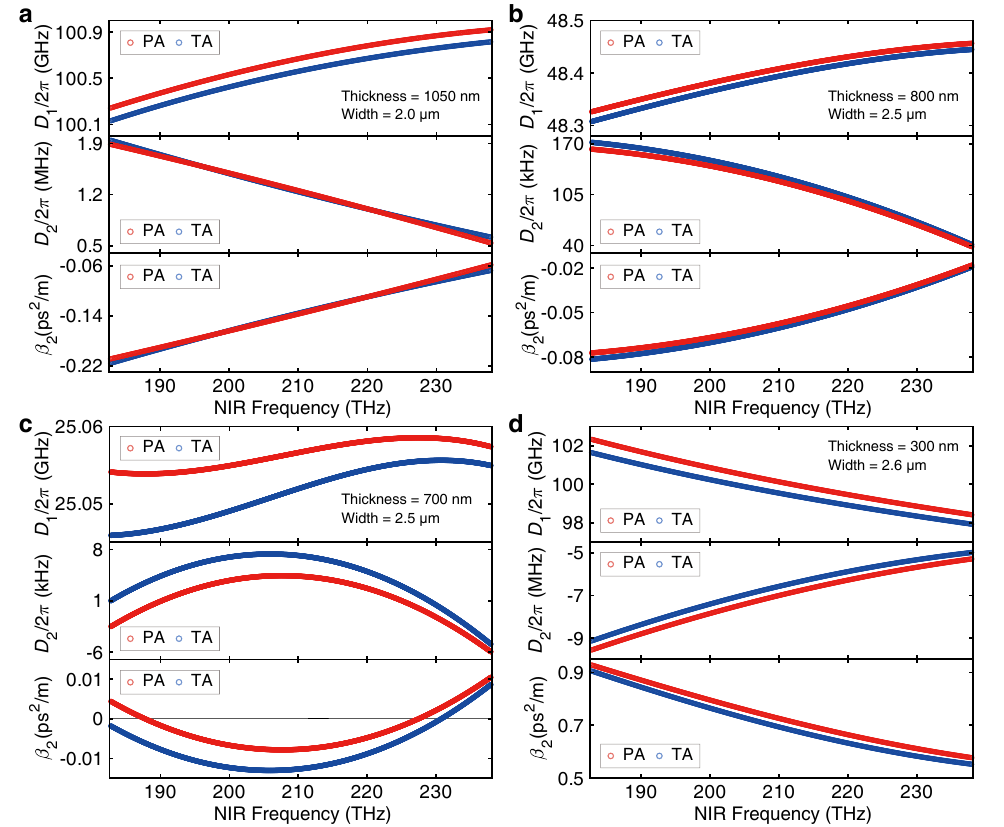}
\caption{
\textbf{Near-infrared characterization of Si$_3$N$_4$ microresonators' dispersion.}
The plots display the measured free spectral range ($D_1/2\pi$), second-order dispersion ($D_2/2\pi$), and derived group-velocity dispersion ($\beta_2$).
Red and blue data points correspond to PA and TA states, respectively. 
The measurements are performed on microresonators with the following dimensions (thickness $\times$ width):
\textbf{a}. 1050 nm $\times$ 2.0 $\mu$m;
\textbf{b}. 800 nm $\times$ 2.5 $\mu$m;
\textbf{c}. 700 nm $\times$ 2.5 $\mu$m; and
\textbf{d}. 300 nm $\times$ 2.6 $\mu$m.
}
\label{Fig:S7}
\end{figure*}

Using the VSA setup described above, we characterized the intrinsic loss rate $\kappa_0/2\pi$ and external coupling rate $\kappa_{\rm{ex}}/2\pi$ for microresonators with various cross-sectional geometries in both PA and TA states.
The results are summarized in Fig. \ref{Fig:S6}.
In the PA state, all microresonators exhibit a strong absorption peak around 197--199 THz, attributed to the first overtone of N--H bond stretching vibrations. 
After TA, these absorption peaks are significantly suppressed. 
Additionally, $\kappa_{\rm{ex}}/2\pi$ remains largely unchanged between the two states, indicating that the geometry of the microresonators is stable during the annealing process.
Furthermore, we characterized the integrated dispersion profiles of the microresonators shown in Fig. \ref{Fig:S6}.
The group velocity dispersion (GVD), $\beta_2$, is derived from the measured $D_1$ and $D_2$ values according to Eq. \ref{Eq:D2}.
The FSR ($D_1/2\pi$), second-order dispersion ($D_2/2\pi$), and derived GVD ($\beta_2$) are plotted in Fig. \ref{Fig:S7}.
Microresonators with various cross-sectional geometries all exhibit a decrease in FSR after TA, indicating an increase in the refractive index, which is consistent with the discussion in Supplementary Note 3.
Finally, the measured $\beta_2$ reveals that the direction of the zero-dispersion wavelength shift (red- or blue-shift) is governed by geometric dispersion, a factor not accounted for in Fig. 1b of the main text.

\clearpage

\section{Characterization of microresonators in the near-visible}
\begin{figure*}[b!]
\renewcommand{\figurename}{Supplementary Figure}
\centering
\includegraphics{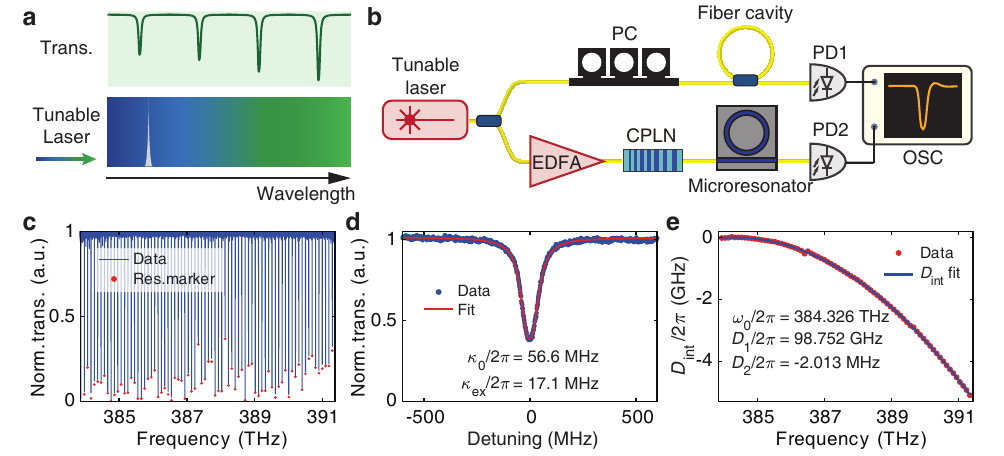}
\caption{
\textbf{Experimental setup to characterize microresonators' loss and dispersion in the near-visible range.} 
\textbf{a,b}. Schematic principle and experimental setup. 
The near-infrared (NIR) tunable laser is power amplified by an erbium-doped fiber amplifier (EDFA).
A chirped periodically poled lithium niobate (CPLN) waveguide is employed to frequency-double the NIR laser and generate the second-harmonic (SH) chirping laser. 
Constrained by the bandwidths of the EDFA and the CPLN waveguide, the wavelength range of the SH laser is 766--781 nm (383.8--391.3 THz). 
The frequencies of both chirping lasers is calibrated by an FSR-calibrated fiber cavity that serves as a time-domain "frequency ruler".
The SH laser output is coupled into optical microresonators under test. 
The transmission spectrum of the microresonator is recorded during the laser chirping.
PC, polarization controller.
PD, photodetector. 
OSC, oscilloscope.
\textbf{c}. A typical microresonator transmission spectrum in the near-visible range. 
Each resonance is identified and fitted as shown in \textbf{d}, whose intrinsic loss $\kappa_0$/2$\pi$ is extracted and analyzed. 
\textbf{e}. Measured integrated microresonator's dispersion profile in the near-visible range.
}
\label{Fig:780setup}
\end{figure*}

The characterization of Si$_3$N$_4$ microresonators in the near-visible regime is performed using a VSA operating in the 766--781 nm wavelength range\cite{Shi:25}.
The principle and experimental setup is illustrated in Fig. \ref{Fig:780setup}a and b, respectively. 
A widely tunable, mode-hop-free, ECDL (Santec TSL 570) in the NIR band is split into two branches.
One branch is directed to a pre-calibrated fiber cavity with a quasi-equidistant grid of fine resonances\cite{Luo:24}, which serves as a time-domain frequency reference. 
The other branch servers as the pump source and is amplified by an erbium-doped fiber amplifier (EDFA) and frequency-doubled via a chirped periodically poled lithium niobate (CPLN) waveguide to generate the second-harmonic (SH) laser.
The output SH laser is then coupled into a microresonator under test.
Figure \ref{Fig:780setup}c presents a typical transmission spectrum in the near-visible wavelength range.
Each resonance within the measurement span is identified and fitted \cite{Li:13} using Eq. \ref{fitres} to extract the intrinsic loss rate $\kappa_0/2\pi$ and the external coupling rate $\kappa_{\text{ex}}/2\pi$, as detailed in Fig. \ref{Fig:780setup}d.
Furthermore, the integrated dispersion profile $D_\text{int}/2\pi$ of the microresonators is characterized using this near-visible VSA setup and fitted according to Eq. \ref{dispersion}.
A representative $D_{\text{int}}$ profile for a 300-nm-thick Si$_3$N$_4$ microresonator is shown in Fig. \ref{Fig:780setup}e.

\begin{figure*}[t!]
\renewcommand{\figurename}{Supplementary Figure}
\centering
\includegraphics{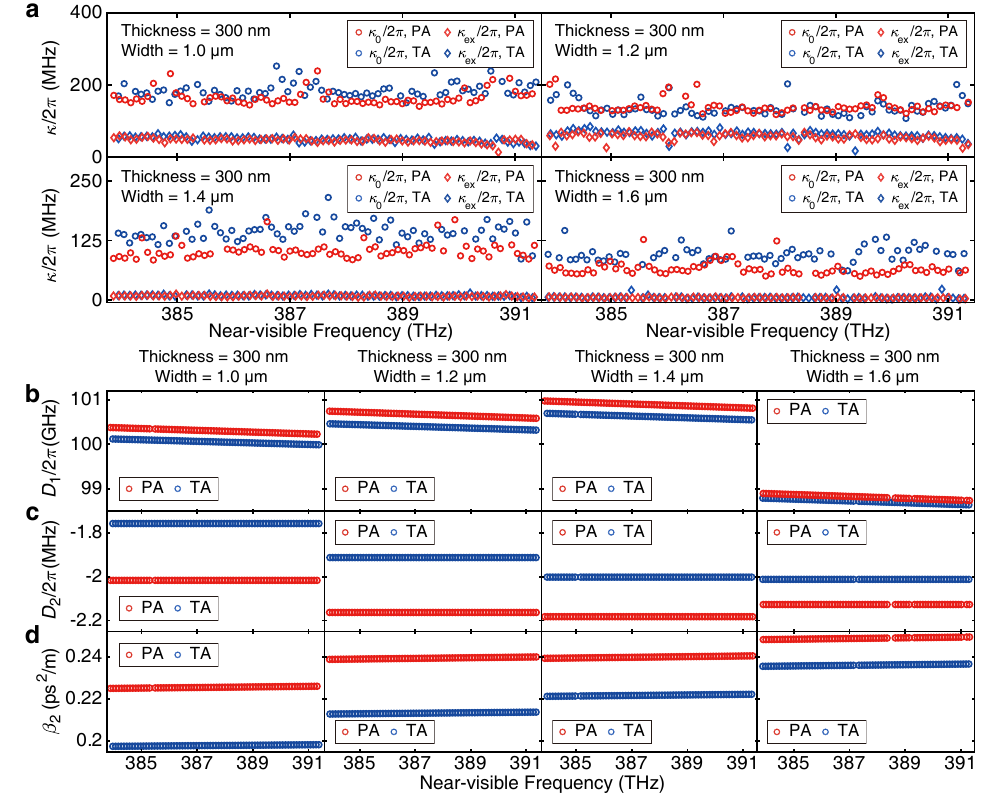}
\caption{
\textbf{Experimental characterization of Si$_3$N$_4$ microresonators in the near visible range.}
Red and blue data represent the partially annealed (PA) and thorough annealed (TA) states, respectively. 
\textbf{a}. Intrinsic loss ($\kappa_0/2\pi$, circles) and external coupling strength ($\kappa_{\text{ex}}/2\pi$, diamonds) for microresonators with width 1.0 $\mu$m, 1.2 $\mu$m, 1.4 $\mu$m and 1.6 $\mu$m, thickness 300 nm. 
\textbf{b}. Free spectral range ($D_1/2\pi$). 
\textbf{c}. Second‑order dispersion ($D_2/2\pi$). 
\textbf{d}. Derived group‑velocity dispersion ($\beta_2$). 
The measurements are performed on microresonators with width 1.0 $\mu$m, 1.2 $\mu$m, 1.4 $\mu$m and 1.6 $\mu$m, thickness 300 nm.
}
\label{Fig:780kappa_D}
\end{figure*}

Using the near-visible VSA setup described above, we characterized the $\kappa_0/2\pi$ and $\kappa_{\rm{ex}}/2\pi$ for 300-nm-thick microresonators with widths ranging from 1.0 to 1.6 $\mu$m in both PA and TA states.
The results are summarized in Fig. \ref{Fig:780kappa_D}a.
For the PA state, the most probable $\kappa_0/2\pi$ values were 154, 130, 95, and 56 MHz (corresponding to increasing widths), which subsequently increased to 178, 130, 143, and 84 MHz after TA, respectively.
This random 0–48 MHz increase in loss is likely attributed to crystallization, thermal stress, or micro-cracking within the Si$_3$N$4$ waveguides induced by the high-temperature ($>1200~^\circ\text{C}$) annealing process \cite{Henry:87}.
Conversely, $\kappa_{\text{ex}}/2\pi$ remains largely unchanged between the PA and TA states, indicating high geometric stability of the microresonators during annealing.
Furthermore, we characterized the integrated dispersion in the near-visible regime of the microresonators.
The free spectral range (FSR, $D_1/2\pi$), second-order dispersion ($D_2/2\pi$), and derived group velocity dispersion (GVD, $\beta_2$) are plotted in Fig. \ref{Fig:780kappa_D}b–d, respectively.
Across all tested geometries, the microresonators consistently exhibit a decreased FSR, an increased $D_2/2\pi$, and a decreased $\beta_2$ after TA.
These trends indicate an elevated refractive index and reduced material dispersion, consistent with the analyses in Supplementary Note 3.

\clearpage
\section{Characterization of microresonators in the MIR}
\begin{figure*}[b!]
\renewcommand{\figurename}{Supplementary Figure}
\centering
\includegraphics{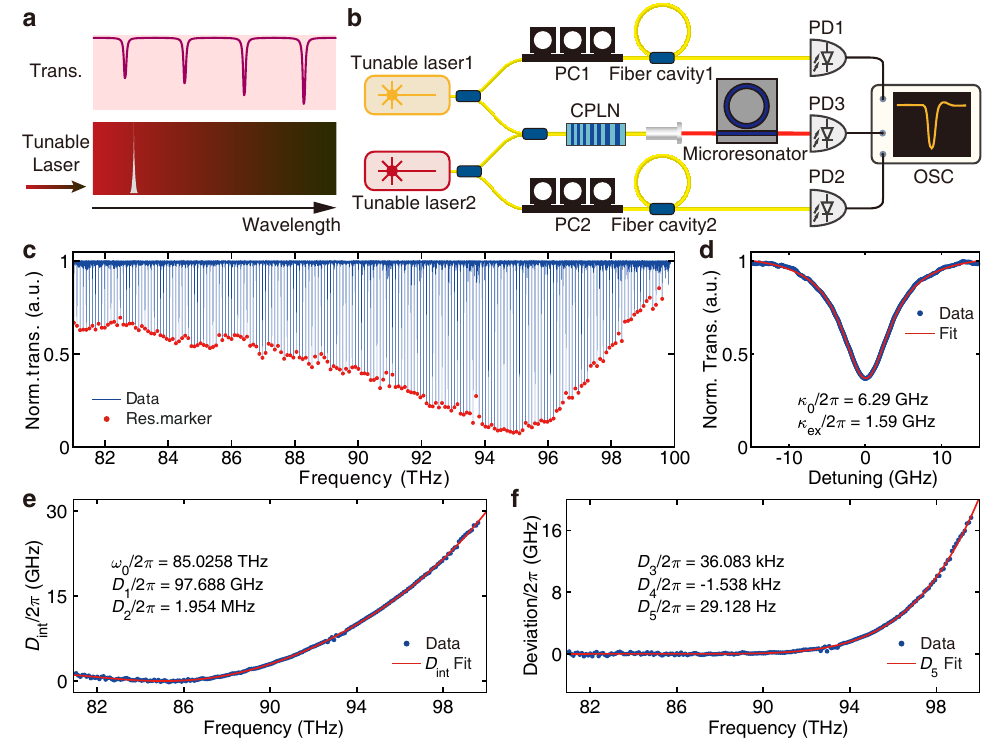}
\caption{
\textbf{Experimental setup to characterize microresonators' loss and dispersion in the MIR.} 
\textbf{a,b}. Principle and experimental setup.
A tunable laser chirping from 3001 to 3711 nm (80.8--99.9 THz) is generated via difference frequency generation (DFG) in a chirped periodically poled lithium niobate (CPLN) waveguide, pumped by two lasers at 1035--1086 nm and 1536--1580 nm. 
The instantaneous frequency of the DFG output is calibrated in real time by stabilizing the two pump lasers with two pre-calibrated fiber cavities separately. 
The DFG output is coupled into the optical microresonators under study. 
PC, polarization controller. 
PD, photodetector. 
OSC, oscilloscope. 
\textbf{c}. A typical microresonator transmission spectrum. 
Each resonance is identified and fitted as shown in \textbf{d}, whose $\kappa_0$/2$\pi$ and $\kappa_{\text{ex}}/2\pi$ are extracted and marked. 
\textbf{e}. Measured integrated microresonator's dispersion profile and fit up to the fifth order, as shown in \textbf{f}. 
}
\label{Fig:S8}
\end{figure*}

\begin{figure*}[t!]
\renewcommand{\figurename}{Supplementary Figure}
\centering
\includegraphics{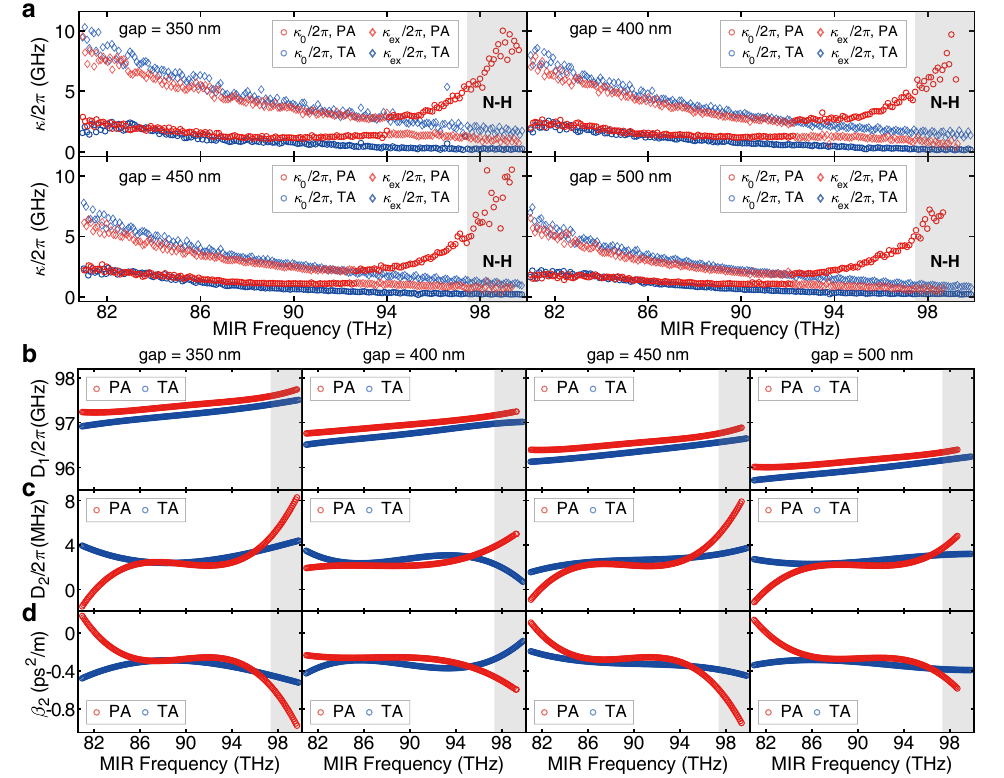}
\caption{
\textbf{Experimental characterization of Si$_3$N$_4$ microresonators in the MIR.}
Red and blue data represent the partially annealed (PA) and thorough annealed (TA) states, respectively. 
Absorption features induced by N--H bonds are indicated by gray shading. 
\textbf{a}. Intrinsic loss ($\kappa_0/2\pi$, circles) and external coupling strength ($\kappa_{\text{ex}}/2\pi$, diamonds) for coupling gaps of 350 nm, 400 nm, 450 nm, and 500 nm. 
\textbf{b}. Free spectral range ($D_1/2\pi$). 
\textbf{c}. Second‑order dispersion ($D_2/2\pi$). 
\textbf{d}. Derived group‑velocity dispersion ($\beta_2$). 
}
\label{Fig:S9}
\end{figure*}

The MIR characterization of the Si$_3$N$_4$ microresonators presented in this work is performed using a self-constructed MIR VSA based on difference-frequency generation (DFG)\cite{Shi:26}. 
The system in Fig. \ref{Fig:S8}b employs two widely tunable NIR ECDLs as primary pumps, with tunable laser 1 spanning 1035-1086 nm (Toptica CTL 1050) and tunable laser 2 spanning 1536-1580 nm (Santec TSL 570). 
Their instantaneous frequencies are rigorously calibrated against fiber ring cavities\cite{Luo:24, Shi:26} and anchored to absolute hyperfine transitions of Iodine, Rubidium, and Potassium\cite{Shi:25, Shi:26}. 
To enable efficient nonlinear conversion, the laser outputs are amplified to 4 W and 5.6 W by a ytterbium-doped fiber amplifier (YDFA, Precilasers FA-SF-1064-15-CW) and an EDFA (Precilasers FA-SF-1550-5-CW), respectively. 
The two amplified lights are then combined via a wavelength division multiplexer (WDM) and coupled into a fiber-pigtailed magnesium-oxide-doped CPLN waveguide module, which is specifically designed for broadband phase-matching\cite{Shi:26}. 
Driven by these frequency-calibrated pumps, the DFG process produces a tunable MIR source with an inherently well-defined frequency. 
After passing through a germanium window filter to suppress the residual pump, this source exhibits the key characteristics essential for high-precision resonator spectroscopy: a continuous, mode-hop-free tuning range of 7.49 THz per sweep---extendable to 19.1 THz (80.8--99.9 THz, 3001--3711 nm) by scan stitching---which enables broad coverage of dispersion profiles; an output power up to 12.6 mW, sufficient for achieving a high signal-to-noise ratio across the entire band; and a narrow dynamic linewidth averaging 242 kHz (100 µs integration time), crucial for precisely resolving individual cavity resonances\cite{Shi:26}. 

For characterization, the generated MIR light is then collimated by a calcium fluoride lens, and coupled into / out of the Si$_3$N$_4$ waveguide using E-coated black diamond-2 lenses. The subsequent data acquisition and processing workflow followed the method described for the NIR band---albeit applied here to the MIR spectra. 
The transmitted signal was recorded with a photodetector and an oscilloscope. 
Figure \ref{Fig:S8}c displays the normalized broadband transmission spectrum of the same cavity characterized in the main text, measured under the PA (high residual hydrogen) condition. 
By fitting each resonance dip in the spectrum with a Lorentzian function, the intrinsic loss and external coupling strength at respective wavelengths were extracted, as depicted in Fig. \ref{Fig:S8}d. 
The measured $D_{\text{int}}$ profile in Fig. \ref{Fig:S8}e was fitted up to the fifth order ($D_5$), yielding parameters including $\omega_0$ and $D_1$--$D_5$. The excellent agreement between the $D_5$ fit and the data in Fig. \ref{Fig:S8}f demonstrates that the broad measurement band enables accurate dispersion calibration up to the fifth order.

Based on this methodology, the extracted $\kappa_0/2\pi$ and $\kappa_{\text{ex}}/2\pi$ are compared in Fig. \ref{Fig:S9}a for microresonators with width 3.0 $\mu$m and coupling gaps of 350, 400, 450, and 500 nm under both PA and TA states. 
The comparison reveals that $\kappa_{\text{ex}}/2\pi$ remains nearly identical for one microresonator, whereas the sharp enhancement in $\kappa_0/2\pi$ near 100 THz under the PA state is eliminated by TA, directly confirming the efficacy of annealing in suppressing N--H bond absorption. 
The free spectral range ($D_1/2\pi$), second-order dispersion ($D_2/2\pi$), and derived group-velocity dispersion ($\beta_2$) are respectively plotted in Fig. \ref{Fig:S9}b--d. 
Within the gray-shaded regions indicating N--H absorption, all three dispersion parameters show clear differences between PA and TA states, providing direct evidence that N--H absorption alters the cavity's effective refractive index and dispersion, in accordance with the KK relations.

\clearpage

\section{Dimensional check of silicon nitride films and waveguides}

\begin{figure*}[b!]
\renewcommand{\figurename}{Supplementary Figure}
\centering
\includegraphics{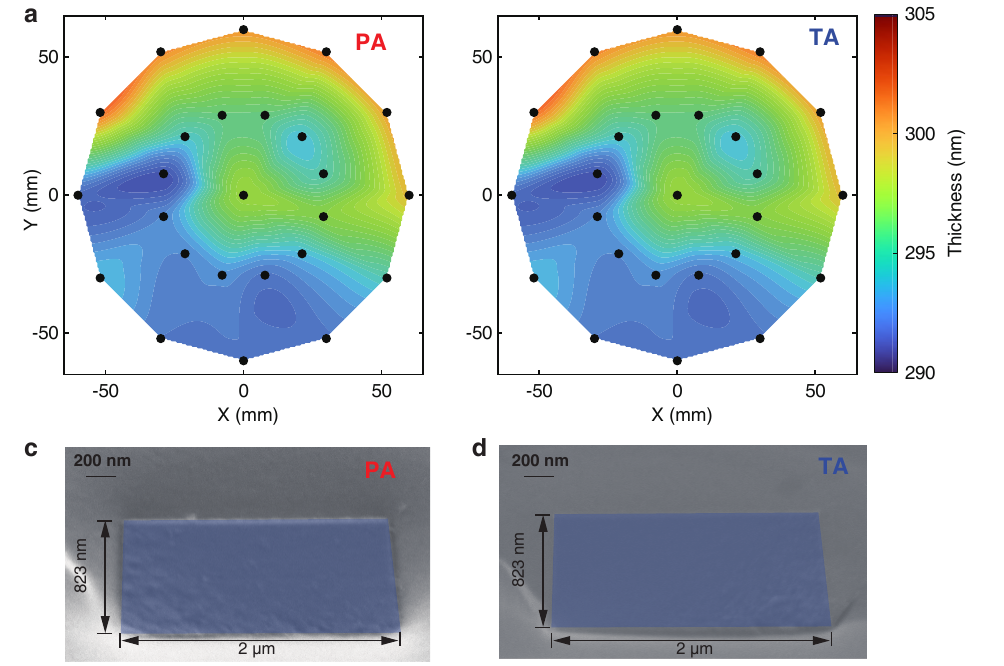}
\caption{
\textbf{Dimensional characterization of Si$_3$N$_4$ films and waveguides.} 
\textbf{a,b}. Ellipsometry measurements of the film thickness after PA and TA, respectively. 
\textbf{c,d}. SEM cross-sectional measurements of the waveguide height after PA and TA, respectively.
}
\label{Fig:S10}
\end{figure*}

To systematically verify the dimensional stability of Si$_3$N$_4$ after annealing, we conducted supplementary experiments at both the thin-film and waveguide device levels.

\vspace{0.2cm}

\textbf{Thermal Stability of Film Thickness:}
A LPCVD Si$_3$N$_4$ film (initial thickness 290.1--290.5 nm) was subjected to a two-stage annealing process. 
First, following a PA step, spectroscopic ellipsometry (Woollam M2000) measurements revealed a thickness reduction to 284.3--285.1 nm, as shown in Fig. \ref{Fig:S10}a. 
Subsequently, the same sample was subjected to TA at a temperature above 1200$^\circ$C.
Ellipsometry measurements performed after this second step confirmed that the film thickness remained stable at 284.3--285.1 nm, as shown in Fig.\ref{Fig:S10}b, with no detectable variation.

\vspace{0.2cm}

\textbf{Thermal Stability of Waveguide Structure:}
To validate these findings at the device level, a subtractively fabricated Si$_3$N$_4$ waveguide was subjected to the same two-step annealing sequence. 
After PA, SEM (Guoyi 5000pro) measurements revealed a waveguide height of 823 nm, as shown in Fig.\ref{Fig:S10}c. Following the subsequent TA ($>$1200$^\circ$C), repeated SEM characterization confirmed that the waveguide height remained precisely 823 nm, with no measurable dimensional change, as shown in Fig.\ref{Fig:S10}d.

\begin{table*}[b!]
\renewcommand{\tablename}{Supplementary Table}
\centering
\caption{\textbf{Evolution of Si$_3$N$_4$ film thickness measured by ellipsometry.}
Measurements compare the thickness of the as-deposited (unannealed), PA, and TA states.}

\setlength{\tabcolsep}{22pt} 
\begin{tabular}{c|c|c|c} 
\hline
Measured Points &  \begin{tabular}{@{}c@{}} Unannealed \end{tabular} & \begin{tabular}{@{}c@{}} Partially annealed \end{tabular} &  \begin{tabular}{@{}c@{}} Thoroughly annealed \end{tabular}\\
\hline
Point 1 & 290.1 nm & 284.3 nm & 284.3 nm\\
Point 2 & 290.5 nm & 285.1 nm & 285.1 nm\\
\hline
\end{tabular}
\label{tab:Ellipsometry}
\end{table*}

\clearpage

\section{Microcomb generation setup and data}
\begin{figure*}[b!]
\renewcommand{\figurename}{Supplementary Figure}
\centering
\includegraphics{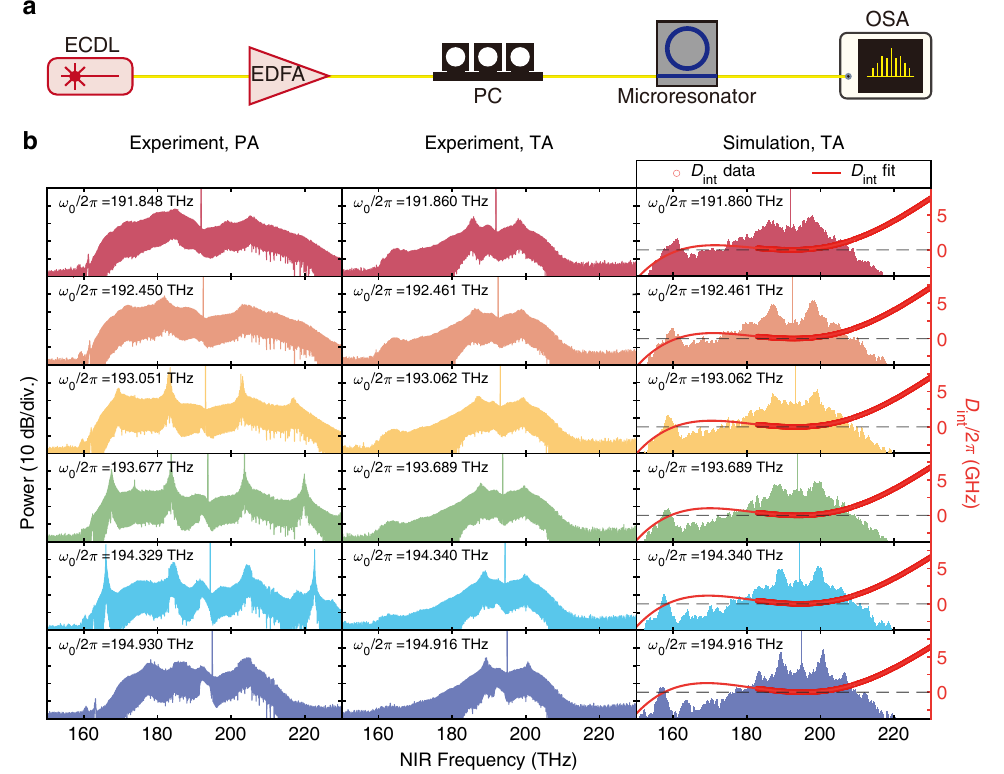}
\caption{
Microcomb generation in a PA/TA Si$_3$N$_4$ microresonator.
\textbf{a}. The experimental setup to generate the microcombs.
ECDL, external cavity diode laser.
EDFA, erbium-doped fiber amplifier.
PC, polarization controller.
OSA, optical spectrum analyzer.
\textbf{b}. Left column: the measured optical spectra of the microcombs in a PA Si$_3$N$_4$ microresonator.
Middle column: the measured optical spectra of the microcombs in the Si$_3$N$_4$ microresonator after TA.
Right column: the simulated optical spectra according to the parameter values of the TA microresonator from the experimental measurement and fitting.
}
\label{Fig:S11}
\end{figure*}

The experimental setup for broadband microcomb generation is shown in Supplementary Fig.~\ref{Fig:S11}a.
An ECDL (Toptica CTL 1550) is power-amplified by an EDFA (Connet MFAS-C-EY-B-5000).
The laser polarization is manipulated by a polarization controller (PC) before coupling into the Si$_3$N$_4$ microresonator chip.
The optical spectrum of the laser coupled out from the chip is measured by an optical spectrum analyzer (OSA, Yokogawa AQ6370D).
The experimental results are shown in Supplementary Fig.~\ref{Fig:S11}b left and middle columns.
The optical spectra of the PA (the left column) and the TA (the middle column) Si$_3$N$_4$ microresonators are measured with varying the pumped microresonator resonance from about 191.8 THz to about 194.9 THz.
For the PA Si$_3$N$_4$ microresonator, the integrated dispersion is near zero at the pumped modes.
After TA, the integrated dispersion has another zero point where it switches from anomalous to normal dispersion and enhanced comb lines are observed in the modulation instability (MI) comb, as described in the main text, also shown here in Supplementary Fig.~\ref{Fig:S11}b middle column.

To confirm the experimental observations, we perform simulations on the MI comb spectra using the Lugiato-Lever Equation (LLE) model\cite{Lugiato:87,Lugiato:18}:
\begin{equation}
\frac{\partial a}{\partial\tau} = -a-i\mathcal F[\zeta_\mu a_\mu](\tau,\phi) + i|a|^2a+f,
\label{eq:LLE}
\end{equation}
where $a=\mathcal F[a_\mu](\tau,\phi)=\sum_\mu a_\mu(\tau)\text e^{i\mu\phi}$ with $a_\mu$ representing the laser amplitude at the $\mu$-th mode and $\phi$ being the azimuthal angle of the microresonator,
$\zeta_\mu= (\omega_0 - \omega_\mathrm{p} + D_n\mu^n/n!)/(\kappa/2)$ with $\omega_0$ ($\omega_\mathrm{p}$) being the microresonator resonance (the pump laser) frequency, $D_{\rm{n}}$ ($n = 2, 3, 4$ in simulation) representing the microresonator dispersion and $\kappa/2\pi$ being the loaded linewidth of the microresonator,
and $f=\sqrt{8g\kappa_\mathrm{ex}P_\mathrm{in}/(\hbar \omega_0 \kappa^3)}$ with $g$ being the nonlinear coefficient, $\kappa_\mathrm{ex}$ being the external coupling efficiency and $P_\mathrm{in}$ being the on-chip pump laser power.
The simulated MI comb spectra are shown in Supplementary Fig.~\ref{Fig:S11}b right column, with the same pumped microresonator resonances as for the TA microresonator and the dispersion values of the TA microresonator from data fitting.
The complete simulation parameter values are displayed in Supplementary Table~\ref{Table:simu}.

\begin{table*}[t!]
\renewcommand\tablename{Supplementary Table}
\renewcommand\arraystretch{1.5}
\center
\caption{Parameter values in simulation}
\begin{tabular}{c|c|c|c|c|c|c|c|c}
\hline
$\omega_0/2\pi$~(THz) &  $\kappa/2\pi$~(MHz) & $\kappa_\mathrm{ex}/2\pi$~(MHz) & $D_1/2\pi$~(GHz) & $D_2/2\pi$~(kHz) & $D_3/2\pi$~(Hz) & $D_4/2\pi$~(mHz) & $g$~(rad/s) & $P_\mathrm{in}$~(mW)\\
\hline
191.860 & 43 & 82 & 25.047 & 5.058 & 8.240 & -15 & 2.177 & 200\\
192.461 & 43 & 82 & 25.047 & 5.252 & 7.880 & -15 & 2.191 & 200\\
193.062 & 43 & 82 & 25.047 & 5.437 & 7.530 & -15 & 2.205 & 200\\
193.689 & 43 & 82 & 25.047 & 5.620 & 7.160 & -15 & 2.219 & 200\\
194.340 & 43 & 82 & 25.048 & 5.802 & 6.780 & -15 & 2.234 & 200\\
194.916 & 43 & 82 & 25.048 & 5.954 & 6.440 & -15 & 2.247 & 200\\
\hline
\end{tabular}
\label{Table:simu}
\end{table*}

\clearpage

\section*{Supplementary References}
\bigskip
\bibliographystyle{apsrev4-1}
\bibliography{bibliography}